\documentclass[floatfix,aps,twocolumn,nofootinbib,superscriptaddress,preprintnumbers,prd,10pt]{revtex4-2}

\usepackage[utf8]{inputenc}  
\usepackage{float}
\usepackage{colortbl}
\usepackage{amsmath,amssymb,amsfonts,bm,bbm,slashed}
\usepackage{dsfont} 
\usepackage{graphicx}
\usepackage{enumitem}
\usepackage{arydshln}
\usepackage{bbold}
\usepackage{braket}
\usepackage{multirow}
\usepackage{booktabs}
\usepackage{soul}
\usepackage[colorlinks=true,breaklinks=true]{hyperref}
\hypersetup{allcolors=[rgb]{0.0 0.0 0.75},linkcolor=[rgb]{0.75 0.05 0.05}}
\usepackage[dvipsnames]{xcolor}
\usepackage{mathtools}
\usepackage{bbold}
\usepackage{upgreek}
\long\def\exclude#1{}
\usepackage{orcidlink}
\usepackage{microtype}
\usepackage{feynmp-auto}
\usepackage{mathrsfs}
\usepackage{euscript}
\usepackage[normalem]{ulem}
\makeatletter
\g@addto@macro\bfseries{\boldmath}
\makeatother

\newcommand{\bp}{{\bf p}}

\newcommand{\bk}{{\bf k}}
\newcommand{\bq}{{\bf q}}

\newcommand{\bw}{{\bf w}}

\newcommand{\bn}{{\bf n}}

\newcommand{\tp}{{\tilde{p}}}

\newcommand{\tq}{{\tilde{q}}}

\newcommand{\cm}{{\rm cm}}
\newcommand{\F}{{\rm F}}
\renewcommand{\P}{{\rm P}}
\newcommand{\g}{{\rm g}}
\newcommand{\keV}{{\rm keV}}
\newcommand{\MeV}{{\rm MeV}}

\allowdisplaybreaks

\begin{document}

\title{Production of Leptophilic Bosons in Ultradegenerate Relativistic Matter}
 
\author{Damiano F.\ G.\ Fiorillo \orcidlink{0000-0003-4927-9850}}
\affiliation{Istituto Nazionale di Fisica Nucleare (INFN), Sezione di Napoli, Complesso Universitario di Monte Sant'Angelo, Via Cintia, 80126 Napoli, Italy}
\affiliation{Gran Sasso Science Institute (GSSI), Viale Francesco Crispi 7, 67100 L’Aquila, Italy}

\author{Alessandro~Lella~\orcidlink{0000-0002-3266-3154}}
\affiliation{Dipartimento di Fisica e Astronomia, Università degli Studi di Padova, Via Marzolo 8, 35131 Padova, Italy}
\affiliation{Istituto Nazionale di Fisica Nucleare (INFN), Sezione di Padova, Via Marzolo 8, 35131 Padova, Italy}%

\author{Georg~G.~Raffelt~\orcidlink{0000-0002-0199-9560}}
\affiliation{Max-Planck-Institut f\"ur Physik, Boltzmannstra\ss e~8, 85748 Garching, Germany}

\author{Nud{\v z}eim~Selimovi{\'c}~\orcidlink{0000-0003-3780-1437}}
\affiliation{Dipartimento di Fisica e Astronomia, Università degli Studi di Padova, Via Marzolo 8, 35131 Padova, Italy}
\affiliation{Istituto Nazionale di Fisica Nucleare (INFN), Sezione di Padova, Via Marzolo 8, 35131 Padova, Italy}%

\author{Edoardo~Vitagliano~\orcidlink{0000-0001-7847-1281}}
\affiliation{Dipartimento di Fisica e Astronomia, Università degli Studi di Padova, Via Marzolo 8, 35131 Padova, Italy}
\affiliation{Istituto Nazionale di Fisica Nucleare (INFN), Sezione di Padova, Via Marzolo 8, 35131 Padova, Italy}

\begin{abstract}
\vspace{3mm}
Neutron stars (NSs) are powerful factories for new particles with masses up to the 100~keV range. These compact stars contain significant amounts of protons, electrons, and muons. We deduce the emission rates for new scalar, vector, and pseudoscalar bosons coupling to electrons and muons. For muonic vector bosons, in addition to direct emission from leptons in incoherent scatterings with the medium, a competing channel is the coherent excitation of the lepton population electromagnetically induced by incoherent hadronic scatterings---an effective in-medium coupling. For muonic scalar bosons, emission from the effective in-medium coupling to electrons is always dominant due to their relativistically boosted emission. Pseudoscalars can only be emitted through incoherent leptonic scatterings. If protons are superconducting, the scalar and vector energy loss rates scale as $T^4$, the pseudoscalar ones as $T^6$, to be compared with $T^8$ for neutrino losses by the modified Urca (MUrca) process. For normal-conducting protons, the damping of transverse photons implies instead scalings with a power \smash{$T^{11/3}$} for scalars, and  \smash{$T^{17/3}$} for pseudoscalars; vectors, emitted by nonrelativistic protons in the nonsuperconducting regime, still scale as $T^4$. Our results can be used to constrain the leptophilic coupling strengths through observed NS cooling ages. 
\vspace{3mm}
\end{abstract}

\maketitle
\tableofcontents

\section{Introduction\label{sec:intro}}

After a neutron star (NS) has formed in a stellar collapse, its subsequent evolution consists mainly of cooling via neutrino emission, and, after some $10^5$~years, surface photon emission \cite{Gamow:1941gis, Yakovlev:2000jp, Yakovlev:2004iq, Page:2005fq}. In a seminal paper in 1984, shortly after ``invisible axions'' had been proposed, Iwamoto \cite{Iwamoto:1984ir} showed that axion production could play a major role in NS cooling, and, conversely, that cooling measurements could constrain the axion-nucleon interaction strength. With the advent of imaging X-ray satellites, thermal NS radiation has been clearly observed, leading to competitive axion constraints \cite{Keller:2012yr, Leinson:2014ioa, Sedrakian:2015krq, Sedrakian:2018kdm, Hamaguchi:2018oqw, Leinson:2021ety, Beznogov:2018fda, Leinson:2019cqv, Buschmann:2021juv, Hamaguchi:2025ztd}.

A few years later, the observation of neutrinos from  SN~1987A allowed for a complementary probe based on the cooling of the nascent NS. The axion bounds are comparable to those from late NS cooling~\cite{Raffelt:1987yt, Turner:1987by, Mayle:1987as, Raffelt:2006cw, Carenza:2019pxu,Fiorillo:2025gnd}, while reaching higher masses due to the hotter core. Within uncertainties, the constraints from both arguments are identical, specifically $g_{aNN}\alt10^{-9}$ for the axion-nucleon coupling \cite{Caputo:2024oqc, Carenza:2024ehj}. The nascent NS is opaque to neutrinos and the bound essentially arises from comparing surface neutrino emission with volume axion emission, whereas for NS cooling, one compares with photon surface emission at a much later stage. Therefore, the similarity of the bounds is mostly fortuitous.

In an ultradegenerate medium, axions primarily emerge from nucleon bremsstrahlung, $NN\to NNa$, leading to an axion luminosity that scales as $L_a\propto T^6$ \cite{Iwamoto:1984ir}, to be compared with $L_\nu\propto T^8$ for neutrinos from the MUrca process. On the other hand, surface photon cooling scales as $L_\gamma\propto T^2$ in terms of the internal NS temperature, so eventually $L_\gamma$ takes over. Once this happens, the temperature drops much more rapidly. Therefore, as $T$ drops, axion emission increases relative to neutrinos until the instant when neutrino and photon cooling are roughly comparable. It is this epoch in NS evolution that leads to the strongest constraints on axion emission; it is reached at some $10^5$~years after formation, when $T\simeq10^8~{\rm K}=8.6~{\rm keV}$.

As some of us have previously argued~\cite{Fiorillo:2025zzx}, these scalings suggest that for any new emission process with a scaling $T^p$ that is weaker than for axions, $p<6$, NS cooling is a more sensitive probe than SN~1987A. In particular, this applies to baryophilic scalar or vector bosons, $\phi$, with
$L_\phi\propto T^4$, leading to a constraint $g_{\phi NN}\alt0.5\times10^{-13}$, once more based on nucleon bremsstrahlung~\cite{Fiorillo:2025zzx}. The weaker scaling arises because scalars couple essentially to the nucleon number, whereas pseudoscalars (axions) to the spin. For nonrelativistic nucleons, an extra factor $\omega^2$ arises (radiation energy $\omega$) and thus an extra power $T^2$ in the energy-loss rate.

While SN cores and NSs are particularly powerful for testing baryo\-philic interactions, one can test couplings to electrons more sensitively with less extreme stars, notably white dwarfs \cite{Raffelt:1985nj, Isern:2022vdx, Fleury:2025ahw, Alberino:2026yxi, Bottaro:2023gep} and the cores of low-mass red giants \cite{Dearborn:1985gp, Grifols:1988fv, Raffelt:1994ry, Hardy:2016kme, Capozzi:2020cbu, Straniero:2020iyi}. On the other hand, for nonuniversal interactions, and notably muonic couplings, SNe and NSs once more stick out because they host large muon populations. The recent availability of muonic SN models has therefore provided the basis for using SN~1987A cooling and all past SNe to constrain muon-philic (hereafter muonic) boson interactions~\cite{Bollig:2020xdr, Croon:2020lrf, Caputo:2021rux, Akita:2023iwq, Blinov:2025aha, Ferreira:2025qui,Das:2026uam}. However, this is another case where NS cooling must be more sensitive, in analogy to baryophilic bosons.

With this motivation in mind, we study the emission rates for muonic bosons from the ultradegenerate medium of a NS, where the charged-particle abundances in a reference case are $Y_p=0.10$ protons per baryon, compensated by $Y_e=0.07$ electrons and $Y_\mu=0.03$ muons. This problem turns out to be surprisingly intricate. As in many other environments, muonic bosons can be emitted from individual muons scattering incoherently with the medium. The target for incoherent scattering, however, depends on the state of the medium, and specifically on whether the protons are superconducting. If this is the case, the gap at the Fermi surface causes a depletion of free proton excitations, effectively removing them as targets for muon scattering. Nonsuperconducting protons instead provide important targets for muonic boson emission. The angular pattern of radiation is quite characteristic, since the mildly relativistic nature of muons---with Fermi momenta of 100--200~MeV and a mass $m_\mu=105.66\,\mathrm{MeV}$---boosts the emissivity. For completeness, we also consider bosons coupling to electrons, for which relativistic effects are even more pronounced.

Our calculations parallel, in many ways, earlier works on NS electron and muon conductive opacities \cite{Gnedin:1995lgf,Shternin:2007ee} and previous calculations of neutrino pair emission by electromagnetic bremsstrahlung processes \cite{Kaminker:1999sd, Jaikumar:2005gm, Shternin:2025swb}. In the earliest of these papers \cite{Gnedin:1995lgf, Kaminker:1999sd}, screening in electromagnetic collisions, in the normal-conducting phase, was implemented in terms of Debye screening of longitudinal fields, leading to a neutrino pair emission rate that scaled as $T^8$. Later, it was realized that actually transverse plasmons, the least damped modes, dominate the collision rate \cite{Shternin:2007ee, Shternin:2006uq}, leading to a bremsstrahlung scaling of $T^{23/3}$ when protons are normal-conducting \cite{Shternin:2025swb}.

Another complication was first identified for neutrino pair emission by the vector-current interaction \cite{Leinson:1997zt, Leinson:1999ut, Leinson:2000ks, Leinson:2000un} and also applies to our case of vector bosons, i.e., the effective in-medium renormalization of the vector charge, caused by the corresponding medium response that causes screening.  This effect is analogous to the reduction in the emission of a boson that mixes with the longitudinal photon, see e.g.~Eq.~(2.6) in Ref.~\cite{Hardy:2016kme}. We unify the treatment of this effect, which appears under different guises in these papers, and provide a simple physical interpretation in terms of the medium screening of the current that sources the vector boson. As a striking result, muonic scalar and vector bosons are mainly radiated by electromagnetic electron scatterings and nuclear proton collisions, respectively. These collisions accelerate electrons and protons over timescales of the order of the inverse temperature, while muons can screen the charge distortion nearly instantaneously, over the much shorter timescale of the inverse plasma frequency. The coherent response of this screening cloud induces a very efficient emission of scalars and vectors.

Particle emission by different interaction structures probes different responses of the dense relativistic medium. For example, up to simple factors, neutrino pair emission by the vector-current interaction depends exactly on the same combination of transverse and longitudinal dynamical structure functions as the emission of a vector boson with mass. The emission of scalars and pseudoscalars, on the other hand, does not directly probe a comparable property to the vector or axial-vector responses that one needs for neutrino pair emission. Therefore, despite the close relation of our work to Shternin's recent calculation of neutrino pair emission from the same environment \cite{Shternin:2025swb}, there is no direct way to translate his results to our cases. Instead, we here directly compute the emission rates for scalar, vector, and pseudoscalar bosons.

Guided by these objectives, we begin in Sec.~\ref{sec:formalism} with the physical setup and formalism. In Sec.~\ref{sec:structure functions}, we briefly touch on the emission rates from the perspective of linear response theory and highlight the similarities to and differences from neutrino pair emission. In Sec.~\ref{sec:renormalization}, we examine the in-medium renormalization of the boson couplings. We then calculate the emission rates for scalar, vector, and pseudoscalar interaction structures in Secs.~\ref{sec:scalar}--\ref{sec:pseudoscalar}.
In Sec.~\ref{sec:one-zone}, we evaluate our results numerically for a representative one-zone model and in this way illustrate the importance of different contributions, and finally conclude in Sec.~\ref{sec:conclusions} with a brief summary. The application of these results to NS cooling is deferred to a companion paper concerning a broader phenomenological assessment of astrophysical tests of muonic bosons~\cite{Fiorillo:2026new}.

\section{Formalism and Physical Setup}
\label{sec:formalism}

\subsection{Leptophilic interaction structures}

In  this work, we consider new low-mass bosons, collectively denoted by $\Phi=\{\phi,V,a\}$, corresponding to scalar, vector, and pseudoscalar states, respectively. These particles are assumed to interact with charged leptons $\ell=e$ or $\mu$ through interaction structures of the form
\begin{subequations}\label{eq:Lagrangians}
    \begin{eqnarray}
        {\cal L}_\phi&\supset& g_{\phi\ell} \,\phi\,\overline\psi_\ell\psi_\ell\,,
        \\
        {\cal L}_V&\supset& g_{V\!\ell}\, V_\nu\,\overline\psi_\ell\gamma^\nu\psi_\ell\,,\label{Eq:L_Vector}
        \\
        {\cal L}_a&\supset& i g_{a\ell}\, a\,\overline\psi_\ell\gamma_5\psi_\ell\,.
        \label{Eq:L_PS}
    \end{eqnarray}
\end{subequations}
In the following, we will focus on tree-level processes, for which the pseudoscalar interaction is equivalent to the derivative form ${\cal L}_a\supset -(g_{a\ell}/2m_\ell)\partial_\nu a\, \overline\psi_\ell\gamma^\nu\gamma_5\psi_\ell$.
The equivalence arises from applying the Dirac equation to an external fermion leg in the given process and therefore assumes the vacuum dispersion relation. While for electrons, the dispersion relation is strongly modified by our NS environment, for muons it remains dominated by the vacuum mass. The coupling strength to electrons is much better tested by less extreme environments as mentioned earlier, where electron dispersion is also dominated by the vacuum mass. Therefore, since muonic bosons are the primary focus of this study, we need not worry about subtleties of interpretation.

The leptophilic couplings imply effective two-photon interactions through a triangle loop. The loop function depends strongly on the assumed interaction structure as reviewed, for example, in Refs.~\cite{Caputo:2021rux,Alda:2024cxn}. As we do not include photonic processes in our calculations, once more we need not worry about such issues and will always use the nonderivative coupling for the pseudoscalar case.

\subsection{Characteristics of the environment}

In this paper, we aim to calculate the emission rates from the ultradegenerate environment of a cooling NS. To develop some intuition for the physical parameters and associated particle properties, we first note that NSs can reach up to a few times the nuclear density of $\rho_0\sim3\times10^{14}~\g~\cm^{-3}$ in their inner core, with typical baryonic densities in the range 2--$12\times10^{14}~\g~\cm^{-3}$.

The corresponding baryon density is $n_B=\rho/m_u=3.61\times10^{38}~{\rm cm}^{-3} \rho_6$, where $\rho_6=\rho/(6\times10^{14}~\g~\cm^{-3})$ and $m_u=1.6605\times10^{-24}~\g$ is the atomic mass unit. (Notice that the baryonic mass density is defined as the baryon number density times~$m_u$.) For a typical profile of physical characteristics we direct the reader to a NS reference model described in our companion paper~\cite{Fiorillo:2026new}.

In a fermion gas at $T=0$, all states are filled up to the Fermi momentum $p_{\F}$, so the number density~is
\begin{equation}
    n=\frac{p_\F^3}{3\pi^2}.
\end{equation}
This expression applies even at nonzero temperatures, provided that $T\ll \sqrt{m^2+p_\F^2}-m$, where $m$ is the mass of the particle, so that the Fermi surface is sharply defined. Numerically, a typical Fermi momentum is
\begin{equation}
 p_{\F}=201.8~\MeV~\left(\frac{Y}{0.10}\,\rho_6\right)^{1/3},
\end{equation}
where $Y$ is the particle abundance per baryon and the reference value of 0.10 would be typical for protons. For our temperatures of interest, $T\simeq10~\keV$, all fermions are extremely degenerate.

We assume the medium to consist of nucleons $n$ and~$p$, and charge neutrality to be established by electrons and muons. (The appearance of hyperons and/or meson condensates would change this simple picture because charged leptons would not be the only populations to take up negative charge, a possibility that we ignore in our calculations.)
Beta equilibrium and the absence of trapped neutrinos imply for the chemical potentials $\mu_e=\mu_\mu=\mu_n-\mu_p$. The Fermi-Dirac distribution for a given species is \smash{$f_\bp=[e^{(E_\bp-\mu)/T}+1]^{-1}$}, which depends on the fermion dispersion relation. In practice, electrons can be taken as massless, $E_\bp=|\bp|$, while muons can be described using their vacuum mass, \smash{$E_\bp=(m_\mu^2+\bp^2)^{1/2}$}. 

On the other hand, the nucleon dispersion relations depend strongly on the properties of the dense medium and therefore on the nuclear equation of state (EoS). In the simplest form, \smash{$E_\bp=U+(m^{*2}+\bp^2)^{1/2}$}, where the energy depends both on the nuclear potential $U$ and the effective mass $m^*$, parameters that are different for $n$ and~$p$ and depend on density and chemical composition. Therefore, determining the proton fraction per baryon, $Y_p$, requires not only the density, but also detailed information from the~EoS. 

Nevertheless, in the bremsstrahlung processes, one may simply neglect $U$ altogether. Since the proton does not turn into a neutron during the collision, a constant energy shift cannot alter the emission rate. Therefore, we will use the simple proton dispersion relation  \smash{$E_\bp=(m^{*2}+\bp^2)^{1/2}$}. Of course, such a simplified approach would not pertain to beta processes as these interconvert protons and neutrons and in the condition for beta equilibrium, $\mu_p+\mu_e=\mu_n$, the chemical potentials must include the nuclear potential, which is different for protons and neutrons.

For a massive particle with a simple dispersion relation, in the ultradegenerate limit, the Fermi energy $E_\F=(m^2+p^2_\F)^{1/2}$ equals the chemical potential $\mu$. All of our results are most economically expressed in terms of the particle velocities at the Fermi surface and the Fermi energy. To avoid notational clutter, we will use the symbol $\varepsilon_j=(m^{*2}_j+p_{\F,j}^2)^{1/2}$ for the kinematical Fermi energy of a species $j$, and \smash{$\beta_j=p_{\F,j}/(m^{*2}_j+p_{\F,j}^2)^{1/2}$} for its velocity at the Fermi surface. For particles with ordinary dispersion relations, $\varepsilon_j=\mu_j$.

In the relativistic and degenerate limit, the effective electron mass, valid for large momenta, is $m_{e}^{*2}=\alpha\mu_e^2/\pi$ \cite{Braaten:1991hg}, which is $m_{e}^{*}\simeq 10~\MeV$ for $p_{\F,e}=180~\MeV$, with $\alpha$ the fine-structure constant. Therefore, the vacuum mass $m_e$ is negligible and even $m_{e}^{*}$ is so small that we can treat electrons essentially as massless. On the other hand, the muon vacuum mass, $m_\mu=105.66~{\rm MeV}$, is so large that we can neglect medium corrections.

However, for bremsstrahlung, the Fermi velocity of the radiating particle appears typically in the form $(1-\beta^2)^{-1}$ so that, for electrons, we cannot simply set $\beta_e$ equal to 1. Instead, we observe that $1-\beta_e^2=m_e^{*2}/\varepsilon_e^2$ and with the earlier expression for the effective mass,
\begin{equation}
    1-\beta_e^2=\alpha/\pi=0.00232.
\end{equation}
Therefore, for our environment, the electron Fermi velocity is a universal number.

With these simplifications, and because $\mu_e=\mu_\mu$, one finds
\begin{equation}
    p_{\F,\mu}=\sqrt{p_{\F,e}^2-m_\mu^2},
\end{equation}
highlighting that for a muon population to exist, one needs $p_{\F,e}>m_\mu$. Charge neutrality implies $p_{\F,p}^3=p_{\F,e}^3+p_{\F,\mu}^3$ so that
\begin{equation}\label{eq:pFp}
    p_{\F,p}=\left[p_{\F,e}^3+\left(p_{\F,e}^2-m_\mu^2\right)^{3/2}\right]^{1/3}.
\end{equation}
Therefore, the proton and muon densities follow in a simple way from that of electrons. For example, $p_{\F,e}=180~\MeV$ implies $p_{\F,\mu}=146~\MeV$ and  $p_{\F,p}=207~\MeV$ and an abundance ratio $Y_\mu/Y_e=p_{\F,\mu}^3/p_{\F,e}^3=0.53$.

Concerning the plasma frequency, near $T\to 0$, the contribution of a single species $i$ of charged particles is $\omega_{\P,i}^2=4\pi\alpha n_i/\varepsilon_{i}$~\cite{Braaten:1993jw}, where we recall that $\varepsilon_i$ is the Fermi energy based on the in-medium Landau mass $m_i^*$. Because $\varepsilon_{\mu}=\varepsilon_{e}=p_{\F,e}$ and the densities obey $n_p=n_\mu+n_e$, the charged lepton contribution is $\omega_{\P,\ell}^2=4\pi\alpha n_p/p_{\F,e}$, which for our reference conditions is $\omega_{\P,\ell}\simeq 10~\MeV$. The proton effective mass is much smaller than its vacuum value, but still far exceeds $m_\mu$, so that the proton contribution to $\omega_\P$ is probably not large. Either way, the thermal population of electromagnetic excitations is hugely suppressed at NS core temperatures $T\simeq 10~\keV$. Therefore, while in a SN core, the dominant emission process of muonic bosons is photoproduction  $\gamma \mu\to\mu \Phi$~\cite{Caputo:2021rux}, this channel is completely suppressed in a cold NS.

Strongly degenerate protons may undergo a phase transition to a $^1S_0$ superfluid state, in which the formation of Cooper pairs occurs due to the long-range attractive nuclear force~\cite{1966ApJ...145..834W,Haskell:2017lkl}. According to cooling simulations, $T$ drops below the estimated critical temperatures of nuclear superfluid phases $T_c\sim3\times10^{9}\,$K within 10--100~years from NS birth~\cite{Chamel:2017wwp}. In a superfluid state, protons become inefficient targets for electromagnetic scattering of charged leptons, as the typical energy transferred in electromagnetic interactions, $\Omega\sim T$, is much lower than the energy required to break Cooper pairs $\Delta_p\sim1\,$MeV. However, at densities exceeding 2--$3\,\rho_0$, $^1S_0$ proton pairing is strongly suppressed and eventually disappears due to many-body effects~\cite{Guo:2018jpw}, leading to a strong reduction of the critical temperature in the densest regions of the NS core. Therefore, $^1S_0$ proton superfluidity is expected to occur only in the outer NS core (see, e.g., Fig.~1 of Ref.~\cite{Kaminker:2001eu}). Protons in the inner core remain viable targets for electromagnetic scatterings.

\subsection{Bremsstrahlung production}

In analogy to the emission of neutrino pairs or axions in degenerate environments, leptophilic bosons can be efficiently emitted through bremsstrahlung in electromagnetic collisions in cold NS environments. Taking a new boson $\Phi$ that couples predominantly to muons as an example, we consider the process
\begin{equation}
\mu(P_1)+X(P_2)\to\mu(P_3)+X(P_4)+\Phi(K)\,,
\label{Eq:bremss}
\end{equation}
where $X=\mu,e,p$ denotes the scattering partners (muons, electrons, or protons), and $\Phi=\phi,V,a$ represents the new boson, corresponding to a scalar, vector, or pseudoscalar particle, respectively.

The general expression for the volumetric energy-loss rate in the $\Phi$ channel is
\begin{align}
    Q_\Phi&=\int \prod_{i=1}^4\frac{d^3\bp_i}{2E_i(2\pi)^3}\frac{d^3\bk}{2\omega(2\pi)^3}\,\omega\,f_\mu(\bp_1)f_X(\bp_2)\nonumber\\[1ex]
    &\times [1-f_\mu(\bp_3)][1-f_X(\bp_4)]\sum_{\rm spins} \mathcal{S}^{-1}|\mathcal{M}|^2\nonumber\\
    &\times (2\pi)^4\delta^4(P_1+P_2-P_3-P_4-K)\,,\label{Eq:Emissivity}
\end{align}
where $P_i = (E_i, \bp_i)$ denote the fermionic four-momenta, with $i=1,2$ for the incoming, $i=3,4$ for the outgoing fermions, and $K=(\omega,\bk)$ the four-momentum of the boson emitted with velocity $\beta_\Phi=|\bk|/\omega$. The functions $f(\bp)$ represent the Fermi–Dirac distributions, while the factors $[1-f(\bp)]$ account for Pauli blocking in the degenerate medium. The symmetry factor $\mathcal{S}$ avoids double counting configurations with identical particles such that $\mathcal{S}=4$ when $X=\mu$, and $\mathcal{S}=1$ otherwise. (Henceforth, lower-case, non-bold symbols will denote the modulus of the corresponding three-vector, $p\equiv|\bp|$.)

In the degenerate NS environment, the boson typically carries an energy of order the temperature, $\omega \sim T$, which is much smaller than the chemical potentials of the participating fermions, $T \ll \mu_X$. This hierarchy justifies evaluating the matrix element in the long-wavelength limit, where the boson momentum can be treated as small compared to the characteristic fermion momenta.

\subsection{In-medium photon propagator and screening}
\label{sec:screening}

In all cases, the squared matrix element, $|\mathcal{M}|^2$, entering Eq.~\eqref{Eq:Emissivity} involves the exchange of a virtual photon between the scattering particles. In the dense medium of a neutron star, medium effects enter through the dressed photon propagator that, in the Lorenz gauge, can be written as
\begin{equation}
    D^{\alpha\beta}(Q) = \frac{P_{\rm T}^{\alpha\beta}}{Q^2-\Pi_{\rm T}} + \frac{P_{\rm L}^{\alpha\beta}}{Q^2-\Pi_{\rm L}}\,,
    \label{Eq:Propagator}
\end{equation}
with $Q=(\Omega,\bq)$ being the four-momentum exchange. The transverse and longitudinal projectors encode the propagation of the corresponding modes, and read
\begin{subequations}\label{eq:TL-Projectors}
\begin{align}
    P_{\rm T}^{ij} & = \delta^{ij}- \frac{q^i q^j}{q^2}\,,\label{Eq:PTij}\\
    P_{\rm T}^{00} & = P_{\rm T}^{0i} = P_{\rm T}^{i0} = 0\,,\\
    P_{\rm L}^{\alpha\beta}&=e_{\rm L}^\alpha e_{\rm L}^\beta\,,
\end{align}    
\end{subequations}
with $e_{\rm L}=(q,\Omega\,\bq/q)/\sqrt{Q^2}$. We mention in passing that the projectors thus defined obey $P^{\alpha\beta}P_\beta^\gamma=-P^{\alpha\gamma}$. They would be strictly idempotent if defined with opposite sign.

The real parts of the polarization functions $\Pi_{\rm T}$ and $\Pi_{\rm L}$ encode the typical screening scale of transverse and longitudinal plasma perturbations, respectively. Longitudinal photons mediating electric currents are screened by charged particles. The imaginary parts of $\Pi_{{\rm T},{\rm L}}$ describe the damping rate of T and L excitations.

For typical NS conditions, most scatterings between charged particles involve momentum exchanges $q$ determined by the screening or damping scale of the exchanged mode, as we show below, which is in the ballpark of $m_{D,M,\Lambda}\sim\mathcal{O}(10)\,$MeV. On the other hand, the amount of energy $\Omega$ exchanged by the virtual photon is of the order of the energy carried away by the radiated particle, which is determined by the much lower temperature of the NS core $T\sim\mathcal{O}(10)\,$keV.
Therefore, for most of the phase space contributing to the emission, it is safe to assume that the virtual photon is static, with $\Omega\ll q$. In this regime, the self-energy for the longitudinal state is given by the static Debye screening scale
\begin{equation}
    \Pi_{\rm L}=m_D^{2}=\frac{4\alpha}{\pi}\!\!\sum_{i=e,\mu,p}\!\!p_{{\F},i}^2\,\frac{\partial p_{{\rm F},i}}{\partial \varepsilon_{i}}
    =\frac{4\alpha}{\pi}\!\!\sum_{i=e,\mu,p}\!\!p_{\F,i}\varepsilon_{i}.
\end{equation}
While the last expression is usually called the Thomas--Fermi scale, we always refer to the Debye scale $m_D$ for static screening. While the Fermi momenta of $p$, $e$, and $\mu$ are comparable, the Fermi energy $\epsilon_p=(m_p^{*2}+p_{\F,p}^2)^{1/2}$ is significantly larger for protons, so the Thomas--Fermi scale is dominated by protons. Numerically, using the Fermi momenta stated after Eq.~\eqref{eq:pFp}, we find $m_D=38~{\rm MeV}$.

The treatment of the transverse polarization function requires more caution. At temperatures below the critical value for  $^1S_0$ nuclear superfluidity, $T\lesssim3\times10^{9}\,$K, protons presumably enter a superconducting phase in which the magnetic flux is partially expelled from the medium~\cite{Meissner:1933ela}. This shows up through the appearance of the Meissner scale $m_M$ in the self-energy of the transverse photons 
\begin{equation}
    \Pi_{\rm T}=m_M^2=\frac{1}{\lambda_M^2}=\frac{4\pi\alpha\,n_p}{m^*_p}\,,
    \label{Eq:Meissner_Scale}
\end{equation}
parametrically determined from the inverse of the London penetration length $\lambda_M$ of the superconductor~\cite{1976itss.book.....K}, where $n_p$ is the proton number density and $m^*_p$ the proton effective mass in the nuclear medium. Using once more the parameters stated after Eq.~\eqref{eq:pFp} and with $m_p^*=400~{\rm MeV}$, we find $m_M=8~{\rm MeV}$. Therefore, there is a mild hierarchy $m_D\gg m_M$ so that bremsstrahlung emission is dominated by transverse photon exchange as emphasized previously in Refs.~\cite{Shternin:2007ee, Shternin:2006uq}.

Nonetheless, as previously discussed, the high density expected in the inner NS core prevents superconductivity. In this case, transverse electromagnetic excitations are not statically screened, and they can only be dynamically absorbed through Landau damping, see Eq.~(2.3) of Ref.~\cite{lindhard1954properties}. Therefore, the transverse photon polarization function reads
\begin{equation}
    \Pi_{\rm T}(\Omega, \bq) = i\frac{\Omega}{q}\alpha\!\!\sum_{i=e,\mu,p}p_{{\rm F},i}^2\,,
    \label{Eq:Landau_Damping}
\end{equation}
which depends explicitly on the energy $\Omega$ and the amount of momentum $q$ exchanged by the transverse photon. Keeping in mind that our main goal is to provide simple expressions for boson emission rates in NS environments, without any claim to precision physics, we can parametrically assume $\Omega\sim T$ in Eq.~\eqref{Eq:Landau_Damping}. Hence, excitations are damped if they have an exchanged momentum below
\begin{equation}
    q\lesssim  m_{\Lambda}=\biggl(\alpha T\!\!\sum_{i=e,\mu,p}p_{{\rm F},i}^2\biggr)^{1/3}\, ,
    \label{Eq:Landau_Scale}
\end{equation}
playing the role of an effective low-energy scale which regulates the infrared divergence of the interaction. Numerically, using the same parameters as in the earlier estimates and $T=10~{\rm keV}$, this is $m_{\Lambda}=2~{\rm MeV}$. Therefore, the hierarchy $m_D\gg m_{\Lambda}$ is even more pronounced and transverse photons strongly dominate bremsstrahlung.

In order to treat transverse photons uniformly across different phases, we define the transverse screening mass~as
\begin{equation}
    m_T =
\begin{cases}
m_{\Lambda}\,, & \langle pp \rangle = 0\,,\\
m_M\,, & \langle pp \rangle \neq 0\,,
\end{cases}
\end{equation}
where $\langle pp \rangle$ denotes the proton condensate. A non\-vanishing condensate signals the superconducting phase.

\begin{figure}[ht]
    \centering
    \includegraphics[width=0.95\columnwidth]{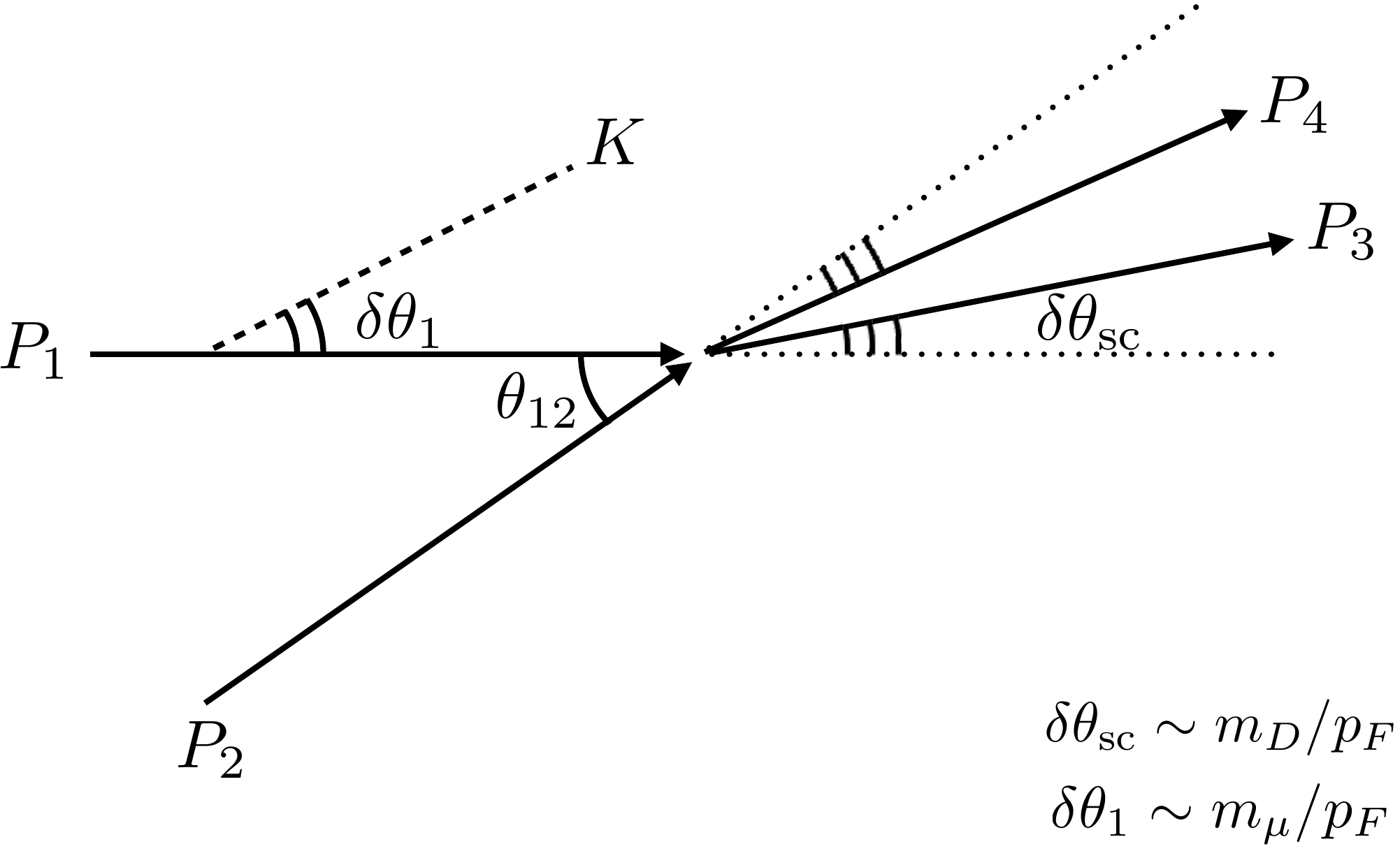}
    \caption{Schematic representation of bremsstrahlung emission by ultradegenerate leptons in the laboratory frame.}
    \label{fig:scheme}
\end{figure}

\subsection{Bremsstrahlung emission at a glance}

The generic structure of bremsstrahlung radiation is schematically shown in Fig.~\ref{fig:scheme}. We use this illustration mostly to characterize the various angular scales that enter the problem, corresponding to different energy scales.

The smallest of these scales is temperature $T$, which characterizes the typical frequency and wavevector of the radiation. This scale is not associated with any small angle. Its smallness with respect to all the other energy scales allows us to neglect it everywhere except in the integration of the Fermi distributions; only particles deviating by an amount $\sim T$ from their Fermi sphere may contribute to the emission. 

One exception to this rule is for transverse excitations in a nonsuperconducting medium, since, as we have seen, they are damped with a rate that depends on the energy exchange among the scattering particles, which in turn is comparable with $T$. We circumvent this problem by simply treating this case parametrically in terms of the typical energy exchange, setting it equal to $T$.

Another small scale is the screening scale $m_{D,M,\Lambda}$; let us consider $m_D$ as an example. Its smallness causes the near-forward nature of the scattering; in Fig.~\ref{fig:scheme}, the initial and final directions of each particle are nearly the same. The typical deviation angle is $\delta \theta_{\rm sc}\sim m_D/p_\F$, where $p_\F$ is the Fermi momentum of the radiating particles. 

Finally, the mass of the radiating particle introduces yet another energy scale. For mildly or ultrarelativistic muons, the typical angle between the radiating particle and the radiation is $\delta \theta_1\sim m_\mu/p_\F$. For electrons, this angle might be much smaller, since their thermal mass $m_e^*\simeq 10~{\rm MeV}$. We generally assume in this work that $\delta\theta_{\rm sc}\lesssim \delta \theta_1$. This is certainly true for muons, and is also true for electrons---at least for the dominant scattering mediated by the transverse electromagnetic field in the nonsuperconducting regions of the star, since $m_\Lambda\ll m_e^*$.

\section{Emission in terms of dynamical structure functions}

\label{sec:structure functions}

\subsection{General formulation}

The viewpoint we have adopted so far is primarily kinetic, focusing on emission from individual pairs of particles as a phase-space integral over a squared matrix element as in 
Eq.~\eqref{Eq:Emissivity}. Nevertheless, it is informative to compare this approach with the more generic relation between particle emission and the dynamical response of the medium. As the main outcome of this more general approach, we will be able to compare the emissivity of bosons with the emissivity of neutrino~pairs.

Let us consider a generic vector or axial vector current coupling to the medium
\begin{subequations}\label{eq:Lagrangians2}
    \begin{eqnarray}
        {\cal L}_V&\supset& g_{V\!\ell}\, V_\nu\,\overline\psi_\ell\gamma^\nu\psi_\ell\,,\label{Eq:L_Vector2}
        \\
        {\cal L}_a&\supset&  -\frac{g_{a\ell}}{2m_\ell}\,\partial_\nu a\,\overline\psi_\ell\gamma^\nu\gamma_5\psi_\ell\,.
    \end{eqnarray}
\end{subequations}
Neutrino pairs, on the other hand, couple to leptons through a Lagrangian that immediately suggests a connection to the boson cases,
\begin{subequations}\label{eq:neutrino-lagrangian}
\begin{eqnarray}
    {\cal L}_{V,\nu}&\supset& -\frac{C_{V\!\ell}G_\F}{\sqrt{2}}\, \overline\psi_\ell\gamma^\alpha\psi_\ell\,\overline\psi_\nu \gamma_\alpha(1-\gamma_5)\psi_\nu\,,\label{Eq:L_Vector3}
        \\
        {\cal L}_{A,\nu}&\supset&  \frac{C_{A\ell}G_\F}{\sqrt{2}}\, \overline\psi_\ell\gamma^\alpha\gamma_5\psi_\ell\,\overline\psi_\nu \gamma_\alpha(1-\gamma_5) \psi_\nu\,.
\end{eqnarray}    
\end{subequations}
For the massless neutrinos assumed here, the factor $(1-\gamma_5)$ could be replaced by 2, while using two-component Weyl spinors. For a neutrino flavor different from the charged lepton, \smash{$C_{V\!\ell}=-\frac{1}{2}+2\sin^2\theta_W=-0.037$} and $C_{A\ell}=-\frac{1}{2}$, whereas for the same flavor, the couplings are \smash{$C_{V\!\ell}=\frac{1}{2}+2\sin^2\theta_W=0.963$} and $C_{A\ell}=\frac{1}{2}$, with $\theta_W$ the Weinberg angle and $G_\F$ the Fermi constant.

We consider first vector-boson emission. The cooling rate can be written, in structural analogy to Eq.~\eqref{Eq:Emissivity}, in the form
\begin{equation}
    Q_V=g_{V\!\ell}^2\int \frac{d^3\bk}{2\omega_\bk(2\pi)^3}\omega_\bk S_{\alpha \beta}R_V^{\alpha \beta}.
\end{equation}
Here $R_{\alpha\beta}$, where $R$ stands for radiation, is the density matrix of the emitted vector summed over all polarizations
\begin{equation}
    R_{V,\alpha\beta}=-g_{\alpha\beta}+\frac{K_\alpha K_\beta}{m_V^2}
\end{equation}
with $K_\alpha$ the four-momentum of the emitted boson. On the other hand, $S_{\alpha\beta}$ is the response function of the medium
\begin{equation}
    S_{\alpha \beta}(K)=\int d^4 x\, e^{iK\cdot x}\langle j_{V,\alpha}(x) j_{V,\beta}(0) \rangle
\end{equation}
with $j_{V,\alpha}=\overline\psi_\ell \gamma_\alpha \psi_\ell$. Thus the emissivity is directly determined by the spectrum of fluctuations of the corresponding fermion current.

Another medium response function for the vector current is the polarization tensor $\Pi_{\alpha\beta}$. In a thermal medium, the connection between them is the fluctuation--dissipation theorem
\begin{equation}
     S_{\alpha\beta}=\frac{2}{e^{-\omega/T}-1}
     {\rm Im}\,\Pi_{\alpha \beta}.
\end{equation}
In other words, the imaginary part of the polarization tensor represents damping or emission of radiation coupled to the vector current.

If the medium is isotropic, the structure function can be given a rather generic form
\begin{equation}
    S^{\alpha \beta}=S_T P_{T}^{\alpha \beta}+S_L P_L^{\alpha\beta}+S_K K^\alpha K^\beta+ S_M P_M^{\alpha \beta},
\end{equation}
where the transverse and longitudinal projectors were already defined in Eq.~\eqref{eq:TL-Projectors} and \smash{$P_M^{\alpha\beta}=K^\alpha U^\beta+K^\beta U^\alpha$} with $U^\alpha=(1,0,0,0)$ the medium four-velocity.

If the vector current is conserved, the structure function must satisfy $K_\alpha S^{\alpha\beta}=0$, which immediately constrains $S_K=S_M=0$. Therefore, the emissivity can finally be written in the form
\begin{equation}
    Q_V=g_{V\!\ell}^2 \int \frac{d^3\bk}{2(2\pi)^3}(2S_T+S_L).
\end{equation}
The functions $S_T$ and $S_L$ depend on both $\omega_\bk$ and $k=|\bk|$ separately in general, where these are frequency and wave number of the radiation in the medium frame.

\subsection{Neutrino pairs}
\label{eq:neutrino-pairs}

Let us now consider the emission of a neutrino pair through the vector current. The cooling rate is
\begin{equation}
    Q_{V,\nu}=\frac{C_{V\!\ell}^2 G_\F^2}{2}\!\int\! \frac{d^3\bk_1}{2\omega_1(2\pi)^3}\frac{d^3\bk_2}{2\omega_2(2\pi)^3}(\omega_1+\omega_2)
    R_{\nu,\alpha \beta}S^{\alpha\beta},
\end{equation}
where for massless neutrinos, $\omega_{1,2}=|\bk_{1,2}|$. Moreover,
\begin{eqnarray}
    R_\nu^{\alpha\beta}&=&8\left(K_1^\alpha K_2^\beta+K_1^\beta K_2^\alpha-K_1\cdot K_2 \,g^{\alpha\beta}\right)
    \nonumber\\
    &=&8\left[\frac{K^\alpha K^\beta-K^2 g^{\alpha\beta}}{2}-2\xi^\alpha \xi^\beta\right],
\end{eqnarray}
where in the second line, it is expressed in terms of $K=K_1+K_2$ and $\xi=(K_1-K_2)/2$.

The integral over the two-neutrino phase space can be partially performed in that the energy-momentum transfer to the medium, $K$, is left open, whereas their relative momentum is integrated out. We use the two integrals
\begin{subequations}
\begin{eqnarray}
    \kern-2em&&\int \frac{d^3\bk_1}{2|\bk_1| (2\pi)^3}\frac{d^3\bk_2}{2|\bk_2| (2\pi)^3}=\int \frac{d^4 K}{128\pi^5},
    \\[2ex]
    \kern-2em&&\int \frac{d^3\bk_1}{2|\bk_1| (2\pi)^3}\frac{d^3\bk_2}{2|\bk_2| (2\pi)^3}\xi^\mu \xi^\nu
    \nonumber\\
    \kern-2em&&\kern4em{}=\int \frac{d^4 K}{1536\pi^5}\left(K^\mu K^\nu - K^2 g^{\mu\nu}\right),
\end{eqnarray}    
\end{subequations}
so that
\begin{equation}
    Q_{V\nu}=\frac{C_{V\!\ell}^2 G_\F^2}{2} \int\frac{d^4 K}{48\pi^5}\omega\left(K^\alpha K^\beta-K^2 g^{\alpha\beta}\right)S_{\alpha\beta},
\end{equation}
where $\omega=K^0$ and the integral is done under the condition $\omega>0$ and $\omega^2>|\bk|^2$.

Since the vector current is still conserved, only the $g^{\alpha\beta}$ term contributes to the projection of $S_{\alpha\beta}$ on the neutrino polarization density and thus returns the same combination $2S_T+S_L$ of structure functions as for the vector boson.

The $(\omega,\bk)$ of a neutrino pair is equivalent to that of a boson with mass \smash{$m_V=\sqrt{\omega^2-|\bk|^2}$}. Therefore, the integral over $\omega$ may be expressed as one over $m_V$, so we finally obtain
\begin{equation}
    Q_{V,\nu}=\frac{C_{V\!\ell}^2G_\F^2}{2} \int_0^{\infty} \frac{dm_V m_V^3}{3\pi^2}\int \frac{d^3\bk}{2(2\pi)^3} 
    \left(2S_T+S_L\right),
\end{equation}
where $S_{T,L}$ are functions of $(\omega,|\bk|)$ and so we need to replace \smash{$\omega=\sqrt{m_V^2+\bk^2}$}.

Thus, we discover that the cooling rate due to emission of neutrinos through the vector current can be directly related to the cooling rate through emission of massive vector bosons
\begin{equation}
    Q_{V,\nu}=\frac{C_{V\!\ell}^2 G_\F^2}{2 g_{V\!\ell}^2} \int_0^{\infty}dm_V\,\frac{m_V^3}{3\pi^2}\, Q_V(m_V) ,
\label{eq: fromVectoNu}
\end{equation}
where $Q_V(m_V)$ is the energy-loss rate for a vector boson with mass $m_V$. The integral will be dominated by $m_V$ of the order of $T$. Therefore, the neutrino emission rate will feature an additional power of $T^4$, as is anyway indicated on purely dimensional grounds.

\subsection{Axial vector coupling }

For the axial current, such a simple relation between the pseudoscalar emission and neutrino pair emission cannot be established in general. Indeed, the emissivity of $a$ is generally proportional to the contraction $K_\alpha K_\beta S_5^{\alpha\beta}$, where $S_5^{\alpha\beta}=\int d^4x\,e^{iK\cdot x}\langle j_5^\alpha(x) j_5^\beta(0)\rangle$ and $j_5^\alpha=\overline\psi_\ell \gamma^\alpha \gamma_5 \psi_\ell$. Instead, for neutrino pair emission, by the same argument as above, the relevant contraction is $(K_\alpha K_\beta-g_{\alpha \beta} K^2) S_5^{\alpha\beta}$. The main difference compared to the vector case is that the axial current is not conserved. Therefore, in the expansion for an isotropic medium, 
\begin{equation}
    S_5^{\alpha \beta}=S_{5,T} P_{T}^{\alpha \beta}+S_{5,L} P_L^{\alpha\beta}+S_{5,K} K^\alpha K^\beta+ S_{5,M} P_M^{\alpha \beta},
\end{equation}
the functions $S_{5,K}$ and $S_{5,M}$ need not vanish. They do not contribute to the emission of neutrinos, whereas $S_{5,K}$ is the only one contributing to the emission of axions. Therefore, the two cases correspond to independent structure functions. 

The situation is different in a medium of nonrelativistic particles, e.g., nucleons with mass $m_N$. In that case, the temporal component $j_5^0$ is smaller than the spatial components $j_5^i$ by a factor $|\bk|/m_N$. Therefore, to first approximation, the structure tensor \smash{$S_5^{\alpha\beta}$} is purely spatial, and for an isotropic medium, in the long-wavelength approximation, where $\bk$ is completely negligible, can only take the form $S_5^{ij}=S_5 \delta^{ij}$. With a single structure function, the emissivity of axions and neutrinos through the axial current can again be related. Such a relation was indeed repeatedly adopted to determine the emission of axions in a dense nuclear medium~\cite{Raffelt:1996wa, Raffelt:2006cw, Lykasov:2008yz, Fiorillo:2025gnd}. So the reason we cannot use this approach here is the relativistic velocity of electrons and muons in the medium.

\section{In-medium renormalization of the boson coupling}

\label{sec:renormalization}

\subsection{Vector current}

The frequency and wavevector of the radiated boson are both of the order of the temperature of the medium $T$, far below the screening scale or plasma frequency. In this regime, charge densities are strongly screened. Our bosons do not directly couple to charge, so they may still retain a strong coupling, but the charge response to any density fluctuation can still renormalize the coupling considerably. In the nonsuperconducting phase, this means that protons, which are accelerated much more rapidly than leptons due to their strong interactions, contribute most to the emission of leptophilic particles: their instant screening by a cloud of leptons offers the radiating population needed for vector emission. 

We now consider this question, focusing on a zero-temperature environment. While the scalar boson case is simplest from the computational perspective, we begin with the vector case, for which the in-medium renormalization is conceptually more transparent. Indeed, it was in the context of vector-current neutrino pair emission where this issue was first brought up~\cite{Leinson:1997zt, Leinson:1999ut, Leinson:2000ks, Leinson:2000un}.

\begin{figure}
  \centering
  \includegraphics[width=0.8\columnwidth]{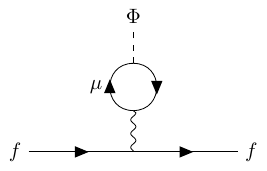}
  \caption{In-medium boson coupling to charged fermions $f=e,\mu,p$ mediated by the muon loop.}\label{fig:muonloop}
\end{figure}

The effective coupling to the muon is renormalized at loop level by the diagram shown in Fig.~\ref{fig:muonloop}. Notice that the photon propagator already includes a full resummation of the intermediate loop in the form of Eq.~\eqref{Eq:Propagator}. To be generic, we assume that the vector boson couples through the combination $(g_{V\!e},\,g_{V\!\mu},\,g_{V\!p})$ to electrons, muons and protons as in Eq.~\eqref{Eq:L_Vector}. We recognize that the fermion loop is simply the self-energy part of the photon field, $\Pi_{f}$, associated with the $f$-th fermion, with $f=(e,\mu,p)$. Then the effective amplitude for coupling to the muon is
\begin{equation}
    {\mathcal{M}}^\alpha=g_{V\!\mu} \gamma^\alpha-\sum_{f=e,\mu,p} g_{V\!f} \Pi_{f}^{\alpha\beta}D_{\beta\delta}\gamma^\delta.
\end{equation}
If we separate out the transverse and longitudinal parts of the self-energy and propagator, this becomes
\begin{equation}
    {\mathcal{M}}^\alpha=\gamma^\beta\!\left[g_{V\!\mu} \delta^\alpha_\beta-\!\!\!\sum_{f=e,\mu,p}\! g_{V\!f}\!\left(\frac{\Pi_{{\rm T},f} P^\alpha_{{\rm T},\beta}}{Q^2-\Pi_{\rm T}}+\frac{\Pi_{{\rm L},f} P^\alpha_{{\rm L},\beta}}{Q^2-\Pi_{\rm L}}\!\right)\!\right]\!.
\end{equation}
If we now take the massless limit, then $Q^2\to 0$ and the longitudinal part decouples by the Ward identity. 

Indeed, in the longitudinal projector $P_{\rm L}^{\alpha\beta}=e_{\rm L}^\alpha e_{\rm L}^\beta$, the vector \smash{$e_{\rm L}^\alpha$} vanishes upon contraction with the transverse polarization vector of the external boson; simultaneously, also $e_{\rm L}^\beta$ vanishes upon contraction with $\gamma^\beta$ between physical external-fermion states. Therefore, the vertex simplifies to
\begin{equation}
    {\mathcal{M}}^\alpha=\gamma^\beta\left[g_{V\!\mu} \delta^\alpha_\beta+\sum_{f=e,\mu,p} g_{V\!f}\frac{\Pi_{{\rm T},f}}{\Pi_{\rm T}}(-\delta^\alpha_\beta)\right];
\end{equation}
we have here used the massless projector $P_{\rm T}^{\alpha\beta}$, removing all terms proportional to $Q^\alpha$ which would vanish upon contraction with the external polarization vector. Hence, the simple expression
\begin{equation}
    {\mathcal{M}}^\alpha=\gamma^\alpha \left[g_{V\!\mu}-\sum_{f=e,\mu,p} \frac{g_{V\!f} \Pi_{{\rm T},f}}{\Pi_{\rm T}}\right]
\end{equation}
is here the renormalization of the vertex.

The self-energies must be evaluated for $Q^2=0$, the on-shell condition of the external particle. This limit pertains even to a massive particle, provided the mass is far below the plasma frequency. Then the transverse self-energy of a degenerate species is
\begin{equation}
    \Pi_{{\rm T},f}=\omega_{{\rm P},f}^2\left(1+\frac{G(\beta_f^2)}{2}\right),
\end{equation}
where $\beta_f$ is the Fermi velocity of the fermion, $\omega_{{\rm P},f}^2=4\alpha p_{{\rm F},f}^3/3\pi \varepsilon_f$ is the contribution to the total plasma frequency from the fermion $f$, and~\cite{Raffelt:1996wa}
\begin{equation}
    G(x)=\frac{3}{x}\left[1-\frac{2x}{3}-\frac{1-x}{2\sqrt{x}}\log\left(\frac{1+\sqrt{x}}{1-\sqrt{x}}\right)\right].
\end{equation}
For $\beta_f\simeq 1$, we simply get $\Pi_{{\rm T},f}=3\omega_{{\rm P},f}^2/2$. 

Since the plasma frequency is suppressed by a power of the Fermi velocity, protons are generally subdominant compared to electrons and muons for the overall self-energy so that $\Pi_{\rm T} \simeq\Pi_{\rm T,e}+\Pi_{\rm T,\mu}$. In this approximation, in the emission rates, we substitute the tree-level vector coupling $g_{V\!\mu}$ with the renormalized in-medium coupling
\begin{equation}
    \hat{g}_{V\!\mu}\simeq g_{V\!\mu}\left(1-\frac{\Pi_{\rm T,\mu} }{\Pi_{{\rm T},\mu}+\Pi_{{\rm T},e}}\right)
    =g_{V\!\mu}\,\frac{\Pi_{\rm T,e} }{\Pi_{{\rm T},\mu}+\Pi_{{\rm T},e}}
    \,,
    \label{eq:treelevelVSinmedium}
\end{equation}
which simultaneously induces an effective in-medium coupling to electrons
\begin{equation}
    \hat{g}_{V\!e}=-g_{V\!\mu}\,\frac{\Pi_{\rm T,\mu} }{\Pi_{{\rm T},\mu}+\Pi_{{\rm T},e}}\,,
\label{eq:gVeEffective}
\end{equation}
and protons
\begin{equation}
    \hat{g}_{V\!p}=-g_{V\!\mu}\,\frac{\Pi_{\rm T,\mu} }{\Pi_{{\rm T},\mu}+\Pi_{{\rm T},e}}\,.
\label{eq:gVpEffective}
\end{equation}
These effective couplings lead to vector boson emission from electron and proton legs in $\mu e$ and $\mu p$ scatterings, respectively, and generate additional contributions to the emission rates from electromagnetic $ee$, $ep$ scatterings and $pp$ and $pn$ nuclear collisions.

By these formulae, we see that the renormalization of the muon coupling is only a numerical factor of order unity. In the limit of ultrarelativistic muons, $\omega_{{\rm P},\mu}^2=\omega_{{\rm P},e}^2$, and so a particle coupling purely to muons would get its coupling lowered by exactly a factor of~2, but it would also gain an in-medium coupling to electrons and protons equal in magnitude to the one to muons. For typical NS conditions, in-medium effects reduce the tree-level coupling of muonic bosons only by $\sim30\%$, since the muon and electron populations differ somewhat. 

However, as we will see, nuclear bremsstrahlung from barely nonrelativistic protons is much more efficient than electromagnetic scattering processes. A subdominant contribution to the total emissivity arises from bremsstrahlung by ultrarelativistic electrons scattering off charged particles, which is itself considerably more efficient than the corresponding process involving semi-relativistic muons. As a result, purely muonic vector bosons are emitted predominantly by protons---with a subdominant contribution from electrons---through a muon-loop-induced effective coupling.

\subsection{Neutrino pair emission}

The in-medium charge renormalization was first studied in the context of neutrino pair emission \cite{Leinson:1997zt, Leinson:1999ut, Leinson:2000ks, Leinson:2000un} and later applied \cite{Jaikumar:2005gm, Shternin:2025swb} to the same NS environment that we are having in mind, where the relevant particles are protons, electrons, and muons. These latter authors assumed that the vector-current pair emission was largely suppressed by the renormalization effect, but this conclusion is not consistent with our findings.

Because of the peculiar value of the weak mixing angle, the vector neutrino neutral-current coupling to charged leptons nearly vanishes (see the coupling constants listed in the text after Eq.~\ref{eq:neutrino-lagrangian}), whereas the charged-current contribution to the effective low-energy neutral-current coupling is large. Therefore, with regard to the vector current, $\nu_e\bar\nu_e$ pairs are almost exclusively emitted by electrons and  $\nu_\mu\bar\nu_\mu$ pairs by muons. In a degenerate electron medium, our results indicate that the effective $\nu_e\bar\nu_e$ coupling will be screened and indeed, their emission will be strongly suppressed. 

However, in a medium of both degenerate electrons and muons, the situation is entirely different in that the renormalized coupling of $\nu_e\bar\nu_e$ gets reduced roughly by a factor of~2, while an equally strong coupling to muons is generated, and vice versa for $\nu_\mu\bar\nu_\mu$. Therefore,  both electrons and muons will emit pairs of both flavors, the various rates depending, in detail, on the Fermi velocities of the charged leptons and the kinematics of the various collision processes.

Ultimately, however, just as for vector bosons, the predominant emission will come from protons. From Eq.~\eqref{eq:gVpEffective}, we see that protons develop an effective coupling to the vector current of neutrinos induced by charged-current interactions with electrons for $\nu_e\overline{\nu}_e$ and with muons for $\nu_\mu\overline{\nu}_\mu$, through the screening cloud of leptons. Therefore, in the nonsuperconducting proton phase, we expect that the vector current emission of neutrino pairs is dominated by neutron-proton bremsstrahlung. Even though ultimately MUrca processes remain the dominant neutrino cooling channel, the dominant vector current emission seems to have been lost from the relevant literature, which does not account for this screening-mediated process. On the other hand, in the superconducting phase, a similar observation has been used by Leinson~\cite{Leinson:2000ks, Leinson:2000un} to infer that proton pair breaking dominates the vector-current emission of neutrino pairs.

\subsection{Physical interpretation}

This quantum field theory calculation may somewhat obscure the underlying physics, so it may be illuminating to consider this effect from the simple perspective of the medium response. In the plasma, the electric potential $A^\mu$ produced in response to an external four-current $j_{\rm ext}^\mu$ obeys the equation
\begin{equation}
    \partial^2A^\alpha=j_{\rm ext}^\alpha-\hat{\Pi}^{\alpha\beta}A_\beta;
\end{equation}
we mark by a hat the self-energy operator in real space, and we use the Lorenz gauge $\partial_\alpha A^\alpha=0$. Therefore, the four-potential in momentum space satisfies
\begin{equation}
    A_\alpha(Q)=-(Q^2 g_{\alpha\beta}-\Pi_{\alpha\beta})^{-1}j_{\rm ext}^\beta.
\end{equation}
If we now introduce a new vector field $V_\alpha$ with couplings $g_{V\!j}=(g_{V\!e},g_{V\!\mu})$, the four-current which sources it must include not only the external one, but also the one sourced by the medium
\begin{equation}
    \partial^2 V^\alpha=\sum_j g_{V\!j}\left(j_{\rm ext,j}^\alpha-\hat{\Pi}_{jj}^{\alpha\beta}A_\beta\right).
\end{equation}
The second term is the origin of the renormalization of the vector coupling. It is the four-current produced by the medium in response to the external electromagnetic field.

\subsection{Scalar coupling}
\label{sec:scalar_coupling}

We next turn to the scalar case, where the loop operator is not the same as the photon self-energy, and must be computed independently. Then the effective amplitude coupling to muons is
\begin{align}
    \mathcal{M}&=g_{\phi\mu}-\sum_j 2m_j e^2 g_{\phi j} \gamma_{\alpha} D^{\alpha\nu}\nonumber\\
    &\kern1em{}\times\int \frac{d^3\bp}{(2\pi)^3}\frac{n_{j,\bp} }{E_\bp}\left[\frac{2 P^\nu+Q^\nu}{Q^2+2P\cdot Q}+\frac{2P^\nu-Q^\nu}{Q^2-2P\cdot Q}\right]\nonumber\\[1.5ex] &=g_{\phi\mu}-\sum_j 2 m_j e^2 g_{\phi j} \gamma_\alpha D^{\alpha\nu}\Gamma_{j,\nu}\,,
\end{align}
where $P^\mu=(E_\bp,\bp)$ is the four-momentum of the intermediate fermion. 

In the definition of the tensor $\Gamma_{j,\nu}$, we may drop the terms proportional to $Q^\nu$, which vanish upon contraction with $D^{\alpha \nu}$. Furthermore, the tensor $\Gamma_{j,\nu}$, apart from terms proportional to $Q^\nu$ which can be neglected, can only be directed along $u^\nu=(1,0,0,0)$, the four-velocity of the plasma. Therefore, only the zeroth component of $D^{\alpha\nu}$ will enter the physical amplitude. This immediately excludes the transverse part of the photon propagator, so we may only consider the longitudinal part, which we write as $D_{\rm L}^{\alpha\nu}=e_{\rm L}^\alpha e_{\rm L}^\nu/(Q^2-\Pi_{\rm L})$. We can simplify further by noting that $\sqrt{Q^2} e_{\rm L}^\alpha=-Q^2u^\alpha/q+(\Omega/q)Q^\alpha$. The term in $Q^\alpha$ vanishes upon contraction with $\gamma_\alpha$ by the Ward identity. Therefore, the expression can be vastly simplified
\begin{equation}\label{eq:vertex}
    \mathcal{M}=g_{\phi\mu}+\sum_j 2m_j e^2 g_{\phi j} \frac{\gamma^0}{Q^2-\Pi_{\rm L}}\frac{E_{\rm L}\cdot \Gamma_j}{q},
\end{equation}
where $E_{\rm L}=\sqrt{Q^2}e_{\rm L}$. 

We may now proceed to compute $E_{\rm L}\cdot \Gamma_j$, which explicitly reads
\begin{equation}
    E_{\rm L}\cdot \Gamma_j=\int \frac{d^3\bp}{(2\pi)^3}\frac{n_{j,\bp}}{E_\bp}\frac{4 P\cdot E_{\rm L}\, Q^2}{Q^4-4(P\cdot Q)^2}.
\end{equation}
Since we only consider the case of massless bosons, in the denominator we may neglect $Q^4$, and in the numerator we may set $E_{\rm L}\to Q$; the term $Q^2$ must be kept since it will cancel with the denominator $Q^2-\Pi_L$. Therefore, we finally find
\begin{equation}
    E_{\rm L}\cdot \Gamma_j=-\frac{Q^2}{\Omega} \int \frac{d^3\bp}{(2\pi)^3}\frac{n_{j,\bp}}{E_\bp(E_\bp-\bp\cdot \bn)}=-\frac{Q^2\xi_j}{\Omega},
\end{equation}
where $\bn$ is the unit vector in the direction of $\bq$. Before substituting into Eq.~\eqref{eq:vertex}, we notice that~\cite{Raffelt:1996wa}
\begin{equation}
    \Pi_{\rm L}=\sum_j \omega_{{\rm P},j}^2 \frac{Q^2}{\Omega^2-\beta^2_j q^2}\left[1-G\left(\frac{\beta^2_j q^2}{\Omega^2}\right)\right].
\end{equation}
Therefore, we finally find
\begin{equation}
    \mathcal{M}=g_{\phi\mu}-2 e^2 \gamma^0 \frac{\sum_j g_{\phi j} m_j\, \xi_j}{\Omega^2-\sum_k\frac{\omega_{{\rm P},k}^2}{1-\beta^2_k}\left[1-G\left(\beta^2_k\right)\right]}\,.
    \label{Eq:induced_amp}
\end{equation}
We have implicitly assumed here that the mass of the scalar particle is much below any other scale, and in particular much below the mass of the electron, so that the deviations from the light cone due to the scalar mass are much smaller than the deviations due to $\Pi_{{\rm L},e}$. 

Notice that $G(\beta^2_k)\to1$ for $\beta_k\to 1$, so for the ultrarelativistic electrons we can use the approximate expression
\begin{equation}
    \frac{1-G(\beta^2_e)}{1-\beta_e^{2}}\simeq-\frac{3\log(1-\beta^{2}_e)}{2}.
\end{equation}
The divergence of this expression for $\beta_e\to 1$ is not surprising, as a plasma of massless particles can freely radiate massless bosons. Of course, this never happens because electrons have a nonzero mass. In fact, the numerator presents a similar behavior
\begin{equation}
    \xi_j=\frac{1}{4\pi^2}\int_0^{p_{\rm F}}\frac{pdp}{E_\bp}\log\left(\frac{E_\bp+p}{E_\bp-p}\right).
\end{equation}
For ultrarelativistic electrons, this becomes
\begin{equation}
    \xi_e\simeq\frac{p_{{\rm F},e}}{2\pi^2}\log\left(\frac{2p_{{\rm F},e}}{m_e^*}\right),
    \label{Eq:xi_e}
\end{equation}
which presents indeed a similar logarithmic enhancement. In particular, for typical neutron star conditions, $p_{{\rm F},e}\simeq180\,{\rm MeV}$, $p_{{\rm F},\mu}\simeq150\,{\rm MeV}$, and $\xi_\mu\sim4\,{\rm MeV}$.
Numerically, the coefficient of the medium-induced $\gamma^0$ interaction in Eq.~\eqref{Eq:induced_amp} is typically at the level of $5\%$ of the tree-level muon coupling.

By the same token, however, also a coupling to protons and electrons is naturally induced. Assuming a purely muonic boson, with $g_{\phi p}=g_{\phi e}=0$, we find an effective operator coupling to electrons
\begin{equation}
    \hat{g}_{\phi e} \gamma^0 \simeq \frac{2 e^2 g_{\phi \mu} m_\mu\, \xi_\mu}{\sum_k\frac{\omega_{{\rm P},k}^2}{1-\beta^2_k}\left[1-G\left(\beta^2_k\right)\right]}\gamma^0\,,
    \label{Eq:hat_g_phie}
\end{equation}
whose matrix element scales with the electron energy rather than its mass, due to the factor $\gamma^0$. In fact, as we will see, electromagnetic scattering involving ultrarelativistic electrons will provide the dominant contribution to the scalar emissivity. Similarly, an effective operator coupling to protons appears
\begin{equation}
    \hat{g}_{\phi p} \gamma^0  \simeq \hat{g}_{\phi p} \simeq\frac{2 e^2 g_{\phi \mu} m_\mu\, \xi_\mu}{\sum_k\frac{\omega_{{\rm P},k}^2}{1-\beta^2_k}\left[1-G\left(\beta^2_k\right)\right]}\,,
    \label{Eq:hat_g_phip}
\end{equation}
where we replaced $\gamma^0\to 1$ for the nonrelativistic proton spinor. This in-medium proton coupling contributes to scalar emissivity through electromagnetic and nuclear collisions.

\subsection{Pseudoscalars}

In this case, mixing with photons is forbidden by parity, so there are no loop-induced corrections to the vertex.

\section{Scalar emission}
\label{sec:scalar}

In this section, we compute the emission rate for a scalar boson $\Phi=\phi$, outlining the key steps and approximations involved. These results provide a general framework that will be extended to the cases of vector and pseudoscalar bosons in the subsequent sections.

\subsection{Amplitude and squared matrix element}

The amplitude for the scalar emission in the general expression
Eq.~\eqref{Eq:bremss} 
for $\mu$--$\mu$ scattering ($X = \mu$) is
\begin{align}
    &\mathcal{M}_\phi^{(\mu\mu)} = - e^2 g_{\phi \mu} m_\mu\left(\frac{1}{K\cdot P_1}{+}\frac{1}{K\cdot P_2} {-} \frac{1}{K\cdot P_3}{-}\frac{1}{K\cdot P_4}\right)\nonumber\\[1ex]
    &\kern2em\times D_{\alpha\beta}(Q)\,j^\alpha(P_1,P_3) \, j^\beta(P_2,P_4)+ \left(P_3\leftrightarrow P_4\right),
    \label{Eq:M_mumu}
\end{align}
where $j^\alpha(P_i,P_j) = \bar{u}(P_i)\gamma^\alpha u(P_j)$ denotes the leptonic vector current, and $D^{\alpha\beta}(Q)$ is the photon propagator transferring the four-momentum $Q=(\Omega,\bq)\approx P_3-P_1 \approx P_2-P_4$ between the two leptonic currents, where the approximation refers to the limit of soft radiation so that $K$ does not contribute to $Q$. The contribution denoted by $(P_3\leftrightarrow P_4)$ denotes the amplitude from the crossed ($u$-channel) diagrams. Here and everywhere, in the denominators we always simplify $(P_i\pm K)^2-m_\mu^2\simeq \pm 2P_i \cdot K$, neglecting the term $K^2=m_\phi^2$. This is justified because the plasma can efficiently emit bosons with masses of order $T$ or smaller, and therefore negligible in comparison to the Fermi momentum $m_\phi\sim T\ll p_{F,\mu}$; we will later account for the corrections in the scalar boson emissivities arising from terms of order $m_\phi/T$.

In the nonrelativistic limit, the amplitudes associated with emission from different external legs cancel, which could raise concerns about neglecting corrections of order $T/\mu_\mu$. However, because $T/\mu_\mu \ll \beta_\mu$, with $\beta_\mu$ the characteristic muon velocity, the leading contributions beyond the cancellation arise from relativistic effects.  In other words, the
soft-radiation approximation is much more appropriate
than using nonrelativistic muons.

The amplitudes for electron-muon and proton-muon scattering are similar, now without crossed diagrams, and emission only from muons (particles 1 and~3)
\begin{align}
    \mathcal{M}_\phi^{(\mu e)} &=  \mathcal{M}_\phi^{(\mu p)} = - e^2 g_{\phi \mu} m_\mu\left(\frac{1}{K\cdot P_1}- \frac{1}{K\cdot P_3}\right)\nonumber\\[1ex]
    &\kern1em\times D_{\alpha\beta}(Q)\,j^\alpha(P_1,P_3) \, j^\beta(P_2,P_4)\,.
    \label{Eq:M_ee}
\end{align}

We include protons only when they are not superconducting and thus do not form Cooper pairs. As discussed in Sec.~\ref{sec:screening}, in this case the transverse photon modes are Landau damped, and the appropriate description is given by Eq.~\eqref{Eq:Landau_Damping}, where $\Pi_{\rm T}$ is dynamical and depends on the energy and momentum transfer. In contrast, in the super\-conducting phase, the population of unpaired protons is exponentially suppressed by a factor $\exp(-\Delta_p/T)$, as most protons are bound into Cooper pairs. Therefore, their contribution to scattering becomes negligible, while transverse photons acquire an effective screening characterized by the Meissner scale of Eq.~\eqref{Eq:Meissner_Scale}.

Lastly, as discussed in Sec.~\ref{sec:scalar_coupling}, a scalar coupled in vacuum only to muons acquires an effective in-medium coupling to electrons given by Eq.~\eqref{Eq:hat_g_phie}. The corresponding amplitude for the scalar emission in the $e-e$ scattering is therefore
\begin{align}
    &\mathcal{M}_\phi^{(ee)} = - e^2 \hat{g}_{\phi e} \,\varepsilon_e\left(\frac{1}{K\cdot P_1}{+}\frac{1}{K\cdot P_2} {-} \frac{1}{K\cdot P_3}{-}\frac{1}{K\cdot P_4}\right)\nonumber\\[1ex]
    &\kern2em\times D_{\alpha\beta}(Q)\,j^\alpha(P_1,P_3) \, j^\beta(P_2,P_4)+ \left(P_3\leftrightarrow P_4\right).
    \label{Eq:M_e_e}
\end{align}
For scattering of an electron with a different charged particle
$X=\mu,p$, assigning $P_{1,3}$ to the electron, the contribution
from radiation off the electron leg reads
\begin{align}
\mathcal{M}_{\phi,e}^{(eX)}
={}&
-e^2\hat g_{\phi e}\varepsilon_e
\left(
\frac{1}{K\cdot P_1}
-\frac{1}{K\cdot P_3}
\right)
\nonumber\\[1ex]
&\times
D_{\alpha\beta}(Q)\,
j^\alpha(P_1,P_3)\,
j^\beta(P_2,P_4).
\label{Eq:M_e_X}
\end{align}
To proceed, we rely on two driving approximations: (i)~Particles lie at the Fermi surface, so their energies and momenta are fixed. Deviations will later be integrated out through the degenerate distribution functions.
(ii)~Near-forward scattering. Typical scattering momenta are of the order of $q\sim m_{D,M,\Lambda}\ll p_{\rm F}$. Under these approximations, the amplitudes for emission from the muon before and after the scattering combine in a particularly simple way
\begin{equation}\label{eq:dipole_expression}
    \frac{1}{K\cdot P_1}-\frac{1}{K\cdot P_3}\simeq -\frac{\bq\cdot\bk}{ (\omega\varepsilon_\mu-\bp_1\cdot\bk)^2},
\end{equation}
where we recall that $\varepsilon_\mu$ is the muon energy at the Fermi surface. We also recall that we are considering bosons with mass so that $\bk^2=\omega^2-m_\phi^2$.

For $\mu$--$\mu$ scattering, a similar expression pertains to the second muon with momenta $P_2$ and $P_4$. In the near-forward scattering limit, we can neglect the interference of the two amplitudes because two incoming muons with different directions cannot interfere in forward scattering. Formally, the terms with $(P_3\leftrightarrow P_4)$ are suppressed by a factor $q^2/p_{{\rm F},\mu}^2$, due to the different denominators in the photon propagator. Hence, it is enough to include the direct diagrams and later multiply by a factor of~2 to account for the two incoherent contributions. 

In Eq.~\eqref{eq:dipole_expression}, we have implicitly assumed that the for\-ward-scattering nature of electromagnetic interaction is much more pronounced than the relativistic beaming by the emitting muons. Indeed, in the denominator we have neglected a term $Q\cdot K\sim\omega\,q \cos\theta_q$ in comparison with $P_1\cdot K\sim\omega \mu_\mu(1-\cos \theta_1)$, where $\theta_q$ and $\theta_1$ are the angles formed by $\bq$ and $\bp_1$ along the direction $\bk$ of the emitted boson, respectively. As we show below, the kinematics of the process enforces $\cos\theta_q\sim\sin\theta_1$~(see Eq.~\eqref{eq:costhetaq}), and a small-angle expansion around $\theta_1\simeq m_\mu/\mu_\mu$ reveals the behavior $Q\cdot K\sim q\,\omega\,m_\mu/\mu_\mu$ and $P_1\cdot K\sim \omega \,m_\mu^2/\mu_\mu\,$. Since the typical scattering momenta are of the order of $q\sim m_{D,M,\Lambda}$, this imposes a condition of applicability
\begin{equation}
    m_{D,M,\Lambda}\ll m_\mu\,.
    \label{Eq:Condition}
\end{equation}
This condition is robustly realized in the mid to outer NS core, where $m_{D}\sim 40\,\mathrm{MeV}$ and $m_M\sim 8\,\mathrm{MeV}$ when protons are superconducting. Moreover, Eq.~\eqref{Eq:Condition} is satisfied even more accurately for transverse photons in the normal-conducting regime, for which $m_{\Lambda} \lesssim m_M$.
On the other hand, the boson emission can also take place from the electron legs due to the non-negligible effective in-medium coupling. Therefore, in order to apply the same approximation to contributions arising from radiating electrons, the condition $m_{D,M,\Lambda}\ll m_e^*$ must also hold. This condition is only satisfied by scatterings mediated by transverse photons in nonsuperconducting regions, for which  $q\sim m_{\Lambda}$, whereas it fails for the typical momentum carried by longitudinal photons $q\sim m_D\simeq40\,{\rm MeV}$. Nevertheless, as discussed above, the contributions from transverse photons largely dominate the emissivities, while terms $\propto 1/m_{D}$ can only provide corrections of at most $10\%$. Therefore, since we aim primarily at a parametric understanding of the emissivity which correctly captures its order of magnitude, we enforce the condition $Q\cdot K\ll P_1\cdot K$ also in diagrams with scalar bosons attached to the electron legs, thereby retaining the dominant scaling even in the ultrarelativistic electron regime.

Finally, if we neglect all factors proportional to $\bq$ in the contraction of the hard part of the scattering given by $D_{\alpha\beta}j^\alpha j^\beta$, assuming near-forward scattering again, the currents are simply $j^\alpha(P_1,P_3)=2P_1^\alpha \delta_{s_1,s_3}$ and $j^\beta(P_2,P_4)=2P_2^\beta \delta_{s_2,s_4}$, where the Kronecker delta enforces conservation of spin between the incoming and outgoing muon. The current-current amplitude can then be written in a rather generic form
\begin{widetext}
\begin{equation}
    D_{\alpha\beta}(Q) j^\alpha(P_1,P_3)j^\beta(P_2,P_4)=4\left[\frac{\varepsilon_{\mu}\varepsilon_{X}}{q^2+\Pi_{\rm L}}-\frac{\bp_1\cdot\bp_2-\frac{\bp_1\cdot \bq\,\bp_2\cdot\bq}{q^2}}{q^2+\Pi_{\rm T}}\right]\simeq 4\varepsilon_{\mu}\varepsilon_{X}\left[\frac{1}{q^2+\Pi_{\rm L}}-\frac{\beta_\mu\beta_{X}\cos\theta_{\bp_1,\bp_2}}{q^2+\Pi_{\rm T}}\right],
\end{equation}
where in the last equality we have used the near-forward approximation, since, as we will see, energy conservation implies $\bp_1\cdot \bq=-\bp_2\cdot\bq=-q^2/2$, so only the product $\bp_1\cdot\bp_2$ needs to be kept to lowest order in $q^2$. Here, $\varepsilon_{X}$ and $\beta_{X}$ indicate the Fermi energy and velocity of fermion $X=\{\mu,e,p\}$. Therefore, after summing over initial and final spins, the squared matrix elements for processes involving radiating muons read
\begin{align}
    \frac{1}{\mathcal{S}}\sum_{\rm spins} \left|\mathcal{M}_\phi^{(\mu\mu)}\right|^2 &= 32 e^4 \hat{g}_{\phi\mu}^2 m_\mu^2 \frac{q^2 x_q^2\beta_\phi^2}{\omega^2}\left(\frac{1}{(1-\beta_\phi\beta_{\mu} x_1)^2}-\frac{1}{(1-\beta_\phi\beta_{\mu} x_2)^2}\right)^2 \left|\frac{1}{q^2+\Pi_{\rm L}}-\frac{\beta_{\mu}^2\,x_{12}}{q^2+\Pi_{\rm T}}\right|^2\,,\label{Eq:M2mumu}\\
    \nonumber\\
    \sum_{\rm spins} \left|\mathcal{M}_\phi^{(\mu X)}\right|^2 &= 64 e^4 \hat{g}_{\phi\mu}^2 m_\mu^2 \left(\frac{\varepsilon_{X}}{\varepsilon_{\mu}}\right)^{\!2}\!\frac{q^2 x_q^2\beta_\phi^2}{\omega^2(1-\beta_\phi\beta_{\mu} x_1)^4} \left|\frac{1}{q^2+\Pi_{\rm L}}-\frac{\beta_{X}\beta_{\mu}\,x_{12}}{q^2+\Pi_{\rm T}}\right|^2\,,\label{Eq:M2ee}
\end{align}
where $X=e,p$ denotes contributions from electrons and protons as scattering partners. Notice that here we introduce an effective muon-scalar coupling 
\begin{equation}
    \hat{g}_{\phi \mu} =g_{\phi\mu}\left(1+\frac{2e^2 m_\mu\, \xi_\mu}{\sqrt{1-\beta_\mu^2}\,\sum_k\frac{\omega_{{\rm P},k}^2}{1-\beta^2_k}\left[1-G\left(\beta^2_k\right)\right]}\right)\,
\end{equation}
which, according to Sec.~\ref{sec:scalar_coupling}, accounts for the medium response to the change in muon velocity. Notice that the $\gamma_0$ in the in-medium muon interaction vertex gives rise to the Lorentz boost factor $1/\sqrt{1-\beta_\mu^2}$ in the effective muonic coupling.

Analogously, the matrix elements arising from emitting electrons read 
    \begin{align}
    \frac{1}{\mathcal{S}}\sum_{\rm spins} \left|\mathcal{M}_\phi^{(ee)}\right|^2 &= 32 e^4 \hat{g}_{\phi e}^2 \varepsilon_e^2 \frac{q^2 x_q^2\beta_\phi^2}{\omega^2}\left(\frac{1}{(1-\beta_\phi\beta_{e} x_1)^2}-\frac{1}{(1-\beta_\phi\beta_{e} x_2)^2}\right)^2 \left|\frac{1}{q^2+\Pi_{\rm L}}-\frac{\beta_{e}^2\,x_{12}}{q^2+\Pi_{\rm T}}\right|^2\,,\\
    \nonumber\\
    \sum_{\rm spins} \left|\mathcal{M}_\phi^{(e X)}\right|^2 &= 64 e^4 \hat{g}_{\phi e}^2 \varepsilon_e^2 \left(\frac{\varepsilon_{X}}{\varepsilon_{e}}\right)^{\!2}\!\frac{q^2 x_q^2\beta_\phi^2}{\omega^2(1-\beta_\phi\beta_{e} x_1)^4} \left|\frac{1}{q^2+\Pi_{\rm L}}-\frac{\beta_{X}\beta_{e}\,x_{12}}{q^2+\Pi_{\rm T}}\right|^2\,,
\end{align}
and here $X=\mu,p$.
\end{widetext}

Additionally, $x_v$ denotes the cosine of the angle between vectors $v = \{\bq,\bp_1,\bp_2,\bp_3,\bp_4\}$ and $\bk$, and $x_{12}$ is the cosine of the angle between $\bp_1$ and $\bp_2$. 

Notice that the squared matrix element for identical scattering particles contains additional terms due to emission from both fermionic legs and is larger by a factor of 2 due to the exchange diagrams. This enhancement is however canceled by the symmetry factor $\mathcal{S}=4$ for identical particles. We underline that the contribution from the transverse modes is proportional to the velocities of the particles involved in the scattering process and becomes negligible for nonrelativistic species.
Moreover, we observe that the squared matrix element factorizes into a term associated with the emission of long-wavelength scalar radiation and a term encoding lepton--lepton or lepton--proton scattering (the squared modulus). In principle, this term corresponds to well-established results derived in the context of thermal conductivity calculations in NS environments~\cite{Gnedin:1995lgf, Shternin:2007ee}.

\subsection{Phase-space integration}

The phase-space integral in Eq.~\eqref{Eq:Emissivity} can be significantly simplified by assuming extremely degenerate fermions, whose distribution functions set their momenta to a narrow region around the Fermi surface. The boson emission spectrum, namely the number of particles emitted per unit volume per unit time per unit energy, is
\begin{widetext}
\begin{eqnarray}\label{eq:master_integral}
    {\cal N}_\phi&=&\frac{dN_\phi}{dt dV d\omega}=\frac{\omega k}{2\pi^2}\prod_i \int \frac{d^3\bp_i}{2E_i(2\pi)^3}\frac{1}{2\omega}(2\pi)^4\delta^3\left(\bp_1+\bp_2-\bp_3-\bp_4\right) \delta\left(E_1+E_2-E_3-E_4-\omega\right)
    \nonumber\\
    &&\kern25em{}\times f_1 f_2 (1-f_3)(1-f_4)\sum_{\rm spins}\frac{|\mathcal{M}|^2}{\mathcal{S}}.
\end{eqnarray}
It can be integrated over the fermion energies by using the universal identity for degenerate fermions, which we have already adopted in previous analogous conditions~\cite{Bottaro:2024ugp, Fiorillo:2025zzx}
\begin{align}
    \int &dE_1dE_2dE_3dE_4\,f_1f_2(1-f_3)(1-f_4)\delta(E_1+E_2-E_3-E_4-\omega)\simeq T^3 J\left(\tilde{\omega}\right)\,,
    \label{Eq:E_integral}
\end{align}
where we introduce
\begin{equation}
    J(x)=\frac{2\pi^2x}{3(e^x-1)}\left(1+\frac{x^2}{4\pi^2}\right)\,,
\end{equation}
and define $\tilde{\omega}=\omega/T$ and, for later use, $\tilde{k}=k/T$. This integration fixes all the momenta involved in the expressions over the corresponding Fermi momentum $p_i\simeq p_{{\rm F},X}$, in which $X=\{\mu,e,p\}$ with the understanding that $p_{1,3}=\{p_{{\rm F},\mu},\, p_{{\rm F},e}\}$ and $p_{2,4}=p_{{\rm F},X}$, so that
\begin{equation}
    \mathcal{N}_\phi=\frac{4\pi\,T^3}{2^5(2\pi)^{11}}\,J(\tilde{\omega})\frac{k}{\omega^2}\int \prod_i p_i d\Omega_i\,\delta^3(\bp_1+\bp_2-\bp_3-\bp_4)\sum_{\rm spins}\mathcal{S}^{-1}|\mathcal{M}|^2\omega^2\,.
\end{equation}
We can also reintroduce the integral over the full three-momenta for $\bp_3$ and $\bp_4$ through the identity $p_i d\Omega_i=2\int d^3\bp_i \delta(\bp_i^2-p_i^2)$, so that the delta enforcing momentum conservation is removed by imposing the condition $\bp_4=\bp_1+\bp_2-\bp_3$, and the integration over $\bp_3$ is converted to an integration over $\bq=\bp_3-\bp_1$. Overall we then find
\begin{equation}\label{Eq:Q_after_Omega}
    \mathcal{N}_\phi=\frac{T^3}{4096\,\pi^{10}}J(\tilde\omega)\frac{k}{\omega^2} \int p_1 d\Omega_1 p_2 d\Omega_2 d^3\bq \delta(|\bp_1+\bq|^2-p_1^2)\delta(|\bp_2-\bq|^2-p_2^2)\sum_{\rm spins}\mathcal{S}^{-1}|\mathcal{M}|^2\omega^2.
\end{equation}
\end{widetext}
This is a particularly symmetric expression that can be further simplified by integrating $\bq$ over the delta functions. Noting that the squared matrix amplitude contains a factor $(\bq\cdot\bn)^2$, where $\bn$ is the direction of the outgoing scalar particle, we need to determine the integral
\begin{equation}\label{Eq:Deltaq}
    I=\int d^3\bq \,\delta(\bq^2+2\bp_1\cdot\bq)\,\delta(\bq^2-2\bp_2\cdot\bq)\, (\bq\cdot\bn)^2.
\end{equation}
In the limit of very small momentum exchange, we may neglect $q^2$ and simply enforce $\bq$ to be orthogonal to both $\bp_1$ and $\bp_2$. In this approximation, a simple geometrical calculation shows that
\begin{equation}\label{eq:costhetaq}
    x_q=\pm\frac{\sqrt{1-x_1^2}\sqrt{1-x_2^2}\sin \phi}{\sqrt{1-x_{12}^2}},
\end{equation}
and the cosine of the angle between $\bp_1$ and $\bp_2$ can be expressed as $x_{12}=x_1 x_2+\sqrt{1-x_1^2}\sqrt{1-x_2^2}\cos\phi$, where $\phi$ is the azimuthal angle of $\bp_2$ around the plane formed by $\bp_1$ and $\bk$. Notice that there are two separate solutions, corresponding to the two signs. Then the delta functions $\delta(2\bp_1\cdot\bq) \delta (2\bp_2\cdot\bq)$ are easily integrated, giving
\begin{equation}\label{eq:def_Upsilon}
    I=\int\!dq\, \frac{q^2(1-x_1^2)(1-x_2^2)\sin^2\phi}{2 p_1 p_2(1-x_{12}^2)^{3/2}}=\int dq \Upsilon,
\end{equation}
already summed over both solutions. The apparent divergence of the integral for $x_{12}\to 1$ is caused by our approximation of very small $q$, requiring $q\ll |\bp_1\times \bp_2|/|\bp_1+\bp_2|$. As we will see, the integral itself is dominated by small values of $q$ for which this approximation is valid. We may therefore extend the integration to $q\to +\infty$.

\begin{widetext}
With this notation, we ultimately find for the emission rate
\begin{equation}
\begin{split}
    \mathcal{N}_\phi=\frac{T^3 p_1 p_2}{2048\pi^9}J(\tilde\omega)\frac{k}{\omega^2}\int dx_1 dx_2 d\phi dq\,\Upsilon\sum_{\rm spins}\frac{\mathcal{S}^{-1}|\mathcal{M}|^2\omega^2}{q^2x_q^2}.
\end{split}
\end{equation}

To proceed further, it is convenient to extract from the squared matrix element the part depending on $\bq$ which can be integrated analytically, albeit through some lengthy expressions. We explicitly introduce therefore
\begin{equation}
    \sum_{\rm spins}\frac{\mathcal{S}^{-1}\left|\mathcal{M}_\phi^{(iX)}\right|^2\omega^2}{q^2x_q^2}=F_\phi^{(iX)}\left[\frac{1}{(q^2+m_D^2)^2}
    +\frac{\beta_i^2\beta_{X}^2x_{12}^2}{(q^2+m_T^2)^2}
    -\frac{2\beta_i\beta_{X}x_{12}}{(q^2+m_D^2)(q^2+m_T^2)}\right],
\end{equation} 
with $i=\mu,e$ denoting the radiating particle and $X=\mu,e,p$ its scattering partner. In particular, for identical-particle scatterings the factors $F_\phi^{(ij)}$ are
\begin{align}  &F_\phi^{(\mu\mu)}=32e^4\hat{g}_{\phi\mu}^2m_\mu^2\beta_\phi^2\!\left(\frac{1}{\!(1-\beta_\mu\beta_\phi x_1)^2}{-}\frac{1}{(1-\beta_\mu\beta_\phi x_2)^2}\!\right)^2\!\!,\\
&F_\phi^{(ee)}=32e^4\hat{g}_{\phi e}^2\varepsilon_e^2\beta_\phi^2\!\left(\frac{1}{\!(1-\beta_e\beta_\phi x_1)^2}{-}\frac{1}{(1-\beta_e\beta_\phi x_2)^2}\!\right)^2\!\!,
\end{align}
while for non-identical scattering partners we have
\begin{align}    
    &F_\phi^{(\mu X)}=64 e^4 \hat{g}_{\phi\mu}^2m_\mu^2\beta_\phi^2\left(\frac{\varepsilon_{X}}{\varepsilon_{\mu}}\right)^2 \frac{1}{(1-\beta_\mu\beta_\phi x_1)^4}\,,\\
    &F_\phi^{(e X)}=64 e^4 \hat{g}_{\phi e}^2\varepsilon_{e}^2\beta_\phi^2\left(\frac{\varepsilon_{X}}{\varepsilon_{e}}\right)^2 \frac{1}{(1-\beta_e\beta_\phi x_1)^4}\,.
\end{align}
where $X\neq\mu$ and $X\neq e$ in the two equations, respectively. 

The integrals over $q$ can now be reduced to a sequence of elementary integrals of the form
\begin{equation}
    \Psi(m_A,m_B)=\int dq \frac{\Upsilon}{(q^2+m_A^2)(q^2+m_B^2)}\,.
    \label{Eq:Psi_SC}
\end{equation}
With the function $\Upsilon$ defined in Eq.~\eqref{eq:def_Upsilon}, we have
\begin{equation}
    \Psi(m_A,m_B)\simeq \frac{\pi(1-x_1^2)(1-x_2^2)\sin^2\phi}{4p_1 p_2(m_A+m_B)(1-x_{12}^2)^{3/2}}.
    \label{Eq:Psi_function}
\end{equation}
Then the emissivities can be expressed in terms of six fundamental integrals
\begin{align}
    J_{n}(\beta_i)&=\int \frac{dx_1 dx_2 d\phi(1-x_1^2)(1-x_2^2)x_{12}^{n}\sin^2\!\phi}{(1-\beta_i\beta_\phi x_1)^4 (1-x_{12}^2)^{3/2}}\,,\\
    K_{n}(\beta_i)&=\int \frac{dx_1 dx_2 d\phi(1-x_1^2)(1-x_2^2)x_{12}^{n}\sin^2\!\phi}{(1-\beta_i\beta_\phi x_1)^2 (1-\beta_i\beta_\phi x_2)^2 (1-x_{12}^2)^{3/2}}\,,\label{eq:K_definition}
\end{align}
with $i=\mu,e$ for radiating muons and electrons, respectively. 
In terms of these integrals, the emission rates become
\begin{align}
    &\mathcal{N}_\phi^{(\mu\mu)}=\frac{\alpha^2 \hat{g}_{\phi\mu}^2 m_\mu^2 T^3}{8\pi^6} J(\tilde\omega) \frac{k}{\omega^2}\beta_\phi^2\left[\frac{J_0-K_0}{2m_D}+\frac{\beta_\mu^4(J_2-K_2)}{2m_T}-\frac{2\beta_\mu^2(J_1-K_1)}{m_D+m_T}\right],\\
    &\mathcal{N}_\phi^{(ee)}=\frac{\alpha^2 \hat{g}_{\phi e}^2 \varepsilon_e^2 T^3}{8\pi^6} J(\tilde\omega) \frac{k}{\omega^2}\beta_\phi^2\left[\frac{J_0-K_0}{2m_D}+\frac{\beta_e^4(J_2-K_2)}{2m_T}-\frac{2\beta_e^2(J_1-K_1)}{m_D+m_T}\right],\\
    &\mathcal{N}_\phi^{(\mu X)}=\frac{\alpha^2 \hat{g}_{\phi\mu}^2 m_\mu^2 T^3}{8\pi^6}J(\tilde\omega) \frac{k}{\omega^2}\beta_\phi^2\left(\frac{\varepsilon_{X}}{\varepsilon_{\mu}}\right)^{\!2}\!\left[\frac{J_0}{2m_D}+\frac{\beta_\mu^2\beta_X^2 J_2}{2m_T}-\frac{2\beta_\mu\beta_X J_1}{m_D+m_T}\right]\,,\\
    &\mathcal{N}_\phi^{(e X)}=\frac{\alpha^2 \hat{g}_{\phi e}^2 \varepsilon_e^2 T^3}{8\pi^6}J(\tilde\omega) \frac{k}{\omega^2}\beta_\phi^2\left(\frac{\varepsilon_{X}}{\varepsilon_{e}}\right)^{\!2}\!\left[\frac{J_0}{2m_D}+\frac{\beta_e^2\beta_X^2 J_2}{2m_T}-\frac{2\beta_e\beta_X J_1}{m_D+m_T}\right].
\end{align}
\end{widetext}
The integrals $J_n$ can be done exactly by introducing as integration variables the angle of $\bp_2$ with respect to $\bp_1$, i.e.\ $\theta_{\bp_1\bp_2}$ and the azimuthal angle $\psi$ of $\bp_2$ around the plane containing $\bp_1$ and $\bk$. One then has the exact identity $\sin\theta_{\bp_1\bp_2}\sin\psi=\sqrt{1-x_2^2} \sin\phi$, so $J_n$ becomes
\begin{align}
    J_n=&\int dx_1 d\theta_{\bp_1\bp_2} d\psi \frac{(1-x_1^2)\,\cos^n\theta_{\bp_1\bp_2}\,\sin^2\psi\,}{(1-\beta_i \beta_\phi x_1)^4\sqrt{1-x_{12}^2}},
    \label{Eq:Jn}
\end{align}
so ultimately we have
\begin{equation}
    J_0= 2J_2= \frac{4\pi^2}{3(1-\beta_i^2 \beta_\phi^2)^2}
    \quad\text{and}\quad 
    J_1=0.
    \label{Eq:J0_J2}
\end{equation}
The region of near-collinear scattering $\theta_{\bp_1\bp_2}\to 0$, where the near-massless approximation for the functions $\Psi$ breaks down, does not lead to any divergence and contributes subdominantly, and therefore the result is not sensitive to our earlier approximation. 

\begin{widetext}
For $K_n$, we may use a similar change to obtain
\begin{equation}
    K_n=\int dx_1 dx_{12}d\psi \frac{(1-x_1^2)(1-x_{12}^2)x_{12}^n \sin^2\psi}{(1-x_{12}^2)^{3/2}(1-\beta_i \beta_\phi x_1)^2}\,\frac{1}{\left[1-\beta_i \beta_\phi(x_1 x_{12}+\sqrt{1-x_1^2}\sqrt{1-x_{12}^2}{\cos\psi})\right]^2},
\end{equation}
which can be integrated exactly over $\psi$ to give
\begin{equation}\label{eq:numerical_K}
    K_n=\int dx_1 dx_{12}\frac{2\pi x_{12}^n}{(1-x_{12}^2)^{3/2}(1-\beta_i \beta_\phi x_1)^2\beta_i^2 \beta_\phi^2}\left[\frac{1-\beta_i \beta_\phi x_1 x_{12}}{\sqrt{(1-\beta_i \beta_\phi x_1 x_{12})^2-\beta_i^2 \beta_\phi^2(1-x_1^2)(1-x_{12}^2)}}-1\right].
\end{equation}
\end{widetext}
Once again, despite appearances, the integral does not have any divergence for $x_{12}\to 1$, since the bracketed expression vanishes in this limit, justifying a posteriori our approximation $q\ll |\bp_1\times \bp_2|/|\bp_1+\bp_2|$ throughout the dominant range of integration.
In the limit $\beta_i\to 1$, a small-angle expansion in the original definition in Eq.~\eqref{eq:K_definition} reveals the asymptotic behavior $K_n\propto (1-\beta_i\beta_\phi)^{-3/2}$. These integrals are then easily obtained numerically and expressed through the simple fitting expressions 
\begin{equation}
    \begin{split}
        K_0&\simeq\frac{2.48}{(1-\beta_i\beta_\phi)^{3/2}}+\frac{2.69}{1-\beta_i\beta_\phi}+6.23\,,\\
        K_1&\simeq\frac{3.08}{(1-\beta_i\beta_\phi)^{3/2}}-\frac{3.06}{1-\beta_i\beta_\phi}-0.46\,,\\
        K_2&\simeq\frac{2.58}{(1-\beta_i\beta_\phi)^{3/2}}-\frac{1.45}{1-\beta_i\beta_\phi}+5.07\,,\\
        K_3&\simeq\frac{2.72}{(1-\beta_i\beta_\phi)^{3/2}}-\frac{3.32}{1-\beta_i\beta_\phi}+0.59\,,\\
    \end{split}
\end{equation}
which reproduce the required asymptotic behavior.

\subsection{Energy-loss rate}

Ultimately, the scalar energy-loss rate is obtained by
integrating the energy carried by the emitted bosons over the emission spectrum   
\begin{equation}
    Q_\phi= \int_{m_\phi}^\infty\,d\omega\,\omega \,\mathcal{N}_\phi\,.
\end{equation}
In the massless limit ($m_\phi\to0$), the boson momentum is $k\to \omega$ and the boson velocity $\beta_\phi\to 1$. The integral can then be evaluated analytically using 
\begin{equation}
     \int_0^{\infty}d{\tilde\omega}\, J(\tilde{\omega})=\frac{11\pi^4}{90}.
     \label{Eq:J_integral}
\end{equation}
Therefore, the massless scalar emissivities are simply
\begin{widetext}
\begin{align}
    Q_{\phi,0}^{(\mu\mu)}&=\frac{11\alpha^2 \hat{g}_{\phi\mu}^2 m_\mu^2 T^4}{720\pi^2}\left[\frac{\pi^2}{3(1-\beta_\mu^2)^2}\left(\frac{2}{m_D}+\frac{\beta_\mu^4}{m_T}\right)-\left(\frac{K_0}{2m_D}-2 \beta_\mu^2\frac{K_1}{m_D+m_T}+\beta_\mu^4\frac{K_2}{2m_T}\right)\right],\\
    Q_{\phi,0}^{(ee)}&=\frac{11\alpha^2 \hat{g}_{\phi e}^2 \varepsilon_e^2 T^4}{720\pi^2}\left[\frac{\pi^2}{3(1-\beta_e^2)^2}\left(\frac{2}{m_D}+\frac{\beta_e^4}{m_T}\right)-\left(\frac{K_0}{2m_D}-2 \beta_e^2\frac{K_1}{m_D+m_T}+\beta_e^4\frac{K_2}{2m_T}\right)\right]\,,
\end{align}
\begin{align}
    Q_{\phi,0}^{(\mu X)}&=\frac{11\alpha^2\hat{g}_{\phi\mu}^2m_\mu^2 T^4}{2160(1-\beta_\mu^2)^2}\left(\!\frac{\varepsilon_{X}}{\varepsilon_{\mu}}\!\right)^{\!2}\!\left[\frac{2}{m_D}+\frac{\beta_\mu^2\beta_X^2}{m_T}\right]\,,\\
    Q_{\phi,0}^{(e X)}&=\frac{11\alpha^2\hat{g}_{\phi e}^2\varepsilon_e^2 T^4}{2160(1-\beta_e^2)^2}\left(\!\frac{\varepsilon_{X}}{\varepsilon_{e}}\!\right)^{\!2}\!\left[\frac{2}{m_D}+\frac{\beta_e^2\beta_X^2}{m_T}\right].\label{Eq:Qphi0_X}
\end{align}
For massive scalars, the energy integral does not admit an analytical solution. However, the exponential suppression expected at large masses $\tilde m_\phi= m_\phi/T\gtrsim 1$ can be captured by the following fitting functions
\begin{align}
    Q_\phi^{(ii)}(\tilde m_\phi,\beta_i)&\simeq Q_{\phi,0}^{(ii)}\,\left\{1-0.71\,\tilde m_\phi^{0.29}
    \left[1-0.97\,\left(1-\beta_i^2\right)\right]\right\}\,e^{-1.06\,\tilde m_\phi}\,,\\
    Q_\phi^{(i X)}(\tilde m_\phi,\beta_i)&\simeq Q_{\phi,0}^{(i X)}\,\left\{1-0.74\,\tilde m_\phi^{0.30}
    \left[1-1.09\,\left(1-\beta_i^2\right)\right]\right\}\,e^{-0.96\,\tilde m_\phi}\,,
\end{align}
The fitting functions agree within 10\% with the numerical result over the range of $\beta_\mu$ values relevant for NS conditions.
\end{widetext}

We emphasize that the dominant contribution to the emissivities arises from scatterings mediated by transverse photons, leading to terms $\propto1/m_T$. Indeed, under typical NS conditions, both electrons and muons are relativistic and the hierarchy $m_T\ll m_D$ always applies.

The power-law dependence on $m_{D,T}$ and $1-\beta_i^2$ is in itself noteworthy. In more conventional settings, bremsstrahlung leads to logarithmic dependence on these small parameters. In our case, however, the power-law dependence stems from the extreme degeneracy of particles involved in the process, leading to the specific hierarchy $T/\varepsilon_{i}\ll m_{D,T}/\varepsilon_{i}\ll \sqrt{1-\beta_i^2}$. Hence, the parameters $m_{D,T}/\varepsilon_{i}$ and $1-\beta_i$ can never truly vanish in the approximation regime we have chosen. If $\sqrt{1-\beta_i^2}\to 0$ more rapidly than $m_{D,T}/\varepsilon_{i}$, the quadratic denominators in Eq.~\eqref{eq:dipole_expression} would not vanish at the same point, regularizing the divergence and leading to a more conventional logarithmic enhancement. Similarly, as $m_{D,T}$ becomes comparable to~$T$, the approximation of assuming all particles exactly at the Fermi surface would also break down. For $\bq\to 0$, bremsstrahlung becomes kinematically forbidden, so one would need to keep the energy conservation as an additional condition on the allowed values of~$\bq$.

As a simple confirmation of our results, we notice that the rate of muon-lepton scattering, as measured by the thermal conductivity, scales as $m_D^{-3}$ (see Eq.~(37) in Ref.~\cite{Gnedin:1995lgf}). For the bremsstrahlung emission rate, this should be multiplied by the square of the momentum transfer---the probability of radiating a boson goes as the square of the velocity change, a well-known property of radiation in electrodynamics, which in our results appears through the $(\bq\cdot\bk)^2$ factor in the squared matrix element---so that we recover the $m_D^{-1}$ scaling of the cooling rate.

In the case of a superconducting medium, the integration region dominating the energy loss is that of unscreened excitations with $q\gtrsim m_M$, i.e., the lowest-momentum excitations which survive screening. Therefore, the transverse excitations with $q\ll m_M$ are screened, and one sets $m_T=m_M$.

Similarly, in a nonsuperconducting medium, the minimum momentum of excitations that are neither screened nor damped is $m_{\Lambda}$, as excitations with longer wavelengths are damped. Therefore, by a direct extension of our previous argument, the energy loss will be dominated by transverse excitations with $q\gtrsim m_\Lambda$. We may therefore obtain the energy loss in the normal state simply by replacing $m_T\to m_{\Lambda}$, and accounting for the $X=p$ contributions in Eq.~\eqref{Eq:Qphi0_X}.

\subsection{Relativistic boosting}

The relativistic velocity of the emitting leptons introduces a peculiar dependence on their Lorentz factors. Our complete calculation reveals that for tree-level processes with radiating muons 
\begin{equation}
    Q_{\phi}^{(\mu)}\sim\frac{\varepsilon_\mu}{1-\beta_\mu^2}\,, 
\end{equation}
while emission from radiating electrons scales as 
\begin{equation}
    Q_{\phi}^{(e)}\sim\frac{\varepsilon_e}{(1-\beta_e^2)^2}\,. 
\end{equation}
These scalings are important enough that they should be understood qualitatively, beyond the somewhat opaque field-theoretic calculation.

From classical electrodynamics, we know that a particle accelerating in a given external electric field radiates a total energy proportional to its squared Lorentz factor. However, the electrons contributing to muonic boson emission are not radiating conventional transverse electromagnetic waves, but rather longitudinal plasma waves that are immediately converted into scalar bosons. For transverse waves, the radiation parallel to the moving particle is suppressed by the angular factor $\sin^2\theta$, where $\theta$ is the angle between the particle acceleration and the radiation. This is the same factor that appears in Larmor’s formula, and simply expresses the fact that light must be polarized transversely to the particle velocity. Since the typical angle of emission of ultrarelativistic particles is $\theta\sim \gamma^{-1}$, this factor gives a characteristic suppression $\gamma^{-2}$ to the overall power. Instead, the radiation of longitudinal waves, which are not polarized, is not suppressed, enhancing its overall power in comparison to transverse radiation by a factor $\gamma^2$, and explaining the characteristic scaling $Q_\phi^{(e)}\propto \gamma_e^4$.

For scalar radiation directly from muons, this enhancement is removed not because of the polarization factor $\sin\theta$, but rather because scalars couple to the rest mass, not the energy, of the radiating particle, which introduces a factor $\gamma^{-2}$. Indeed, we find $Q_\phi^{(\mu)}\propto \gamma_\mu^2$. 

In the field-theoretic language, the additional relativistic enhancement emerges from the in-medium scalar–electron coupling itself: mediated by the longitudinal electromagnetic field, this interaction ties scalar radiation directly to the electron's own charge density. As the medium breaks Lorentz invariance and defines a physically preferred rest frame, the electron charge density is genuinely boosted in this frame, giving rise to the structure $\sim\bar{u}\gamma^0u$. 

The same enhancement is absent for nonrelativistic protons, and scalar radiation from electromagnetic proton scattering would be parametrically suppressed by the coefficient in front of the in-medium coupling in Eq.~\eqref{Eq:hat_g_phip}. We therefore neglect these processes in the following.

\section{Vector emission from electromagnetic scatterings}

\label{sec:vector}

\subsection{Amplitude and squared matrix element}

As in the scalar case, muonic vector bosons ($\Phi=V$) are predominantly produced via bremsstrahlung
\begin{equation}
\mu(P_1)+X(P_2)\to\mu(P_3)+X(P_4)+V(K)\,,
\label{Eq:bremss_Vec}
\end{equation}
where again $X=\{\mu,e,p\}$. The corresponding emissivity is obtained from Eq.~\eqref{Eq:Emissivity}, the only modification being the appropriate matrix element. For $\mu$--$\mu$ scattering ($X = \mu$), the amplitude takes the form
\begin{align}
    \mathcal{M}_V^{(\mu\mu)} &= - e^2 g_{V\!\mu}\left(\frac{P_1^\alpha}{K\cdot P_1}+\frac{P_2^\alpha}{K\cdot P_2} - \frac{P_3^\alpha}{K\cdot P_3}-\frac{P_4^\alpha}{K\cdot P_4}\right)\nonumber\\
    &\times \epsilon_{\alpha}^*(K) D_{\mu\nu}(Q)j^\mu(P_1,P_3) \, j^\nu(P_2,P_4)\nonumber\\
    &+ \left(P_3\leftrightarrow P_4\right)\,,
    \label{Eq:M_mumu_Vec}
\end{align}
where the only new quantity is the polarization vector $\epsilon(K)$. Likewise, when $X=e$ or $p$, the amplitude is 
\begin{align}
    \mathcal{M}_V^{(\mu e)} = \mathcal{M}_V^{(\mu p)} &= - e^2 g_{V\!\mu}\left(\frac{P_1^\alpha}{K\cdot P_1} - \frac{P_3^\alpha}{K\cdot P_3}\right)\epsilon_{\alpha}^*(K)\nonumber\\
    &\times D_{\mu\nu}(Q)\,j^\mu(P_1,P_3) \, j^\nu(P_2,P_4) \,.
    \label{Eq:M_mue_Vec}
\end{align}
Noting that $Q\simeq P_3 -P_1 \simeq P_2-P_4$, we can express the soft factors in terms of $P_1$, $P_2$ and $Q$ as
\begin{align}
&\kern-2em\frac{P_1^\alpha}{K\cdot P_1}+\frac{P_2^\alpha}{K\cdot P_2} - \frac{P_3^\alpha}{K\cdot P_3}-\frac{P_4^\alpha}{K\cdot P_4}\nonumber\\
&\kern3em\simeq - Q^\alpha\left(\frac{1}{P_1\cdot K} - \frac{1}{P_2\cdot K}\right)\nonumber\\
&\kern3em{}+Q\cdot K \left(\frac{P_1^\alpha}{(P_1\cdot K)^2}-\frac{P_2^\alpha}{(P_2\cdot K)^2}\right)
\end{align}
and
\begin{equation}
\frac{P_1^\alpha}{K\cdot P_1} - \frac{P_3^\alpha}{K\cdot P_3}\simeq - \frac{Q^\alpha}{P_1\cdot K} + P_1^\alpha\frac{Q\cdot K}{(P_1\cdot K)^2}\,,
\end{equation}
where we have neglected $\mathcal{O}(q^2/p^2)$ terms. Similarly to the scalar case, this last step implies the additional assumption $Q\cdot K\ll P_1\cdot K$, for which not only should $q\ll p_1$, but also $m_{D,M,\Lambda}\ll m_\mu$. 

Since the in-medium vector coupling to electrons is simply related to the tree-level muon coupling by a parametric rescaling, the matrix element for vector-boson radiation in electron bremsstrahlung has the same structure as that for processes involving radiating muons. Therefore, under the assumption $m_{D,M,\Lambda}\ll m^*_e$--which is verified only for the dominant contribution from transverse-photon-mediated scatterings in nonsuperconducting regions--the expressions for electron bremsstrahlung can be obtained from those for muon bremsstrahlung through straightforward substitutions.

Summing over spins and vector polarizations, and performing the same contractions and approximations as in the scalar case, we obtain
\begin{widetext}
\begin{align}
    \frac{1}{\mathcal{S}}\!
    \sum_{\rm sp,\, pol} \left|\mathcal{M}_V^{(\mu\mu)}\right|^2 &=32 e^4 \hat{g}_{V\!\mu}^2 \frac{\varepsilon_{\mu}^2}{\omega^2}\left|\frac{1}{q^2+\Pi_{\rm L}}-\frac{\beta_{\mu}^2 x_{12}}{q^2+\Pi_{\rm T}}\right|^2\times\Bigg\{(q^2+m_V^2)\left(\frac{1}{1-\beta_{\mu} \beta_{V} x_1}-\frac{1}{1-\beta_{\mu} \beta_{V}  x_2}\right)^2\nonumber\\
    &+\frac{2(\beta_V\omega q x_q+m_V^2)}{\omega\varepsilon_{\mu}}\left[\frac{\bp_1\cdot \bq}{(1-\beta_\mu\beta_{V}x_1)^2}-\frac{\bp_2\cdot \bq}{(1-\beta_\mu \beta_{V} x_2)^2}\right]\left(\frac{1}{1-\beta_{\mu} \beta_{V} x_1}-\frac{1}{1-\beta_{\mu} \beta_{V}  x_2}\right)\nonumber\\
    & - \frac{(\beta_V\omega q x_q+m_V^2)^2}{\omega^2\varepsilon_{\mu}^2}\left[\frac{m_\mu^2}{(1-\beta_{\mu} \beta_{V} x_1)^4} - \frac{2\varepsilon_{\mu}^2(1-\beta_\mu^2x_{12})}{(1-\beta_{\mu} \beta_{V} x_1)^2 (1-\beta_{\mu} \beta_{V} x_2)^2}+\frac{m_\mu^2}{(1-\beta_{\mu} \beta_{V} x_2)^4}\right]\Bigg\}\label{Eq:M2mumu_Vec}\,,\\
    \nonumber\\
    \sum_{\rm sp,\, pol} \left|\mathcal{M}_V^{(\mu X)}\right|^2 &=64 e^4 \hat{g}_{V\!\mu}^2 \frac{\varepsilon_{X}^2}{\omega^2}\left|\frac{1}{q^2+\Pi_{\rm L}}-\frac{\beta_X\beta_{\mu} x_{12}}{q^2+\Pi_{\rm T}}\right|^2\times\Bigg\{\frac{q^2+m_V^2}{(1-\beta_{\mu} \beta_{V} x_1)^2}+\frac{2\bp_1\cdot \bq(\beta_V\omega q x_q+m_V^2)}{\omega\varepsilon_{\mu}(1-\beta_\mu\beta_{V}x_1)^3}\nonumber\\
    & - \frac{m_\mu^2(\beta_V\omega q x_q+m_V^2)^2}{\omega^2\varepsilon_{\mu}^2(1-\beta_{\mu} \beta_{V} x_1)^4}\Bigg\}\,,\label{Eq:M2ee_Vec}
\end{align}
\end{widetext}
where $\hat{g}_{V\!\mu}$ is the in-medium effective coupling to muons, related to the tree-level coupling by Eq.~\eqref{eq:treelevelVSinmedium}, and $X=e,p$ denotes electron or proton targets.

In computing the squared amplitude, we use the polarization sum $\sum \epsilon_\alpha \epsilon^*_\beta = - g_{\alpha\beta} + K_\alpha K_\beta/m_{V}^2$. However, the contribution from $K_\alpha K_\beta/m_V^2$ identically vanishes upon contraction with the soft factor in Eqs.~\eqref{Eq:M_mumu_Vec} and~\eqref{Eq:M_mue_Vec}. Instead, we keep the corrections depending on $\beta_V$, since in these terms the corrections due to $m_V$ become relevant already when $m_V\sim \omega\sim T$ and therefore cannot be neglected.

The effective coupling to electrons $\hat{g}_{V\!e}$ in Eq.~\eqref{eq:gVeEffective} yields additional contributions to the emissivities. Neglecting interference between diagrams in the near-forward limit, $\mu e$ scatterings receive an additional contribution arising from radiation from the electron legs. Similarly, $ee$ and $ep$ scatterings also contribute. Since our approximations remain valid for the leading contributions from transverse photons, the matrix element for these additional processes can be obtained by the substitutions $\hat{g}_{V\!\mu}\to\hat{g}_{V\!e}$, $\beta_\mu\leftrightarrow\beta_e$ and $\varepsilon_\mu\leftrightarrow\varepsilon_e$ in Eq.~\eqref{Eq:M2mumu_Vec} and Eq.~\eqref{Eq:M2ee_Vec}.

\subsection{Energy-loss rate}

Next, we compute the emissivity 
in analogy to the scalar case, leading to Eq.~\eqref{Eq:Q_after_Omega}. The matrix element for the vector emission is technically more involved as it depends on the cosine of the angle between $\bq$ and the momentum of the emitted particle, $x_q$, in addition to $x_q^2$. However, after neglecting terms of order \smash{$q^2$}, the delta function in Eq.~\eqref{Eq:Deltaq} enforces $\bp_1$ and $\bp_2$ to be orthogonal to $\bq$. In the same approximation, the term proportional to $\cos\theta_\bq$ is even smaller, since upon averaging over the azimuthal angle $\phi$ we have $\langle x_q\rangle=0$. Therefore, this term can be safely neglected.

Accordingly, the only new contributions compared to the scalar case are those independent of $\theta_\bq$, which trivially come from Eq.~\eqref{eq:def_Upsilon} once the factor $\cos^2\theta_{\bq}$ is left out
\begin{equation}
    \int d^3\bq \delta(\bq^2+2\bp_1\cdot\bq)\delta(\bq^2-2\bp_2\cdot\bq)\simeq \int \frac{dq}{2p_1 p_2 \sqrt{1-x_{12}^2}}.
    \label{Eq:Delta_p1.q}
\end{equation}
The integral  over $q$ reduces to 
\begin{align}
    X(m_A,m_B) &= \frac{1}{2|\bp_1\times\bp_2|} \int_{0}^{\infty} \!\!\!\frac{q^2 dq}{(q^2+m_A^2)(q^2+m_B^2)}\,,
    \label{Eq:X_funct}
\end{align}
which becomes
\begin{equation}
    X(m_A,m_B) = \frac{\pi}{4 p_1 p_2 (m_A+m_B) (1-x_{12}^2)^{1/2}}\,.
\end{equation}
The emissivities can be written in terms of the angular integrals
\begin{align}
    M_n &= \int  \frac{dx_1 dx_2 d\phi\, x_{12}^n}{(1-\beta_\mu \beta_Vx_1)^2 \sqrt{1-x_{12}^2}}\,,\label{Eq:Mn}\\[2ex]
    N_n &= \int \frac{dx_1 dx_2 d\phi\, x_{12}^n}{(1-\beta_\mu \beta_Vx_1)(1-\beta_\mu \beta_Vx_2) \sqrt{1-x_{12}^2}}\,,\label{Eq:Nn}
\end{align}
with $n=0,1,2$. Using the same change of angular variables as in Eq.~\eqref{Eq:Jn}, the integrals $M_n$ can be evaluated 
\begin{widetext}
\noindent analytically, yielding
\begin{equation}
    M_0 = 2M_2 = \frac{4\pi^2}{1-\beta_\mu^2\beta_V^2}
    \quad\text{and}\quad
    M_1 = 0\,,
\end{equation}
while the $N_n$ functions read
\begin{equation}
    N_n=\int dx_1 dx_{12}d\psi \frac{x_{12}^n}{(1-\beta_\mu \beta_V x_1)\,\sqrt{1-x_{12}^2}}\\ \frac{1}{1-\beta_\mu \beta_V\left(x_1 x_{12}+\sqrt{1-x_1^2}\sqrt{1-x_{12}^2}{\cos\psi}\right)}\,.
\end{equation}
Also in this case, the integral over $\psi$ can be carried out analytically, giving
\begin{equation}\label{Eq:NnPsiIntegrated}
N_n=\int dx_1 dx_{12} \frac{2\pi\,x_{12}^n}{(1-\beta_\mu \beta_V x_1)\sqrt{1-x_{12}^2}}
    \frac{1}{\sqrt{\left(1-\beta_\mu \beta_V x_1x_{12}\right)^2 - \beta_\mu^2 \beta_V^2(1-x_1^2)(1-x_{12}^2)}}\,.
\end{equation}
\end{widetext}
In the ultrarelativistic limit ($\beta_\mu\to1$, $\beta_V\to1$), the small-angle expansion of Eq.~\eqref{Eq:Nn} reveals the asymptotic behavior $N_n\sim(1-\beta_\mu\beta_V)^{-1/2}$. Once again, after integrating Eq.~\eqref{Eq:NnPsiIntegrated} numerically, these functions can be fitted through simple fitting expressions
\begin{align}
        N_0&\simeq-31.83+\frac{43.06}{\sqrt{1-\beta_\mu\beta_V}}+26.38\,\sqrt{1-\beta_\mu\beta_V}\,,\\
        N_1&\simeq-58.88+\frac{24.92}{\sqrt{1-\beta_\mu\beta_V}}+36.95\,\sqrt{1-\beta_\mu\beta_V}\,,\\
        N_2&\simeq-33.90+\frac{27.44}{\sqrt{1-\beta_\mu\beta_V}}+27.07\,\sqrt{1-\beta_\mu\beta_V}\,,
\end{align}
which reproduce the required asymptotic behavior.

Finally, the emission spectrum for $\mu\mu\to\mu\mu V$ production takes the form
\begin{widetext}
\begin{align}
\label{eq:NV_mu}
    \mathcal{N}_V^{(\mu\mu)}=\frac{\alpha^2 \hat{g}_{V\!\mu}^2 \varepsilon_{\mu}^2 T^3}{8\pi^6}J(\tilde\omega)\frac{k}{\omega^2}&\Bigg\{\left[\frac{M_0-N_0+\beta_V^2K_0}{2m_D}-\frac{2\beta_\mu^2(-N_1+\beta_V^2K_1)}{m_D+m_T}+\frac{\beta_\mu^4(M_2-N_2+\beta_V^2K_2)}{2m_T}\right]\nonumber\\
     &- \frac{m_\mu^2}{\varepsilon_{\mu}^2}\beta_V^2\left[\frac{J_0}{2m_D}+\frac{\beta_\mu^4 J_2}{2m_T}\right]-\beta_\mu^2 \beta_V^2\left[\frac{K_1}{2m_D}-\frac{2\beta_\mu^2 K_2}{m_D+m_T}+\frac{\beta_\mu^4 K_3}{2m_T}\right]\Bigg\}\,,
\end{align}
while for $\mu X\to \mu X V$, with $X=e,p$, it reads
\begin{align}
\label{eq:NV_e}
    \mathcal{N}_V^{(\mu X)}&=\frac{\alpha^2 \hat{g}_{V\!\mu}^2 \varepsilon_{X}^2 T^3}{8\pi^6}J(\tilde\omega)\frac{k}{\omega^2}\Bigg[\frac{M_0}{2m_D}-\frac{2\beta_\mu\beta_X M_1}{m_D+m_T}+\frac{\beta_\mu^2\beta_X^2 M_2}{2m_T}
    -\frac{m_\mu^2}{\varepsilon_{\mu}^2}\beta_V^2\left(\!\frac{J_0}{2m_D}-\frac{2\beta_\mu\beta_X J_1}{m_D+m_T}+\frac{\beta_\mu^2\beta_X^2 J_2}{2m_T}\!\right)\!\Bigg]\,.
\end{align}

The emissivities can be computed analytically in the massless vector-boson limit, yielding
\begin{align}
    Q_{V,0}^{(\mu\mu)}&=\frac{11\alpha^2 \hat{g}_{V\!\mu}^2 \varepsilon_{\mu}^2 T^4}{720\pi^2}\Bigg\{\left[\frac{M_0-N_0+K_0}{2m_D}-\frac{2\beta_\mu^2(-N_1+K_1)}{m_D+m_T}+\frac{\beta_\mu^4(M_2-N_2+K_2)}{2m_T}\right]\nonumber\\ 
     &\qquad\qquad\qquad- \frac{m_\mu^2}{\varepsilon_{\mu}^2}\left[\frac{J_0}{2m_D}+\frac{\beta_\mu^4 J_2}{2m_T}\right]-\beta_\mu^2 \left[\frac{K_1}{2m_D}-\frac{2\beta_\mu^2 K_2}{m_D+m_T}+\frac{\beta_\mu^4 K_3}{2m_T}\right]\Bigg\}\,,
     \label{Eq:QV_mu}\\[0.5em]
     Q_{V,0}^{(\mu X)}&=\frac{11\alpha^2 \hat{g}_{V\!\mu}^2 \varepsilon_{X}^2 T^4}{720\pi^2}\Bigg[\frac{M_0}{2m_D}+\frac{\beta_\mu^2\beta_X^2 M_2}{2m_T}-\frac{m_\mu^2}{\varepsilon_{\mu}^2}\left(\!\frac{J_0}{2m_D}+\frac{\beta_\mu^2\beta_X^2 J_2}{2m_T}\!\right)\!\Bigg]\,.
     \label{Eq:QV_e}
\end{align}
In-medium contributions to the emissivities arising from radiating electrons and protons in $\mu e$, $\mu p$, $ep$ and $ee$ electromagnetic scatterings are obtained through the trivial substitutions $\hat{g}_{V\!\mu}\to\hat{g}_{V\!e},\hat{g}_{V\!p}$\,, $\beta_\mu\rightarrow\beta_e,\beta_p$ and $\varepsilon_\mu\rightarrow\varepsilon_e,\varepsilon_p$. However, we highlight that processes in which the proton radiates the vector boson do not benefit from the relativistic enhancement $\propto1/(1-\beta^2)$, since protons in the NS plasma are barely nonrelativistic. Consequently, these channels are subdominant with respect to those involving radiating muons or electrons and will be neglected in the following.

For the massive case, numerical evaluation of the exact expressions shows that the emissivities are accurately reproduced by the following fitting formulas
\begin{align}
    Q_V^{(\mu\mu)}(\tilde m_V,\beta_\mu)&\simeq Q_{V,0}^{(\mu\mu)}\,\left[1+0.36\,\tilde m_V^{1.89}
    \left(1-\beta_\mu^2\right)\right]\,e^{-1.26\,\tilde m_V}\,,\\
    Q_V^{(\mu X)}(\tilde m_V,\beta_\mu)&\simeq Q_{V,0}^{(\mu X)}\,\left[1+0.47\,\tilde m_V^{1.66}
    \left(1-\beta_\mu^2\right)\right]\,e^{-1.09\,\tilde m_V}\,\,,
\end{align}
where $\tilde m_V=m_V/T$.
\end{widetext}

\subsection{Comparison with neutrino pair emission}

Given the close correspondence between the matrix elements for vector emission and the vector-current contribution to neutrino pair emission as discussed in Sec.~\ref{eq:neutrino-pairs},
we conclude this section by comparing our results with those of Shternin~\cite{Shternin:2025swb}. The difference in the scaling with temperature in the emission rates is due to an additional $K^2$ in the neutrino matrix element, and an extra energy integral $\int dK_0 K_0 $ reflecting the presence of an additional final-state particle. Since radiated energies and momenta are set by the temperature, these considerations lead to the scaling of the $\nu\bar{\nu}$ emissivity as~$T^8$. 

At the same time, the structure of the momentum-exchange integrals remains unchanged, implying the same inverse dependence on the Debye and Meissner screening masses. On the other hand, while in $\mu e$ and $\mu p$ we recover the result that interference terms between transverse and longitudinal photon exchange vanish---consistent with $M_1=J_1=0$ in Eqs.~\eqref{Eq:QV_mu} and~\eqref{Eq:QV_e}---we still observe the presence of nonvanishing interference terms for scatterings involving identical particles. Furthermore, following the discussion in Sec.~\ref{sec:structure functions}, the neutrino pair emissivities can be obtained by integrating the vector boson emissivities over the vector mass. 

To validate this result, let us consider only the contribution due to longitudinal photons, in which the screening scale involved is not dynamical, thereby allowing a direct comparison between our results and those derived in~\cite{Shternin:2025swb}. Moreover, we set the muonic couplings  in both vector and neutrino emissivities to their tree-level value. From Eq.~\eqref{eq:NV_e}, the corresponding term is then given by
\begin{widetext}
\begin{equation}
    \omega\mathcal{N}_{V,L}^{(\mu e)}=\frac{\alpha^2 g_{V\mu}^2 \varepsilon_e^2 T^3 J(\tilde{\omega})\beta_V}{4\pi^4 m_D(1-\beta_\mu^2 \beta_V^2)}\!\left[\!1-\frac{ \beta_V^2 m_\mu^2}{3\varepsilon_\mu^2(1-\beta_\mu^2 \beta_V^2)}\!\right]\!.
\end{equation}
The neutrino emissivity can be obtained from Eq.~\eqref{eq: fromVectoNu}; it is convenient to replace the integral over the vector boson mass $m_V$ by an integral over the velocity $\beta_V=\sqrt{1-m_V^2/\omega^2}$, so that we obtain
\begin{equation}
    Q_{V,L,\nu}^{(\mu e)}=\frac{C_{V\mu}^2 G_F^2\alpha^2 \varepsilon_e^2 T^3}{24\pi^6 m_D}\int_0^{\infty}d\omega\,\omega^4 J(\omega/T) \int_0^1 d\beta_V\frac{\beta_V^2 (1-\beta_V^2)}{1-\beta_\mu^2 \beta_V^2}\left[1-\frac{\beta_V^2m_\mu^2}{3\varepsilon_\mu^2(1-\beta_\mu^2 \beta_V^2)}\right].
\end{equation}
Upon integration, this becomes
\begin{equation}
    Q_{V,L,\nu}^{(\mu e)}=\frac{41\pi^2 C_{V\mu}^2 G_F^2 \alpha^2 \varepsilon_e^2 T^8}{5670m_D}\left[\frac{15-\beta_\mu^2-8\beta_\mu^4}{18\beta_\mu^6}+\frac{-5+2\beta_\mu^2+3\beta_\mu^4}{6\beta_\mu^7}{\rm ArcTanh\beta_\mu}\right]\,,
\end{equation}
which, in the limit $p_{F,\mu}\gg m_D$, identically coincides with the longitudinal photon contribution to $\mu e$ scatterings in neutrino pair emissivities of Ref.~\cite{Shternin:2025swb}.
\end{widetext}

\section{Boson emission in nuclear scatterings}

As discussed in Sec.~\ref{sec:renormalization}, both scalar and vector bosons acquire an in-medium coupling to protons through the response of the medium charge distribution to muon density fluctuations induced by an external probe. In the nonsuperconducting regime, this effective coupling enables protons to emit scalar and vector radiation when scattering off neutrons via strong interactions. To compute the contributions arising from nuclear scatterings, we closely follow the derivation of Ref.~\cite{Fiorillo:2025zzx}, inspired by the approach of Friman and Maxwell for neutrino pair emission~\cite{Friman:1979ecl}, which models the interaction in the spirit of a Fermi liquid through a short-range potential
\begin{eqnarray}\label{Eq:GenPotential}
  V(\mathbf{q}) &=& f + f' \boldsymbol{\tau}_1 \cdot \boldsymbol{\tau}_2 + g \boldsymbol{\sigma}_1 \cdot \boldsymbol{\sigma}_2 + g'_\bq \boldsymbol{\tau}_1 \cdot \boldsymbol{\tau}_2 \, \boldsymbol{\sigma}_1 \cdot \boldsymbol{\sigma}_2  
  \nonumber\\
  &&+\,h'_\bq(\boldsymbol{\sigma}_1 \cdot \boldsymbol{\hat{\bq}}) (\boldsymbol{\sigma}_2 \cdot \boldsymbol{\hat{\bq}}) \boldsymbol{\tau}_1 \cdot \boldsymbol{\tau}_2,  
\end{eqnarray}
where $\bq$ is the nucleon three-momentum exchange. In this framework, short-range interactions are expressed through Landau parameters inferred from measurements of the response functions of heavy nuclei, defined as $\{f, f', g, g'\} = \frac{\pi^2}{2 \, m_{\rm N} \, p_{{\rm F},n}} \{F_0, F_0', G_0, G'_0\}$, where $F_0=0.5, F_0' = 0.7, G_0 = G_0' = 1.1$ and where $p_{{\rm F}, n}$ is the neutron Fermi momentum. We stress that these parameters are derived from measurements usually performed on symmetric nuclear matter. Therefore, they should be regarded as indicative values when applied to the highly asymmetric nuclear matter in neutron stars, where neutrons constitute the dominant species.
\begin{figure}
  \centering
  \includegraphics[width=1\columnwidth]{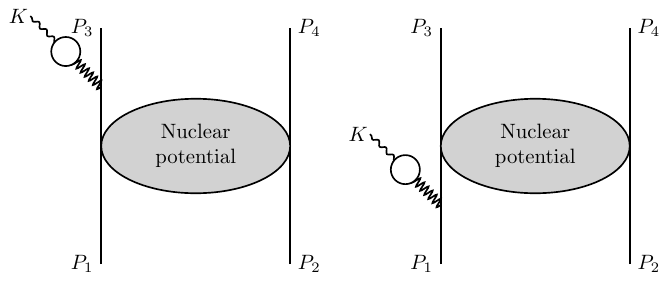}
  \caption{Feynman diagrams for the emission of vector bosons via $pn$ bremsstrahlung. Here $P_1$ and $P_3$ identify the four-momenta of the incoming and outgoing protons, respectively.}\label{fig:nucbrem}
\end{figure}

Long-range interactions are primarily mediated by $\pi$ and $\rho$ meson exchange. The tensor interaction is given by
\begin{equation}\label{eq:hk}
    h'_\bq \equiv - \frac{f_\pi^2}{m_\pi^2} \Big(\frac{|\bq|^2}{|\bq|^2 +m_\pi^2} - C_\rho \frac{|\bq|^2}{|\bq|^2 +m_\rho^2}\Big), 
\end{equation}
while the spin-spin interaction $g'_\bq$ reads
\begin{equation}\label{eq:gk}
    g'_\bq \equiv g' - C_\rho \, \frac{f_\pi^2}{m_\pi^2} \frac{|\bq|^2}{|\bq|^2+m_\rho^2},
\end{equation}
where we take $m_\rho = 769 \, \rm MeV$, $C_\rho = 1.4$, and  $f_\pi \simeq 1$ (not to be confused with the pion decay constant). Referring to Fig.~\ref{fig:nucbrem}, we identify the nucleons carrying four-momenta $P_1$ and $P_3$ as the radiating proton legs. Assuming nucleons to be nonrelativistic, $v_\F\sim p_\F/m_N\ll1$, $P_1^\mu\simeq(m_N, \bp_1)$ and $P_3^\mu\simeq(m_N, \bp_3)$, and $Q^\mu\simeq(0,\bq)$ with $\bq=\bp_3-\bp_1$. Notice that this approximation is only marginally verified in the inner regions of the NS core, where $p_{{\rm F},p}\simeq200\,{\rm MeV}$ and the effective proton mass $m_p^*\simeq400\,{\rm MeV}$. Neglecting corrections of the order of the nucleon velocity, the matrix element for scalar emission coincides with the one already obtained in Ref.~\cite{Fiorillo:2025zzx}
\begin{widetext}
\begin{equation}
    \mathcal{M}^{(pn)}_\phi=\hat{g}_{\phi p} \sum_\mathcal{O}V_{\mathcal{O}}(\bq) \overline{u}_4 \mathcal{O} u_2
    \left[\frac{\overline{u}_3\mathcal{O}(\slashed{P}_1-\slashed{K}+m_N) u_1}{(P_1-K)^2-m_N^2}+\frac{\overline{u}_3 (\slashed{P}_3+\slashed{K}+m_N) \mathcal{O} u_1}{(P_3+K)^2-m_N^2}\right]\,.
\end{equation}
Referring to the diagrams in Fig.~\ref{fig:nucbrem}, the matrix element for vector emission reads
\begin{equation}
    \mathcal{M}^{(pn)}_V=-i\hat{g}_{V\!p}\epsilon^\mu(K) \sum_\mathcal{O}V_{\mathcal{O}}(\bq) \overline{u}_4 \mathcal{O} u_2
    \left[\frac{\overline{u}_3\mathcal{O}(\slashed{P}_1-\slashed{K}+m_N)\gamma_\mu u_1}{(P_1-K)^2-m_N^2}+\frac{\overline{u}_3 \gamma_\mu(\slashed{P}_3+\slashed{K}+m_N) \mathcal{O} u_1}{(P_3+K)^2-m_N^2}\right],
\end{equation}
where $\epsilon^\mu$ is the boson polarization four-vector and we summed over all the possible operator structures $\mathcal{O}$ in Eq.~\eqref{Eq:GenPotential}. Neglecting $\slashed{K}$ in the numerators and $K^2=m_V^2$ in the denominator as they are of order $T^2\ll p_{{\rm F},p}^2$, we can recast the scattering amplitude as
\begin{equation}
    \mathcal{M}^{(pn)}_V=-i\hat{g}_{V\!p}\,\mathcal{S}\,\sum_\mathcal{O}V_{\mathcal{O}}(\bq) \overline{u}_3\mathcal{O}u_1\overline{u}_4\mathcal{O}u_2,
\end{equation}
where $\mathcal{S}$ is the soft factor for vector boson emission
\begin{equation}
    \mathcal{S}=\epsilon_\mu(K)\left(\frac{P_1^\mu}{K\cdot P_1} - \frac{P_3^\mu}{K\cdot P_3}\right)\\
    \simeq-\frac{\epsilon_\mu(K)}{m_N\omega}\left(Q^\mu+\frac{\bp_3\cdot\mathbf{n}}{m_N}P_3^\mu-\frac{\bp_1\cdot\mathbf{n}}{m_N}P_1^\mu\right) \,,
\end{equation}
and we introduced $\mathbf{n}=\bk/|\bk|$. 
After squaring the soft factor and summing over the polarization states we obtain
\begin{equation}
\sum_\epsilon\mathcal{S}\,\mathcal{S}^\dagger=-\frac{g_{\mu\nu}}{m_N^2\omega^2}\left(Q^\mu+\frac{\bp_3\cdot\mathbf{n}}{m_N}P_3^\mu-\frac{\bp_1\cdot\mathbf{n}}{m_N}P_1^\mu\right)
\left(Q^\nu+\frac{\bp_3\cdot\mathbf{n}}{m_N}P_3^\nu-\frac{\bp_1\cdot\mathbf{n}}{m_N}P_1^\nu\right)\simeq\frac{|\bq|^2}{m_N^2\omega^2}(1-\cos^2\theta)\,,
\end{equation}
where we only kept terms of zeroth order in the nucleon velocities and introduced the angle $\theta$ between $\bk$ and $\bq$. 
Ultimately, averaging over the emission angle, the spin-summed matrix element is given by 
\begin{equation}
    |\mathcal{M}^{(pn)}_V|^2=\frac{\hat{g}_{V\!p}^2}{m_N^2 \omega^2 } \frac{2}{3}|\bq|^2\sum_\sigma\Bigl|\sum_\mathcal{O}V_{\mathcal{O}}(\bq) \overline{u}_3\mathcal{O}u_1\overline{u}_4\mathcal{O}u_2\Bigr|^2,
\end{equation}
which coincides with twice the result of Ref.~\cite{Fiorillo:2025zzx} for scalar emission in $pn$ bremsstrahlung at dipole level, as expected in the nonrelativistic nucleon limit. The derivation of the vector and scalar emission spectrum can then be carried out in close analogy to Ref.~\cite{Fiorillo:2025zzx}. The final expression can be expanded in terms of the coefficients 
    \begin{eqnarray}
    &&C_0=4\left[f^2+7f^{'2}-3f'(g-4g')+9(g^2+g g'+g^{'2})+f(f'-3(2g+g'))\right],\\ 
    \nonumber && C^k_\rho=2 C_\rho C^k_\pi=\frac{8C_\rho f_\pi^2 (3g-f-2f')}{m_\pi^2},\; C^l_\rho=2 C_\rho C^l_\pi=\frac{16 C_\rho f_\pi^2 (f-f'-3(g+g'))}{m_\pi^2},\\ \nonumber &&
    C^{ll}_{\pi\pi}=4C^{kk}_{\pi\pi}=-12 C^{kl}_{\pi\pi}=\frac{16 f_\pi^4}{m_\pi^4},\; C^{ll}_{\rho\rho}=4 C^{kk}_{\rho\rho}=-6C^{kl}_{\rho\rho}=\frac{32 C_\rho^2 f_\pi^4}{m_\pi^4}, C^{kl}_{\pi \rho}=C^{lk}_{\pi\rho}=C^{kl}_{\rho\pi}=C^{lk}_{\rho\pi}=-\frac{1}{2}\frac{16 C_\rho f_\pi^4}{3 m_\pi^4}\,,
\end{eqnarray}
and the following fundamental integrals
\begin{eqnarray}
    I^k_\mu&&=I_k(\alpha_\mu),\; I^l_\mu=I_0\frac{\tp_n^2}{\tp_n^2+\alpha_\mu},\; \\ \nonumber I^{kk}_{\mu\nu}&&=I_{kk}(\alpha_\mu,\alpha_\nu),\; I^{ll}_{\mu\nu}=I_0 \frac{\tp_n^4}{(\tp_n^2+\alpha_\mu)(\tp_n^2+\alpha_\nu)},\;  I^{kl}_{\mu\nu}=I^{lk}_{\nu\mu}= I_k(\alpha_\mu) \frac{\tp_n^2}{\tp_n^2+\alpha_\nu}\,, 
\end{eqnarray}
where
\begin{eqnarray}
    &&I_0=\frac{8\tp_p^3}{3},\\ \nonumber 
    &&I_k(\alpha)=\frac{8\tp_p^3}{3}-2\tp_p \alpha+\alpha^{3/2}\arctan\left(\frac{2\tp_p}{\sqrt{\alpha}}\right),\\ \nonumber
    &&I_{kk}(\alpha,\beta)=\frac{2\tp_p(\alpha-\beta)\left[4\tp_p^2-3(\alpha+\beta)\right]+3\alpha^{5/2}\arctan\left(\frac{2\tp_p}{\sqrt{\alpha}}\right)-3\beta^{5/2}\arctan\left(\frac{2\tp_p}{\sqrt{\beta}}\right)}{3(\alpha-\beta)}.
\end{eqnarray}
In these expressions, we defined $\tq=q/\sqrt{m_NT}$, $\tp_p=p_{{\rm F},p}/\sqrt{m_NT}$, $\tp_n=p_{{\rm F}, n}/\sqrt{m_NT}$, $\alpha_{\pi}=m_\pi^2/(m_N T)$ and $\alpha_{\rho}=m_\rho^2/(m_N T)$. Ultimately, the spectrum of vector and scalar bosons produced in $pn$ nuclear scatterings reads 
\begin{equation}\label{eq:scalar_emission_np}
    \mathcal{N}_{\Phi}^{(pn)}=\eta_\Phi\frac{\hat{g}_{\Phi p}^2\,m_N^{7/2}\,T^{9/2} J(\tilde{\omega})}{192\pi^7}\frac{k}{\omega^2}\left[C_0 I_0+\sum_{i=k,l} \sum_{\mu=\pi,\rho} C^i_\mu I^i_\mu+\sum_{i,j=k,l} \sum_{\mu,\nu=\pi,\rho} C^{ij}_{\mu\nu} I^{ij}_{\mu\nu}\right]\,,
\end{equation}
where $\Phi=\phi,V$ and $\eta_\Phi=1,2$ for scalar and vector emission, respectively. 
The energy loss rate per unit volume then reads
\begin{equation}\label{eq:scalar_emission_np-2}
     Q_\Phi^{(pn)}=\eta_\Phi\frac{\hat{g}_{\Phi\!p}^2\,m_N^{7/2}\,T^{11/2} }{192\pi^7}B(\tilde{m}_\Phi)\left[C_0 I_0+\sum_{i=k,l} \sum_{\mu=\pi,\rho} C^i_\mu I^i_\mu+\sum_{i,j=k,l} \sum_{\mu,\nu=\pi,\rho} C^{ij}_{\mu\nu} I^{ij}_{\mu\nu}\right]\,,
\end{equation}
with $\tilde{m}_\Phi=m_\Phi/T$ and
\begin{equation}
    B(\tilde{m}_\Phi)=\int_{\tilde{m}_\Phi}^{\infty}\,d\tilde{\omega}\, \tilde{\omega} \sqrt{\tilde{\omega}^2-\tilde{m}_\Phi^2}\,\frac{J(\tilde{\omega})}{\tilde{\omega}^2}\simeq\frac{11}{90}\pi^4\,e^{-0.7606\,\tilde{m}_\Phi}\,. 
\end{equation}
We highlight that, once the underlying temperature dependence in the integrals involved in the expansion has been taken into account, this emissivity retains the scaling $Q_V\sim T^4$.
\end{widetext}

Finally, we note that bremsstrahlung emission of light bosons can also take place from $pp$ nuclear scatterings. However, for identical particles the emission can only proceed at the quadrupole level, leading to an additional suppression by powers of the nucleon velocity. Assuming $\beta_p\sim0.5$ in the inner core, the results of Ref.~\cite{Fiorillo:2025zzx} suggest that vector emission from $pp$ scatterings is a factor of $\sim20$ less efficient than that from $pn$ scatterings. Consequently, we neglect this contribution in the following. 

Overall, at the parametric level we expect vector boson emission to be enhanced by neutron-proton bremsstrahlung, although not by several orders of magnitude. From Eq.~\eqref{eq:scalar_emission_np}, we see that the parametric scaling of the bremsstrahlung emissivity with the internal conditions is roughly
\begin{equation}
    Q_V^{(pn)}\sim \hat{g}_{Vp}^2 T^4 p_{F,p};
\end{equation}
due to charge neutrality we can approximately set here $p_{F,p}\sim \varepsilon_e$. Meanwhile, bremsstrahlung from ultrarelativistic electrons, according to Eq.~\eqref{Eq:QV_e} in which we keep only the terms proportional to $m_T^{-1}$, scales as
\begin{equation}
    Q_V^{(e)} \sim \alpha^2 \hat{g}_{Ve}^2 \varepsilon_e^2 \frac{T^4}{m_T}\frac{1}{1-\beta_e^2} \sim \frac{\alpha \varepsilon_e^2 \hat{g}_{Ve}^2 T^4}{m_T}.
\end{equation}
In the nonsuperconducting phase, with $m_T\sim (\alpha T \varepsilon_e^2)^{1/3}$, we see that
\begin{equation}
    \frac{Q_V^{(pn)}}{Q_V^{(e)}}\sim \alpha^{-2/3}\left(\frac{T}{\varepsilon_e}\right)^{1/3}\sim 1
\end{equation}
for $T\sim 10\,\mathrm{keV}$ and $\varepsilon_e\sim 100\,\mathrm{MeV}$. Such an estimate is of course very approximate and is only meant to show that neutron-proton bremsstrahlung is expected to be ever more important as the environment is less degenerate. When all prefactors are taken into account, we find that $pn$ bremsstrahlung can dominate even by a factor of 10 over electron bremsstrahlung.

On the other hand, as observed in Sec.~\ref{sec:scalar}, scalar radiation from electrons via the in-medium induced coupling is characterized by a further relativistic enhancement factor $(1-\beta_e^2)^{-1}\sim\alpha^{-1}$, leading to 
\begin{equation}
        \frac{Q_\phi^{(pn)}}{Q_\phi^{(e)}}\sim \alpha^{1/3}\left(\frac{T}{\varepsilon_e}\right)^{1/3}\sim 10^{-2}\,.
\end{equation}
Consequently, $pn$ scatterings are expected to play a subdominant role in the total scalar boson emissivities, dominated by ultrarelativistic electron bremsstrahlung via the in-medium coupling.

\section{Pseudoscalar emission}
\label{sec:pseudoscalar}

\subsection{Amplitude and squared matrix element}

Adapting the same notation for the quantities as in the scalar and vector cases, the amplitude for the pseudo-scalar ($\Phi=a$) emission in the soft-radiation limit for $X = \mu$ reads
\begin{widetext}
\begin{eqnarray}
    \mathcal{M}_a^{(\mu\mu)} &=& \frac{g_{a\mu} e^2}{2} \Bigg[\frac{\bar{u}(P_3) \gamma^\mu \slashed{K}\gamma_5 u(P_1) \bar{u}(P_4) \gamma^\nu u(P_2)}{K\cdot P_1}
    + \frac{\bar{u}(P_3) \gamma_5 \slashed{K}\gamma^\mu u(P_1) \bar{u}(P_4) \gamma^\nu u(P_2)}{K\cdot P_3}\nonumber\\
    &&{}+\frac{\bar{u}(P_3) \gamma^\mu u(P_1) \bar{u}(P_4) \gamma^\nu \slashed{K}\gamma_5 u(P_2)}{K\cdot P_2}
    + \frac{\bar{u}(P_3) \gamma^\mu u(P_1) \bar{u}(P_4) \gamma_5 \slashed{K}\gamma^\nu u(P_2)}{K\cdot P_4}\Bigg]
     D_{\mu\nu}(Q) + (P_3\leftrightarrow P_4)\,.
\end{eqnarray}
Similarly, when $X=e$ or $p$, the amplitude is
\begin{equation}
    \mathcal{M}_a^{(\mu e)} = \mathcal{M}_a^{(\mu p)} = \frac{g_{a\mu} e^2}{2} \Bigg[\frac{\bar{u}(P_3) \gamma^\mu \slashed{K}\gamma_5 u(P_1) \bar{u}(P_4) \gamma^\nu u(P_2)}{K\cdot P_1}
    + \frac{\bar{u}(P_3) \gamma_5 \slashed{K}\gamma^\mu u(P_1) \bar{u}(P_4) \gamma^\nu u(P_2)}{K\cdot P_3}\Bigg]D_{\mu\nu}(Q)\,.
\end{equation}
\end{widetext}
To obtain the squared matrix element, we aim to keep only the leading terms in $Q=P_3-P_1$,
since, as we argued above, in the relevant regime $Q$ is very small. We can therefore use the identity $\slashed{K}\gamma^\mu=-\gamma^\mu \slashed{K}+2K^\mu$ to rewrite
\begin{equation}
    \mathcal{M}_a^{(\mu\mu)}=\frac{g_{a\mu}e^2}{2}\left[F^\mu_1 j_2^\nu+j_1^\mu F_2^\nu\right]D_{\mu\nu}(Q)+(P_3\leftrightarrow P_4)\,.
\end{equation}
Here we have evaluated the free current of the non-radiating lepton as $j_2^\mu=2P_2^\mu \delta_{s_2,s_4}$ for vanishing transfer momentum, while for the radiating lepton to lowest order in $Q$ we have
\begin{equation}\label{eq:definition_F_current}
    F_1^\mu=2K^\mu \frac{\overline u_3\gamma_5 u_1}{P_1\cdot K}+\overline{u}_3 \gamma_5 \gamma^\mu \slashed{K} u_1 \frac{Q\cdot K}{(P_1\cdot K)^2}.
\end{equation}
We use here the more compact notation $u(P_i)=u_i$. As in the case of the scalar and vector bosons, we have here assumed $|Q\cdot K|\ll |P_1\cdot K|$. For scattering on $X=e,p$, the amplitude is identical except that it accounts only for the term with a single radiating lepton; we do not reproduce the resulting simple expressions. 

Notice that, while the first term is not evidently proportional to $Q$, in practice it is, since for vanishing momentum transfer the matrix element $\overline u_3 \gamma_5 u_1=0$. We can see this more explicitly by using the identity $2m_\mu \overline u_3\gamma_5 u_1=Q_\mu \overline{u}_3 \gamma^\mu \gamma_5 u_1$, so that

\begin{equation}
    F_1^\mu=\frac{K^\mu}{m_\mu (P_1\cdot K)}\overline{u}_3 \slashed{Q}\gamma_5 u_1+\frac{Q\cdot K}{(P_1\cdot K)^2}\overline{u}_3 \gamma_5 \gamma^\mu \slashed{K}u_1.
\end{equation}
In this form, we may assume that the spinors both refer to the four-momentum $P_3\simeq P_1$; the small effects of the momentum transfer already appear in the terms in front of the spinor contraction.
\begin{widetext}
With this notation, the squared matrix element takes a rather simple form
\begin{equation}
    |\mathcal{M}_a^{(\mu\mu)}|^2=\frac{g_{a\mu}^2 e^4}{4}D_{\mu\nu}D_{\alpha\beta}\left\{8P_2^\nu P_2^\beta\mathrm{Tr}(F_1^\mu F_1^{*,\alpha})+8P_1^\mu P_1^\alpha \mathrm{Tr}(F_2^\nu F_2^{*,\beta})+2\mathrm{Re}\left[ \mathrm{Tr}(F_1^{*,\mu} j_1^\alpha) \mathrm{Tr}(F_2^\beta j_2^{*,\nu})\right]\right\}+(P_3\leftrightarrow P_4).
\end{equation}
\end{widetext}
When it comes to the emission integrated over the phase space of the leptons, clearly the emission from both legs will give the same contribution; furthermore, the switch $(P_3\leftrightarrow P_4)$ similarly gives a factor of 2, so that we may in practice replace
\begin{align}
    |\mathcal{M}_a^{(\mu\mu)}|^2&=\frac{g_{a\mu}^2 e^4}{2}D_{\mu\nu}D_{\alpha\beta}\left\{16P_2^\nu P_2^\beta\mathrm{Tr}(F_1^{*,\mu} F_1^\alpha)\right.\nonumber\\
    &\left.{}+2 \mathrm{Re}\left[\mathrm{Tr}(F_1^{*,\mu} j_1^\alpha) \mathrm{Tr}(F_2^{\beta} j_2^{*,\nu})\right]\right\}.
\end{align}
The mixed terms $\mathrm{Tr}(F_1^{*,\mu} j_1^\alpha)$ might be thought to be of order $Q$, since $F_1$ contains one power of $Q$ in its definition. However, one can easily see that to this order the trace actually vanishes; the current $j_1^\alpha=2P_1^\alpha \delta_{s_1 s_3}$ is diagonal in spin, and therefore the trace over spin of the spin matrices $\gamma^5 \gamma^\mu \gamma^\nu$ vanishes. Thus, we may neglect the mixed terms
\begin{equation}
    |\mathcal{M}^{(\mu\mu)}_a|^2=8g_{a\mu}^2 e^4 D_{\mu\nu} D_{\alpha \beta}P_2^\nu P_2^\beta \mathrm{Tr}(F_1^{*,\mu} F_1^\alpha);
\end{equation}
for $|\mathcal{M}_a^{(\mu X)}|^2$, this is a factor 4 smaller. (We recall again that $|\mathcal{M}_a^{(\mu\mu)}|^2$ should be divided by a symmetry factor $4$ which is absent for scatterings of non-identical particles.)

The last remaining step is to compute $\mathrm{Tr}(F_1^\mu F_1^{\alpha,*})$, which gives
\begin{widetext}
\begin{eqnarray}
    \mathrm{Tr}(F_1^\mu F_1^{\alpha,*})&=&\frac{8K^\mu K^\alpha [(P_1\cdot Q)^2-m_\mu^2 Q^2]}{m_\mu^2(P_1\cdot K)^2}+\frac{8(Q\cdot K)^2[P_1\cdot K (K^\mu P_1^\alpha+K^\alpha P_1^\mu)-m_a^2 P_1^\mu P_1^\alpha -g^{\mu\alpha}(P_1\cdot K)^2]}{(P_1\cdot K)^4}\nonumber\\
    &+&\frac{8Q\cdot K }{(P_1\cdot K)^3}[K\cdot P_1 (K^\mu Q^\alpha+K^\alpha Q^\mu)-Q\cdot K (P_1^\mu K^\alpha+P_1^\alpha K^\mu)].
\end{eqnarray}
The terms proportional to $Q^\mu$ vanish upon projection on the photon propagator $D_{\mu\nu}$. Keeping only the relevant terms, the matrix element can finally be simplified to
\begin{equation}
    |\mathcal{M}_{a}^{(\mu\mu)}|^2=\frac{64g_{a\mu}^2 e^4 D_{\mu\nu}D_{\alpha\beta}P_2^\nu P_2^\beta}{(P_1\cdot K)^2 m_\mu^2}\left[K^\mu K^\alpha [(P_1\cdot Q)^2-m_\mu^2 Q^2]-\frac{(Q\cdot K)^2 m_\mu^2(m_a^2 P_1^\mu P_1^\alpha+g^{\mu\alpha}(P_1\cdot K)^2)}{(P_1\cdot K)^2}\right].
\end{equation}
Performing the Lorentz contractions, we find the following matrix element for the $\mu\mu\to\mu\mu a$ process
\begin{align}
    \frac{1}{\mathcal{S}}\sum \left|\mathcal{M}_a^{(\mu\mu)}\right|^2 &= \frac{64 g_{a\mu}^2 e^4}{(1-\beta_a\beta_\mu x_1)^2} q^2 \left\{\left(\frac{1}{q^2+\Pi_{\rm L}}-\frac{\beta_a\beta_\mu x_2}{q^2+\Pi_{\rm T}}\right)^2\left(1+\frac{q^2}{4m_\mu^2}\right) + \beta_a^2 x_q^2 \left[\frac{\beta_\mu^2}{(q^2+\Pi_{\rm T})^2}-\frac{1}{(q^2+\Pi_{\rm L})^2}\right.\right.\nonumber\\
    &\left.\left.-\frac{m_a^2}{\omega^2(1-\beta_a\beta_\mu x_1)^2}\left(\frac{1}{q^2+\Pi_{\rm L}}-\frac{\beta_\mu^2 x_{12}}{q^2+\Pi_{\rm T}}\right)^2\right]\right\}\,,
    \label{Eq:M2_mu_PS}
\end{align}
while for the $\mu X \to \mu X a$, with $X=e,p$, we have
\begin{align}
    \sum \left|\mathcal{M}_a^{(\mu X)}\right|^2 &= \frac{64 g_{a\mu}^2 e^4}{(1-\beta_a\beta_\mu x_1)^2} \left(\frac{\varepsilon_X}{\varepsilon_\mu}\right)^2 q^2 \left\{\left(\frac{1}{q^2+\Pi_{\rm L}}-\frac{\beta_a\beta_X x_2}{q^2+\Pi_{\rm T}}\right)^2\left(1+\frac{q^2}{4m_\mu^2}\right) + \beta_a^2 x_q^2 \left[\frac{\beta_X^2}{(q^2+\Pi_{\rm T})^2}-\frac{1}{(q^2+\Pi_{\rm L})^2}\right.\right.\nonumber\\
    &\left.\left.{}-\frac{m_a^2}{\omega^2(1-\beta_a\beta_\mu x_1)^2}\left(\frac{1}{q^2+\Pi_{\rm L}}-\frac{\beta_X\beta_\mu x_{12}}{q^2+\Pi_{\rm T}}\right)^2\right]\right\}\,.\label{Eq:M2_PS_X}
\end{align}
To arrive at Eqs.~\eqref{Eq:M2_mu_PS} and~\eqref{Eq:M2_PS_X} we have used the following contractions
\begin{align}
    &K^\mu P^{\nu} D_{\mu\nu}(Q) = \frac{k^0 p^0}{q^2+\Pi_{\rm L}}- \frac{\bk\cdot\bp - \frac{(\bk\cdot\bq) (\bp\cdot\bq)}{q^2}}{q^2+\Pi_{\rm T}}\,,\\
    &P^\beta P^\nu D_{\mu\nu}(Q) D^\mu_{\beta}(Q)= \frac{(p^0)^2}{(q^2+\Pi_{\rm L})^2}-\frac{\bp^2 - \frac{(\bp\cdot\bq)^2}{q^2}}{(q^2+\Pi_{\rm T})^2}\,,
\end{align}
together with $\bp_1\cdot\bq = - q^2/2$, imposed by the phase space, analogously to Eq.~\eqref{Eq:Delta_p1.q}.

\subsection{Phase-space integration}
Following the procedure for the phase-space integration already introduced above, we can write
\begin{align}
    \mathcal{N}_a^{(\mu\mu)}&=\frac{T^3 J(\tilde\omega)k}{4096\pi^{9}}\,\!\int\! \frac{dx_1 dx_2 d\phi dq}{\sqrt{1-x_{12}^2}}\frac{1}{\mathcal{S}} \sum\left|\mathcal{M}_a^{(\mu\mu)}\right|^2.\label{Eq:Emiss_PS}
\end{align}
The integration over $q$ can be performed as before, leaving out the angular integrals. The pseudoscalar emission spectrum from $\mu\mu$ scatterings then reads
\begin{equation}
    \mathcal{N}_a^{(\mu\mu)}=\frac{g_{a\mu}^2 \alpha^2T^3}{16\pi^{6}} J(\tilde\omega)k\left\{\frac{L_0}{m_D}+\frac{\beta_a^2\beta_\mu^2 L_2}{m_{T}}+\beta_a^2\left[\left(-\frac{1}{m_D}+\frac{\beta_\mu^2}{m_{T}}\right)H_0 - \frac{m_a^2}{\omega^2}\left(\frac{J_0}{m_D}+\frac{\beta_\mu^4 J_2}{m_{T}}\right)\right]\right\}\,,\nonumber
\end{equation}
\end{widetext}
with $J_{0,2}$ defined in Eq.~\eqref{Eq:J0_J2} replacing $\beta_\phi\to\beta_a$, and the following functions
\begin{align}
    H_{0}&=\int \frac{dx_1 dx_2 d\phi(1-x_1^2)(1-x_2^2)\sin^2\!\phi}{(1-\beta_\mu\beta_a x_1)^2 (1-x_{12}^2)^{3/2}}\,,\\
    L_n&=\int dx_1 dx_2 d\phi \frac{x_2^n}{(1-\beta_\mu\beta_a x_1)^2 (1-x_{12}^2)^{1/2}}\,.
\end{align}
Both $H_0$ and $L_n$ can be analytically computed. For the relevant values of $n$, they read
\begin{align}
    H_0 &= \frac{4\pi^2}{\beta_a^3\beta_\mu^3}\left({\rm ArcTanh}(\beta_a\beta_\mu)-\beta_a\beta_\mu\right)\,,\\
    L_0 &= M_0 = \frac{4\pi^2}{1-\beta_a^2\beta_\mu^2}\,,\\
    L_2 &= \frac{2\pi^2}{\beta_a^3\beta_\mu^3} \left(\frac{\beta_a\beta_\mu}{1-\beta_a^2\beta_\mu^2}-{\rm ArcTanh}(\beta_a\beta_\mu)\right)\,,
\end{align}
while they vanish for $n$ odd, and we already neglected such contributions. Moreover, we have neglected the contribution from the $q^2/4m_\mu^2$ term in the matrix elements, as it gives rise to terms suppressed by $m_\mu^{-1}$ in the emissivities, and $m_\mu\gg m_{T}$. 

\begin{widetext}
Analogously, the emission spectra from $\mu X$ scatterings are given by 
\begin{equation}
    \mathcal{N}_a^{(\mu X)}=\frac{g_{a\mu}^2 \alpha^2 T^3 }{16\pi^{6}}J(\tilde\omega)k\left(\!\frac{\varepsilon_X}{\varepsilon_\mu}\!\right)^{\!2}\! \left\{\frac{L_0}{m_D}+\frac{\beta_a^2\beta_X^2 L_2}{m_{T}}+\beta_a^2\left[\left(\frac{\beta_X^2}{m_{T}}-\frac{1}{m_D}\right)H_0 - \frac{m_a^2}{\omega^2}\left(\frac{J_0}{m_D}+\frac{\beta_\mu^2\beta_X^2 J_2}{m_{T}}\right)\right]\right\},
\end{equation}
and protons contribute only in the nonsuperconducting phase with $m_T=m_\Lambda$.\\

\subsection{Energy-loss rate}

Once again, the pseudoscalar emissivities can be computed analytically in the massless limit. Compared to the scalar and vector boson emission, the matrix elements in this case do not depend on the energy of the emitted pseudoscalar. In consequence, the emission rates scale as $T^6$ and the integration over the energy of the emitted boson has an additional factor of $\tilde{\omega}^2$ compared to Eq.~\eqref{Eq:J_integral}
\begin{equation}
    \int_0^{\infty}d\tilde{\omega}\,\tilde{\omega}^2 J(\tilde{\omega})= \frac{62 \pi^6}{945}\,.\label{Eq:J_integral_PS}
\end{equation}
Ultimately, the pseudoscalar emissivities in the massless limit can be expressed as
\begin{eqnarray}
    \kern-2.2em Q_{a,0}^{(\mu\mu)}&=&\frac{31\pi^2 g_{a\mu}^2 \alpha^2 T^6}{1890(1-\beta_\mu^2)}\left[\frac{1}{2m_T}\left(2\beta_\mu^2-1+\frac{1-\beta_\mu^2}{\beta_\mu}{\rm ArcTanh}\,\beta_\mu\right)+\frac{1}{m_D}\left(\frac{1}{\beta_\mu^2}-\frac{1-\beta_\mu^2}{\beta_\mu^3}{\rm ArcTanh}\,\beta_\mu\right)\right]\!,\label{eq:Qa_mu}\\
    \kern-2.2em Q_{a,0}^{(\mu X)}&=&\frac{31\pi^2 g_{a\mu}^2 \alpha^2 T^6}{1890(1-\beta_\mu^2)}\!\left(\!\frac{\varepsilon_X}{\varepsilon_\mu}\!\right)^{\!2}\!\left[\frac{1}{2m_T}\frac{\beta_X^2}{\beta_\mu^2}\left(\!2\beta_\mu^2-1+\frac{1-\beta_\mu^2}{\beta_\mu}{\rm ArcTanh}\,\beta_\mu\right)+\frac{1}{m_D}\left(\frac{1}{\beta_\mu^2}-\frac{1-\beta_\mu^2}{\beta_\mu^3}{\rm ArcTanh}\,\beta_\mu\right)\right]\!,\label{eq:Qa_e}  
\end{eqnarray}
with $X=e,p$. The results in the massive case can be recovered through the following fitting expressions ($\tilde m_a=m_a/T$)
\begin{eqnarray}
   Q_a^{(\mu\mu)}(\tilde m_a,\beta_\mu)&\simeq&Q_{a,0}^{(\mu\mu)}\,\left[1+1.03\,\tilde m_a^{0.86}\left(1-\beta_\mu^2\right)\right] e^{-0.74\,\tilde m_a}\,,\\
    Q_a^{(\mu X)}(\tilde m_a,\beta_\mu)&\simeq& Q_{a,0}^{(\mu X)}\,\left[1+0.95\,\tilde m_a^{0.78}\left(1-\beta_\mu^2\right)\right] e^{-0.73\,\tilde m_a}\,.
\end{eqnarray}
\end{widetext}

\section{One-zone numerical model}
\label{sec:one-zone}

To illustrate these results numerically for a simple case, we consider a one-zone model for a NS of baryonic mass $1.8\,M_\odot$, for which we assume the APR equation of state. Averaging over the mass distribution in the stellar profile, we consider an average density $\rho\simeq8.5\times10^{14}~{\rm g}~{\rm cm}^{-3}$ and electron abundance $Y_e=0.06$, implying for the other parameters the values $1-\beta_e\simeq10^{-3}$, $\beta_\mu=0.83$, and $\beta_p=0.51$, where the effective proton mass is $m_p^*=370\,{\rm MeV}$, and the kinematical Fermi energies $\varepsilon_e\simeq\varepsilon_\mu=190\,{\rm MeV}$, $\varepsilon_p=430\,{\rm MeV}$. For the bosons, we use the purely muonic case, without direct coupling to electrons.

In a superconducting medium, the luminosity associated with massless-particle emission from such a schematic star, namely the energy radiated in bosons per unit time, is found to be
\begin{eqnarray}\label{Eq:luminosity_SC}
    \kern-2em L_\phi&\simeq&1.4\times10^{25}L_\odot\,T_8^{4}g_{\phi\mu}^2(F_{\mu \mu}+F_{\mu e}+F_{ee}),
    \\[0.3cm]
    \kern-2em L_V&\simeq& 1.2\times10^{24}L_\odot\,T_8^4g_{V\!\mu}^2(F_{\mu \mu}+F_{\mu e}+F_{e e})\,,
     \\[0.3cm]
    \kern-2em L_a&\simeq&6.3\times10^{15}L_\odot\,T_8^6g_{a\mu}^2 (F_{\mu \mu}+F_{\mu e})\,, 
\end{eqnarray}
where $T_8=T/10^8~{\rm K}$ and $F_{ij}$ encodes the relative contribution from scatterings $i+j\to i+j+X$, $i,j=e,\mu,p,n$. Notice that in these expressions, protons do not appear as targets because they are here in a superconducting state. 

\begin{table*}
\centering
\renewcommand{\arraystretch}{1.5}
\setlength{\tabcolsep}{12pt}
\begin{tabular}{lccc cccccc}
\toprule
& \multicolumn{3}{c}{Superconducting}
& \multicolumn{6}{c}{Nonsuperconducting} \\
\cmidrule(lr){2-4}
\cmidrule(lr){5-10}

& $F_{\mu\mu}$ 
& $F_{ee}$ 
& $F_{e \mu}$ 

& $F_{\mu\mu}$ 
& $F_{ee}$ 
& $F_{e \mu}$ 
& $F_{\mu p}$ 
& $F_{ep}$
& $F_{pn}$\\

\midrule

Scalar
& $\sim10^{-3}$
& $0.56$
& $0.44$
& $\sim10^{-3}$
& $0.28$
& $0.21$
& $\sim10^{-3}$
& $0.50$
& $0.01$\\

Vector
& $0.01$
& $0.55$
& $0.44$
& $\sim10^{-3}$
& $0.08$
& $0.06$
& $0.01$
& $0.09$
& $0.76$\\

Pseudoscalar
& 0.42
& ---
& 0.58
& 0.20
& ---
& 0.30
& 0.50
& ---
& ---\\

\bottomrule
\end{tabular}
\caption{Relative contributions to the emissivities $F_{ij}$ from the scattering channels $i+j\to i+j+X$, where $i,j=e,\mu,p,n$ and $X=\phi,V,a$ denotes a purely muonic boson, for superconducting and nonsuperconducting proton matter, to be used in Eqs.~\eqref{Eq:luminosity_SC}--\eqref{Eq:luminosity_axion}.}
\label{tab:Fij}

\end{table*}

If protons are not superconducting, they contribute as scattering targets and the screening of transverse photons occurs not with the Meissner scale, but instead with a Landau damping scale given in Eq.~\eqref{Eq:Landau_Scale}. In addition, protons can radiate scalar and vector bosons in $pn$ nuclear bremsstrahlung processes. The luminosities are now parametrically given by
\begin{eqnarray}\label{Eq:luminosity_NSC}
    \kern-3.5em L_\phi&\simeq&8.3\times10^{25}L_\odot\,T_8^{11/3}\,g_{\phi\mu}^2\\\nonumber
    \kern-3.5em\kern3em{}&\times&(F_{\mu\mu}+F_{e\mu }+F_{\mu p}+F_{ee}+F_{ep}+T_8^{1/3} F_{pn} ),
    \\[0.3cm]
    \kern-3.5em L_V\!&\simeq&2.6\times10^{25}L_\odot\,T_8^{11/3}g_{V\!\mu}^2  \\\nonumber
    \kern-3.5em\kern3em{}&\times&(F_{\mu\mu}+F_{e\mu }+F_{\mu p}+F_{ee}+F_{ep}+T_8^{1/3} F_{pn}),\label{Eq:luminosity_vector}
     \\[0.3cm]
    \kern-3.5em L_a&\simeq&4.1\times10^{16} L_\odot\,T_8^{17/3}\,g_{a\mu}^2 (F_{\mu\mu}+F_{e\mu}+F_{\mu p}).\label{Eq:luminosity_axion}
\end{eqnarray}    
Values for the coefficients $F_{ij}$ for both superconducting and nonsuperconducting protons are reported in Table~\ref{tab:Fij}. 

Intriguingly, the dominant contributions to both scalar and vector boson emissivities arise from processes induced by the response of the muon charge distribution to external electromagnetic excitations generated by accelerated electrons and protons. Vector boson emissivities are dominated by $pn$ nuclear scatterings enabled by a nonvanishing vector in-medium coupling to protons $\hat{g}_{V\!p}/\g_{V\!\mu}\sim0.33$. Enhanced by the efficient nuclear collision rate, this process turns out to be roughly one order of magnitude more efficient than any of the viable electromagnetic scattering processes. The remaining contribution to the vector boson emissivity arises from processes involving  ultrarelativistic electron bremsstrahlung, as the energy radiated per unit time by a charged particle moving in an electromagnetic field benefits from a typical relativistic Lorentz enhancement $\sim(1-\beta_e^2)^{-1}\sim\pi/\alpha$. Consequently, scattering channels involving radiating electrons largely dominate over those involving radiating muons, which are only mildly relativistic in the NS environment.

In contrast, scalar boson emissivities turn out to be dominated by processes involving radiating electrons. Although the parametric suppression of the renormalized in-medium coupling is even stronger than in the vector case $\hat{g}_{\phi e}/\hat{g}_{\phi\mu}\sim0.05$, the corresponding emission rates receive a significantly larger enhancement $\sim(1-\beta_e^2)^{-2}\sim(\pi/\alpha)^2$ from the additional Lorentz factor generated by the boosted electron charge distribution generating the in-medium interaction. This enhancement allows electron-radiating processes to dominate over both tree-level electromagnetic processes involving radiating muons and proton bremsstrahlung induced by $pn$ nuclear collisions.

To assess the relevance of these luminosities for NS cooling, we compare them with the standard luminosity around the epoch of neutrino--photon equality. As demonstrated in our companion paper~\cite{Fiorillo:2026new}, requiring the luminosity of a new cooling channel to satisfy
\begin{equation}
    L_X\lesssim L_\nu\,,
\end{equation}
at this epoch provides an excellent approximate criterion for avoiding an
appreciable modification of the observed cooling evolution. For the
reference NS model considered here, neutrino--photon equality occurs at a
core temperature of order $T\simeq10^8~{\rm K}$, where
\begin{equation}
    L_\nu\simeq L_\gamma\simeq1.4\times10^{33}\,
    \mathrm{erg}\,\mathrm{s}^{-1}
    \simeq0.37\,L_\odot\,.
\end{equation}
Equating the luminosities in Eqs.~\eqref{Eq:luminosity_NSC}--\eqref{Eq:luminosity_axion}, evaluated at $T_8=1$, to the standard luminosity
above gives
\begin{align}
    g_{\phi\mu}&\simeq6.7\times10^{-14}\,,\\
    g_{V\mu}&\simeq1.2\times10^{-13}\,,\\
    g_{a\mu}&\simeq3.0\times10^{-9}\,.
\end{align}
Thus, couplings of these sizes are expected to produce a luminosity comparable to the standard neutrino luminosity and can appreciably modify the late-time thermal evolution. Clearly, since $L_X\propto g_{X\mu}^2$, increasing any of these couplings by a factor of three enhances the corresponding luminosity by approximately one order of magnitude.

A quantitative implementation of the emission rates in the \texttt{NSCool} code~\cite{2016ascl.soft09009P}, including the resulting cooling curves and their comparison with observations, is presented in the companion paper~\cite{Fiorillo:2026new}. The limits obtained there, $g_{\phi\mu}\lesssim7.4\times10^{-14}$, $g_{V\mu}\lesssim1.4\times10^{-13}$, and $g_{a\mu}\lesssim2.9\times10^{-9}$, agree closely with the simple one-zone estimates above.

We finally remark that the comparatively weaker sensitivity to the pseudoscalar coupling follows from the steeper
temperature dependence of its luminosity compared
with both scalar and vector boson emission.

\section{Conclusions}\label{sec:conclusions}

The large populations of electrons and muons in NS interiors can abundantly produce light leptophilic bosons through bremsstrahlung emission from charged particles. Therefore, observed NS cooling ages can provide constraints, notably on muonic couplings, that are far more restrictive than other arguments such as the SN~1987A cooling limit. With this motivation in mind, we have computed the emissivities of new scalar ($\phi$), vector ($V$), and pseudoscalar ($a$) particles in ultradegenerate relativistic $ep\mu$ matter, representative of NS conditions. In our calculations, we do not consider whether hyperons or charged-meson condensates might exist in~a~NS.

Electromagnetic processes do not suffer from the notorious nuclear-physics uncertainties that bedevil, for example, axion emission rates from nuclear processes~\cite{Fiorillo:2025gnd}. Nevertheless, they involve their own complications, mainly caused by the response of the plasma to perturbations. First, the boson effective couplings are renormalized by the screening populations, which is important for vectors and scalars, but absent for pseudoscalars due to parity selection rules. As a consequence, purely muonic scalar and vector bosons are predominantly emitted by electromagnetic electron scatterings and proton nuclear collisions, respectively. We have provided a physical picture of this effect, in comparison to earlier mathematical treatments~\cite{Leinson:1997zt, Leinson:1999ut}.

Second, the electromagnetic interaction responsible for lepton acceleration and consequent bremsstrahlung is screened, a particularly relevant effect in our regime, where the momentum exchanged among leptons is orders of magnitude larger than the exchanged energy. Another in-medium effect is that the electron Fermi velocity depends on the effective mass $m_e^*\simeq10$~MeV, so electrons are less ultrarelativistic than they would otherwise be. 

The emission rates for different coupling structures depend on various combinations of medium response functions. In this way, they are partly related to neutrino pair-emission rates for the same NS conditions. Our vector-boson emission rate, after a suitable phase-space transformation, agrees with the earlier vector-current neutrino emission rate~\cite{Shternin:2025swb}. On the other hand, we disagree with the earlier finding that the vector-current emission rate was suppressed by the weak charge renormalization---this suppression only pertains to a purely degenerate electron gas, but not to a mixture of electrons and muons. For our scalar and pseudoscalar emission rates, there is no immediate correspondence to neutrino rates.

Longitudinal electromagnetic modes are screened by the static Debye scale, whereas the behavior of transverse modes depends on proton superconductivity. In the superconducting phase, protons form Cooper pairs and effectively decouple from scattering, while endowing transverse photons with an effective mass set by the Meissner scale, in turn fixed by the inverse London penetration length. In the normal phase, transverse modes are instead subject to dynamical Landau damping.

In analogy to neutrino pair emission, we find that transverse photon exchange always dominates, as these modes are less efficiently screened or damped in either proton phase. 
Before in-medium renormalization of the coupling, the emissivities for purely muonic bosons parametrically scale as
\begin{equation}
    Q_{\phi,V}\propto \frac{\varepsilon_{\mu}^2 T^4}{(1-\beta_\mu^2) m_T}
    \quad\text{and}\quad
    Q_a \sim \frac{T^6}{(1-\beta_\mu^2) m_T},
\end{equation}
with $m_T$ denoting the relevant scale of the exchanged momentum, set by either Meissner screening in the superconducting phase, or Landau damping in the normal phase. In the latter case, the dynamical nature of the damping, as encoded in Eq.~\eqref{Eq:Landau_Scale}, effectively reduces the temperature dependence of the emissivities by a factor $T^{1/3}$, in agreement with previous findings~\cite{Shternin:2025swb}.
Once in-medium renormalization is taken into account, scalar bosons are predominantly radiated in electromagnetic scatterings involving ultrarelativistic electrons, retaining the same $T^{11/3}$ scaling as in the corresponding muon processes. At the same time, vector bosons are mainly produced through nuclear bremsstrahlung in proton-neutron collisions, for which the emission rate scales as $Q_V\sim T^4$.

Notice that the relativistic enhancement by a factor $(1-\beta_X^2)^{-1}$, which plays a key role for the dominance of ultrarelativistic electron bremsstrahlung, is in fact simply the traditional enhancement factor in the intensity radiated by an ultrarelativistic particle. In the instantaneous rest frame of the particle, its radiated energy per unit time is proportional to the square of its acceleration $|\bw|^2$ by Larmor's formula. The covariant generalization shows that in the laboratory frame the radiated energy per unit time is proportional to $-w_\mu w^\mu$, where $w^\mu=du^\mu/d\tau$ is the four-acceleration, $u^\mu$ is the four-velocity, and $\tau$ is the proper time. In a given electromagnetic field that scatters the charge, $-w^2=e^2 F^{\mu\nu} F_{\alpha \mu} u_\nu u^\alpha/m^2$ and is thus proportional to the squared Lorentz factor.

Our results also reveal a power-law dependence on the screening scale and kinematic factors, with $Q\propto [(1-\beta_\mu^2)\,m_T]^{-1}$. This behavior originates from the extreme degeneracy, where the exchanged momentum among scattering particles is much larger than the exchanged energy. This hierarchy prevents the possibility of very forward scatterings, in which the more familiar logarithmic divergence typical of less degenerate systems would emerge.

Intriguingly, we observe that muonic scalar and vector bosons are primarily produced by electrons and protons, while the large muon population hosted in the NS interior mainly acts as a mediator for boson emission processes. In fact, in the most efficient scalar and vector emission channels, boson radiation arises from the coherent response of the muon population to electromagnetic fluctuations induced by efficient collisions involving electrons and protons, through the screening cloud that accompanies any charge accelerated on timescales longer than the inverse plasma frequency.

Our conclusion extends to the Standard Model case, where neutrino pair emission through vector currents is dominated by $pn$ bremsstrahlung in the nonsuperconducting phase, a point that seems to have gone unnoticed (see, however, Refs.~\cite{Leinson:2000ks, Leinson:2000un} for a similar observation in the superconducting phase). Still, this process remains subdominant compared to the standard~MUrca.

Overall, our framework provides a systematic approach to computing emissivities of light states in dense matter, and can be readily applied to a broad class of models in which new light degrees of freedom couple nontrivially to leptons. In a companion paper, we use these results to derive leading constraints on muonic bosons from NS cooling ages~\cite{Fiorillo:2026new}.

\vspace{0.5cm}
\section*{Acknowledgments}
This article is based upon work from COST Action COSMIC WISPers (CA21106), supported by COST (European Cooperation in Science and Technology). 
DFGF acknowledges support by the TAsP (Theoretical Astroparticle Physics) project. GGR acknowledges partial support by the German Research Foundation (DFG) through the Collaborative Research Centre ``Neutrinos and Dark Matter in Astro- and Particle Physics (NDM),'' Grant SFB--1258--283604770 and under Germany’s Excellence Strategy through the Cluster of Excellence ORIGINS EXC--2094--390783311.
EV acknowledges support by the Italian MUR Departments of Excellence grant 2023-2027 ``Quantum Frontier''. AL, NS and EV are supported by the Italian MUR through the FIS 2 project FIS-2023-01577 (DD n. 23314 10-12-2024, CUP C53C24001460001), and by Istituto Nazionale di Fisica Nucleare (INFN) through the Theoretical Astroparticle Physics (TAsP) project.

\bibliographystyle{bibi}
\bibliography{references}

@article{Leinson:1999ut,
    author = "Leinson, L. B.",
    title = "{Neutrino pair emission due to electron phonon scattering in a neutron star crust: A reappraisal}",
    eprint = "hep-ph/0009049",
    archivePrefix = "arXiv",
    doi = "10.1016/S0370-2693(99)01278-2",
    journal = "Phys. Lett. B",
    volume = "469",
    pages = "166",
    year = "1999"
}

@article{Das:2026uam,
    author = "Das, Debottam and Ghosh, Purusottam and Puri, Rahul",
    title = "{SN1987A Constraints of Light $\boldsymbol{Z'}$ with Non-Mixing Polarisations}",
    eprint = "2606.23769",
    archivePrefix = "arXiv",
    primaryClass = "hep-ph",
    month = "6",
    year = "2026"
}

@article{Leinson:2000ks,
    author = "Leinson, L. B.",
    title = "{Neutrino emission due to Cooper pairing of protons in cooling neutron stars: Collective effects}",
    eprint = "hep-ph/0009050",
    archivePrefix = "arXiv",
    doi = "10.1016/S0370-2693(99)01473-2",
    journal = "Phys. Lett. B",
    volume = "473",
    pages = "318",
    year = "2000"
}

@article{Leinson:2000un,
    author = "Leinson, L. B.",
    title = "{Collective neutrino pair emission due to Cooper pairing of protons in superconducting neutron stars}",
    eprint = "hep-ph/0009052",
    archivePrefix = "arXiv",
    doi = "10.1016/S0375-9474(00)00575-3",
    journal = "Nucl. Phys. A",
    volume = "687",
    pages = "489--511",
    year = "2001"
}

@article{Lykasov:2008yz,
    author = "Lykasov, G. I. and Pethick, C. J. and Schwenk, A.",
    title = "{A Unified approach to structure factors and neutrino processes in nucleon matter}",
    eprint = "0808.0330",
    archivePrefix = "arXiv",
    primaryClass = "nucl-th",
    reportNumber = "NORDITA-2008-30",
    doi = "10.1103/PhysRevC.78.045803",
    journal = "Phys. Rev. C",
    volume = "78",
    pages = "045803",
    year = "2008"
}

@article{Fiorillo:2025gnd,
    author = "Fiorillo, Damiano F. G. and Gil Muyor, {\'A}ngel and Janka, Hans-Thomas and Raffelt, Georg G. and Vitagliano, Edoardo",
    title = "{Axion-photon conversion in transient compact stars: Systematics, constraints, and opportunities}",
    eprint = "2509.13322",
    archivePrefix = "arXiv",
    primaryClass = "hep-ph",
    doi = "10.1088/1475-7516/2026/03/053",
    journal = "JCAP",
    volume = "03",
    pages = "053",
    year = "2026"
}

@article{Leinson:1997zt,
    author = "Leinson, L. B. and Perez, A.",
    title = "{Collective effects in neutrino anti-neutrino synchrotron radiation from neutron stars}",
    eprint = "astro-ph/9710257",
    archivePrefix = "arXiv",
    doi = "10.1103/PhysRevD.59.043002",
    journal = "Phys. Rev. D",
    volume = "59",
    pages = "043002",
    year = "1999"
}

@article{Gnedin:1995lgf,
    author = "Gnedin, O. Y. and Yakovlev, D. G.",
    title = "{Thermal conductivity of electrons and muons in neutron star cores}",
    doi = "10.1016/0375-9474(94)00503-F",
    journal = "Nucl. Phys. A",
    volume = "582",
    pages = "697--716",
    year = "1995"
}

@article{Shternin:2006uq,
    author = "Shternin, P. S. and Yakovlev, Dima G.",
    title = "{Electron thermal conductivity owing to collisions between degenerate electrons}",
    eprint = "astro-ph/0608371",
    archivePrefix = "arXiv",
    doi = "10.1103/PhysRevD.74.043004",
    journal = "Phys. Rev. D",
    volume = "74",
    pages = "043004",
    year = "2006"
}

@article{Shternin:2007ee,
    author = "Shternin, P. S. and Yakovlev, D. G.",
    title = "{Electron-muon heat conduction in neutron star cores via the exchange of transverse plasmons}",
    eprint = "0705.1963",
    archivePrefix = "arXiv",
    primaryClass = "astro-ph",
    doi = "10.1103/PhysRevD.75.103004",
    journal = "Phys. Rev. D",
    volume = "75",
    pages = "103004",
    year = "2007"
}

@article{Shternin:2025swb,
    author = "Shternin, Peter S.",
    title = "{Neutrino-Pair Bremsstrahlung Due to Electromagnetic Collisions in Neutron Star Cores Revisited}",
    eprint = "2512.08780",
    archivePrefix = "arXiv",
    primaryClass = "astro-ph.HE",
    doi = "10.3390/particles8040100",
    journal = "Particles",
    volume = "8",
    number = "4",
    pages = "100",
    year = "2025"
}

@article{Kaminker:1999sd,
    author = "Kaminker, A. D. and Haensel, P.",
    title = "{Neutrino emission due to electron bremsstrahlung in superfluid neutron star cores}",
    eprint = "astro-ph/9908249",
    archivePrefix = "arXiv",
    journal = "Acta Phys. Polon. B",
    volume = "30",
    pages = "1125",
    year = "1999"
}

@article{Jaikumar:2005gm,
    author = "Jaikumar, Prashanth and Gale, Charles and Page, Dany",
    title = "{Bremsstrahlung neutrinos from electron-electron scattering in a relativistic degenerate electron plasma}",
    eprint = "hep-ph/0508245",
    archivePrefix = "arXiv",
    doi = "10.1103/PhysRevD.72.123004",
    journal = "Phys. Rev. D",
    volume = "72",
    pages = "123004",
    year = "2005"
}

@ARTICLE{Braaten:1991hg,
       author = {{Braaten}, Eric},
        title = "{Neutrino Emissivity of an Ultrarelativistic Plasma from Positron and Plasmino Annihilation}",
      journal = {Astrophys. J.},
         year = 1992,
        month = jun,
       volume = {392},
        pages = {70},
          doi = {10.1086/171405}
}

@article{Braaten:1993jw,
    author = "Braaten, Eric and Segel, Daniel",
    title = "{Neutrino energy loss from the plasma process at all temperatures and densities}",
    eprint = "hep-ph/9302213",
    archivePrefix = "arXiv",
    reportNumber = "NUHEP-TH-93-1",
    doi = "10.1103/PhysRevD.48.1478",
    journal = "Phys. Rev. D",
    volume = "48",
    pages = "1478--1491",
    year = "1993"
}

@article{Raffelt:1985nj,
    author = "Raffelt, Georg G.",
    title = "{Axion Constraints From White Dwarf Cooling Times}",
    reportNumber = "MPI-PAE/PTh 67/85",
    doi = "10.1016/0370-2693(86)91588-1",
    journal = "Phys. Lett. B",
    volume = "166",
    pages = "402--406",
    year = "1986"
}

@article{Isern:2022vdx,
    author = "Isern, Jordi and Torres, Santiago and Rebassa-Mansergas, Alberto",
    title = "{White Dwarfs as Physics Laboratories: Lights and Shadows}",
    eprint = "2202.02052",
    archivePrefix = "arXiv",
    primaryClass = "astro-ph.HE",
    doi = "10.3389/fspas.2022.815517",
    journal = "Front. Astron. Space Sci.",
    volume = "9",
    pages = "815517",
    year = "2022"
}

@article{Fleury:2025ahw,
    author = "Fleury, Leesa and Obertas, Alysa and Richer, Harvey and Heyl, Jeremy",
    title = "{Axion Constraints from White Dwarf Cooling in 47 Tucanae}",
    eprint = "2511.21676",
    archivePrefix = "arXiv",
    primaryClass = "astro-ph.SR",
    month = "11",
    year = "2025"
}

@article{Alberino:2026yxi,
    author = "Alberino, Mart{\'\i}n L. and Miller Bertolami, Marcelo M. and Camisassa, Mar{\'\i}a E. and Caputo, Andrea and Torres, Santiago",
    title = "{New axion bounds derived from the 100-parsec Gaia DR3 white dwarf luminosity function}",
    eprint = "2603.00901",
    archivePrefix = "arXiv",
    primaryClass = "astro-ph.SR",
    month = "3",
    year = "2026"
}

@article{Raffelt:1987yt,
    author = "Raffelt, Georg and Seckel, David",
    title = "{Bounds on Exotic Particle Interactions from SN 1987A}",
    reportNumber = "SCIPP-87/107",
    doi = "10.1103/PhysRevLett.60.1793",
    journal = "Phys. Rev. Lett.",
    volume = "60",
    pages = "1793",
    year = "1988"
}

@article{Dearborn:1985gp,
    author = "Dearborn, Daivd S. P. and Schramm, David N. and Steigman, Gary",
    title = "{Astrophysical Constraints on the Couplings of Axions Majorons and Familons}",
    reportNumber = "BA-85-54",
    doi = "10.1103/PhysRevLett.56.26",
    journal = "Phys. Rev. Lett.",
    volume = "56",
    pages = "26",
    year = "1986"
}

@article{Raffelt:1994ry,
    author = "Raffelt, Georg and Weiss, Achim",
    title = "{Red giant bound on the axion-electron coupling revisited}",
    eprint = "hep-ph/9410205",
    archivePrefix = "arXiv",
    doi = "10.1103/PhysRevD.51.1495",
    journal = "Phys. Rev. D",
    volume = "51",
    pages = "1495--1498",
    year = "1995"
}

@article{Straniero:2020iyi,
    author = "Straniero, O. and Pallanca, C. and Dalessandro, E. and Dominguez, I. and Ferraro, F. R. and Giannotti, M. and Mirizzi, A. and Piersanti, L.",
    title = "{The RGB tip of galactic globular clusters and the revision of the axion-electron coupling bound}",
    eprint = "2010.03833",
    archivePrefix = "arXiv",
    primaryClass = "astro-ph.SR",
    doi = "10.1051/0004-6361/202038775",
    journal = "Astron. Astrophys.",
    volume = "644",
    pages = "A166",
    year = "2020"
}

@article{Turner:1987by,
    author = "Turner, Michael S.",
    title = "{Axions from SN 1987A}",
    reportNumber = "FERMILAB-PUB-87-202-A",
    doi = "10.1103/PhysRevLett.60.1797",
    journal = "Phys. Rev. Lett.",
    volume = "60",
    pages = "1797",
    year = "1988"
}

@article{Mayle:1987as,
    author = "Mayle, Ron and Wilson, James R. and Ellis, John R. and Olive, Keith A. and Schramm, David N. and Steigman, Gary",
    title = "{Constraints on Axions from SN 1987A}",
    reportNumber = "FERMILAB-PUB-87-225-A, EFI-87-104-CHICAGO, UMN-TH-637-87, CERN-TH-4887-87",
    doi = "10.1016/0370-2693(88)91595-X",
    journal = "Phys. Lett. B",
    volume = "203",
    pages = "188--196",
    year = "1988"
}

@article{Raffelt:2006cw,
    author = "Raffelt, Georg G.",
    editor = "Kuster, Markus and Raffelt, Georg and Beltran, Berta",
    title = "{Astrophysical axion bounds}",
    eprint = "hep-ph/0611350",
    archivePrefix = "arXiv",
    reportNumber = "MPP-2006-172",
    doi = "10.1007/978-3-540-73518-2_3",
    journal = "Lect. Notes Phys.",
    volume = "741",
    pages = "51--71",
    year = "2008"
}

@article{Carenza:2019pxu,
    author = "Carenza, Pierluca and Fischer, Tobias and Giannotti, Maurizio and Guo, Gang and Mart{\'\i}nez-Pinedo, Gabriel and Mirizzi, Alessandro",
    title = "{Improved axion emissivity from a supernova via nucleon-nucleon bremsstrahlung}",
    eprint = "1906.11844",
    archivePrefix = "arXiv",
    primaryClass = "hep-ph",
    doi = "10.1088/1475-7516/2019/10/016",
    journal = "JCAP",
    volume = "10",
    number = "10",
    pages = "016",
    year = "2019",
    note = "Erratum
    \href{https://doi.org/10.1088/1475-7516/2020/05/E01}{{\em JCAP} {\bf 05} (2020) E01}"
}

@article{Yakovlev:2000jp,
    author = "Yakovlev, D. G. and Kaminker, A. D. and Gnedin, Oleg Y. and Haensel, P.",
    title = "{Neutrino emission from neutron stars}",
    eprint = "astro-ph/0012122",
    archivePrefix = "arXiv",
    doi = "10.1016/S0370-1573(00)00131-9",
    journal = "Phys. Rept.",
    volume = "354",
    pages = "1",
    year = "2001"
}

@article{Page:2005fq,
    author = "Page, Dany and Geppert, Ulrich and Weber, Fridolin",
    title = "{The Cooling of compact stars}",
    eprint = "astro-ph/0508056",
    archivePrefix = "arXiv",
    doi = "10.1016/j.nuclphysa.2005.09.019",
    journal = "Nucl. Phys. A",
    volume = "777",
    pages = "497--530",
    year = "2006"
}

@article{Gamow:1941gis,
    author = "Gamow, G. and Schoenberg, M.",
    title = "{Neutrino Theory of Stellar Collapse}",
    doi = "10.1103/PhysRev.59.539",
    journal = "Phys. Rev.",
    volume = "59",
    number = "7",
    pages = "539",
    year = "1941"
}

@article{Caputo:2024oqc,
    author = "Caputo, Andrea and Raffelt, Georg",
    title = "{Astrophysical Axion Bounds: The 2024 Edition}",
    eprint = "2401.13728",
    archivePrefix = "arXiv",
    primaryClass = "hep-ph",
    reportNumber = "MPP-2024-13, CERN-TH-2024-013",
    doi = "10.22323/1.454.0041",
    journal = "PoS",
    volume = "COSMICWISPers",
    pages = "041",
    year = "2024"
}

@article{Blinov:2025aha,
    author = "Blinov, Nikita and Fox, Patrick J. and Kelly, Kevin J. and Plestid, Ryan and Zhou, Tao",
    title = "{$L_{\mu}-L_{\tau}$ gauge bosons in beam dumps and supernovae}",
    eprint = "2511.09619",
    archivePrefix = "arXiv",
    primaryClass = "hep-ph",
    reportNumber = "FERMILAB-PUB-25-0818-T, MI-HET-869, CALT-TH/2025-034, CERN-TH-2025-229",
    doi = "10.1007/JHEP06(2026)155",
    journal = "JHEP",
    volume = "06",
    pages = "155",
    year = "2026"
}

@article{Akita:2023iwq,
    author = "Akita, Kensuke and Im, Sang Hui and Masud, Mehedi and Yun, Seokhoon",
    title = "{Limits on heavy neutral leptons, {$Z'$} bosons and majorons from high-energy supernova neutrinos}",
    eprint = "2312.13627",
    archivePrefix = "arXiv",
    primaryClass = "hep-ph",
    reportNumber = "CTPU-PTC-23-55",
    doi = "10.1007/JHEP07(2024)057",
    journal = "JHEP",
    volume = "07",
    pages = "057",
    year = "2024"
}

@article{Croon:2020lrf,
    author = "Croon, Djuna and Elor, Gilly and Leane, Rebecca K. and McDermott, Samuel D.",
    title = "{Supernova Muons: New Constraints on $Z'$ Bosons, Axions and ALPs}",
    eprint = "2006.13942",
    archivePrefix = "arXiv",
    primaryClass = "hep-ph",
    reportNumber = "MIT-CTP/5214, FERMILAB-PUB-20-246-A-T",
    doi = "10.1007/JHEP01(2021)107",
    journal = "JHEP",
    volume = "01",
    pages = "107",
    year = "2021"
}

@article{Iwamoto:1984ir,
    author = "Iwamoto, N.",
    title = "{Axion Emission from Neutron Stars}",
    doi = "10.1103/PhysRevLett.53.1198",
    journal = "Phys. Rev. Lett.",
    volume = "53",
    pages = "1198--1201",
    year = "1984"
}

@article{Leinson:2021ety,
    author = "Leinson, Lev B.",
    title = "{Impact of axions on the Cassiopea A neutron star cooling}",
    eprint = "2105.14745",
    archivePrefix = "arXiv",
    primaryClass = "hep-ph",
    doi = "10.1088/1475-7516/2021/09/001",
    journal = "JCAP",
    volume = "09",
    pages = "001",
    year = "2021"
}

@article{Beznogov:2018fda,
    author = "Beznogov, Mikhail V. and Rrapaj, Ermal and Page, Dany and Reddy, Sanjay",
    title = "{Constraints on Axion-like Particles and Nucleon Pairing in Dense Matter from the Hot Neutron Star in HESS J1731-347}",
    eprint = "1806.07991",
    archivePrefix = "arXiv",
    primaryClass = "astro-ph.HE",
    reportNumber = "INT-PUB-18-029",
    doi = "10.1103/PhysRevC.98.035802",
    journal = "Phys. Rev. C",
    volume = "98",
    number = "3",
    pages = "035802",
    year = "2018"
}

@article{Leinson:2019cqv,
    author = "Leinson, Lev B.",
    title = "{Constraints on axions from neutron star in HESS J1731-347}",
    eprint = "1909.03941",
    archivePrefix = "arXiv",
    primaryClass = "hep-ph",
    doi = "10.1088/1475-7516/2019/11/031",
    journal = "JCAP",
    volume = "11",
    pages = "031",
    year = "2019"
}

@article{Keller:2012yr,
    author = "Keller, Jochen and Sedrakian, Armen",
    title = "{Axions from cooling compact stars}",
    eprint = "1205.6940",
    archivePrefix = "arXiv",
    primaryClass = "astro-ph.CO",
    doi = "10.1016/j.nuclphysa.2012.11.004",
    journal = "Nucl. Phys. A",
    volume = "897",
    pages = "62--69",
    year = "2013"
}

@article{Sedrakian:2018kdm,
    author = "Sedrakian, Armen",
    title = "{Axion cooling of neutron stars. II. Beyond hadronic axions}",
    eprint = "1810.00190",
    archivePrefix = "arXiv",
    primaryClass = "astro-ph.HE",
    doi = "10.1103/PhysRevD.99.043011",
    journal = "Phys. Rev. D",
    volume = "99",
    number = "4",
    pages = "043011",
    year = "2019"
}

@article{Hamaguchi:2025ztd,
    author = "Hamaguchi, Koichi and Nagata, Natsumi and Zheng, Jiaming",
    title = "{Axion emission from proton Cooper pairs in neutron stars}",
    eprint = "2502.18931",
    archivePrefix = "arXiv",
    primaryClass = "hep-ph",
    doi = "10.1088/1475-7516/2025/06/038",
    journal = "JCAP",
    volume = "06",
    pages = "038",
    year = "2025"
}

@article{Buschmann:2021juv,
    author = "Buschmann, Malte and Dessert, Christopher and Foster, Joshua W. and Long, Andrew J. and Safdi, Benjamin R.",
    title = "{Upper Limit on the QCD Axion Mass from Isolated Neutron Star Cooling}",
    eprint = "2111.09892",
    archivePrefix = "arXiv",
    primaryClass = "hep-ph",
    doi = "10.1103/PhysRevLett.128.091102",
    journal = "Phys. Rev. Lett.",
    year = "2022",
    volume = "128",
    number = "9",
    pages = "091102"
}

@book{lindhard1954properties,
  title={On the properties of a gas of charged particles},
  author={Lindhard, J.},
  series={Kgl. Danske Videnskabernes Selskab, Matematisk-fysiske Meddelelser},
  url={https://books.google.it/books?id=yfC9tAEACAAJ},
  year={1954},
  publisher={Munksgaard},
  note="\href{https://gymarkiv.sdu.dk/MFM/kdvs/mfm%2020-29/mfm-28-8.pdf}{{\em Dan. Mat. Fys. Medd.} {\bf 28} (1954) 1--57}"
}

@article{Haskell:2017lkl,
    author = "Haskell, Brynmor and Sedrakian, Armen",
    title = "{Superfluidity and Superconductivity in Neutron Stars}",
    eprint = "1709.10340",
    archivePrefix = "arXiv",
    primaryClass = "astro-ph.HE",
    doi = "10.1007/978-3-319-97616-7_8",
    journal = "Astrophys. Space Sci. Libr.",
    volume = "457",
    pages = "401--454",
    year = "2018"
}

@article{Kaminker:2001eu,
    author = "Kaminker, A. D. and Yakovlev, D. G. and Gnedin, Oleg Y.",
    title = "{Three types of cooling superfluid neutron stars: theory and observations}",
    eprint = "astro-ph/0111429",
    archivePrefix = "arXiv",
    doi = "10.1051/0004-6361:20011797",
    journal = "Astron. Astrophys.",
    volume = "383",
    pages = "1076",
    year = "2002"
}

@ARTICLE{1966ApJ...145..834W,
       author = {{Wolf}, Richard A.},
        title = "{Some Effects of the Strong Interactions on the Properties of Neutron-Star Matter}",
      journal = {\apj},
         year = 1966,
        month = sep,
       volume = {145},
        pages = {834},
          doi = {10.1086/148829},
       adsurl = {https://ui.adsabs.harvard.edu/abs/1966ApJ...145..834W}
}

@article{Meissner:1933ela,
    author = "Meissner, W. and Ochsenfeld, R.",
    title = {{Ein neuer Effekt bei Eintritt der Supraleitf{\"a}higkeit}},
    doi = "10.1007/BF01504252",
    journal = "Naturwiss.",
    volume = "21",
    number = "44",
    pages = "787--788",
    year = "1933"
}

@BOOK{1976itss.book.....K,
       author = {{Kittel}, Charles},
        title = "{Introduction to Solid State Physics}",
         year = 1976,
         publisher="Wiley",
       adsurl = {https://ui.adsabs.harvard.edu/abs/1976itss.book.....K}
}

@article{Guo:2018jpw,
    author = "Guo, Wenmei and Dong, J. M. and Shang, X. and Zhang, H. F. and Zuo, W. and Colonna, M. and Lombardo, U.",
    title = "{Proton-proton $^1S_0$ pairing in neutron star}",
    eprint = "1810.02709",
    archivePrefix = "arXiv",
    primaryClass = "nucl-th",
    doi = "10.1016/j.nuclphysa.2019.02.008",
    journal = "Nucl. Phys. A",
    volume = "986",
    pages = "18--25",
    year = "2019"
}

@article{Chamel:2017wwp,
    author = "Chamel, N.",
    title = "{Superfluidity and Superconductivity in Neutron Stars}",
    eprint = "1709.07288",
    archivePrefix = "arXiv",
    primaryClass = "astro-ph.HE",
    doi = "10.1007/s12036-017-9470-9",
    journal = "J. Astrophys. Astron.",
    volume = "38",
    pages = "43",
    year = "2017"
}

@article{Grifols:1988fv,
    author = "Grifols, J. A. and Mass\'o, E. and Peris, S.",
    title = "{Energy Loss From the Sun and Red Giants: Bounds on Short Range Baryonic and Leptonic Forces}",
    reportNumber = "UAB-FT-161-REV, UAB-FT-161",
    doi = "10.1142/S0217732389000381",
    journal = "Mod. Phys. Lett. A",
    volume = "4",
    pages = "311",
    year = "1989"
}

@article{Fiorillo:2025zzx,
    author = "Fiorillo, Damiano F. G. and Lella, Alessandro and O'Hare, Ciaran A. J. and Vitagliano, Edoardo",
    title = "{Leading Bounds on Micrometer to Picometer Fifth Forces from Neutron Star Cooling}",
    eprint = "2506.19906",
    archivePrefix = "arXiv",
    primaryClass = "hep-ph",
    reportNumber = "BARI-TH/776-25",
    doi = "10.1103/tlqz-713s",
    journal = "Phys. Rev. Lett.",
    volume = "135",
    number = "21",
    pages = "211003",
    year = "2025"
}

@article{Fiorillo:2026new,
    author = "Fiorillo, Damiano F. G. and Lella, Alessandro and Raffelt, Georg G. and Selimovi{\'c}, Nud{\v z}eim and  Vitagliano, Edoardo",
    title = "{Neutron Star Bounds on Muonic Fifth Forces from Picometer to Meter Scales}",
    eprint = "2605.24094",
    archivePrefix = "arXiv",
    primaryClass = "hep-ph",
    month = "5",
    year = "2026"
}

@article{Capozzi:2020cbu,
    author = "Capozzi, Francesco and Raffelt, Georg",
    title = "{Axion and neutrino bounds improved with new calibrations of the tip of the red-giant branch using geometric distance determinations}",
    eprint = "2007.03694",
    archivePrefix = "arXiv",
    primaryClass = "astro-ph.SR",
    reportNumber = "MPP-2020-106",
    doi = "10.1103/PhysRevD.102.083007",
    journal = "Phys. Rev. D",
    volume = "102",
    number = "8",
    pages = "083007",
    year = "2020"
}

@article{Hardy:2016kme,
    author = "Hardy, Edward and Lasenby, Robert",
    title = "{Stellar cooling bounds on new light particles: plasma mixing effects}",
    eprint = "1611.05852",
    archivePrefix = "arXiv",
    primaryClass = "hep-ph",
    doi = "10.1007/JHEP02(2017)033",
    journal = "JHEP",
    volume = "02",
    pages = "033",
    year = "2017"
}

@book{Raffelt:1996wa,
  author    = {Georg G. Raffelt},
  title     = {Stars as Laboratories for Fundamental Physics},
  publisher = {University of Chicago Press},
  address   = {Chicago},
  year      = {1996},
  isbn      = {0-226-70272-3}
}

@article{Bottaro:2023gep,
    author = "Bottaro, Salvatore and Caputo, Andrea and Raffelt, Georg and Vitagliano, Edoardo",
    title = "{Stellar limits on scalars from electron-nucleus bremsstrahlung}",
    eprint = "2303.00778",
    archivePrefix = "arXiv",
    primaryClass = "hep-ph",
    reportNumber = "CERN-TH-2023-035",
    doi = "10.1088/1475-7516/2023/07/071",
    journal = "JCAP",
    volume = "07",
    pages = "071",
    year = "2023"
}

@article{Yakovlev:2004iq,
    author = "Yakovlev, Dima G. and Pethick, C. J.",
    title = "{Neutron star cooling}",
    eprint = "astro-ph/0402143",
    archivePrefix = "arXiv",
    doi = "10.1146/annurev.astro.42.053102.134013",
    journal = "Ann. Rev. Astron. Astrophys.",
    volume = "42",
    pages = "169--210",
    year = "2004"
}

@software{2016ascl.soft09009P,
       author = {{Page}, Dany},
        title = "{NSCool: Neutron star cooling code}",
 howpublished = {Astrophysics Source Code Library, record ascl:1609.009},
         year = 2016,
        month = sep,
          eid = {ascl:1609.009},
       adsurl = {https://ui.adsabs.harvard.edu/abs/2016ascl.soft09009P}
}

@article{Friman:1979ecl,
    author = "Friman, B. L. and Maxwell, O. V.",
    title = "{Neutron Star Neutrino Emissivities}",
    doi = "10.1086/157313",
    journal = "Astrophys. J.",
    volume = "232",
    pages = "541--557",
    year = "1979"
}

@article{Bottaro:2024ugp,
    author = "Bottaro, Salvatore and Caputo, Andrea and Fiorillo, Damiano F. G.",
    title = "{Neutrino emission in cold neutron stars: Bremsstrahlung and modified urca rates reexamined}",
    eprint = "2406.18640",
    archivePrefix = "arXiv",
    primaryClass = "hep-ph",
    reportNumber = "CERN-TH-2024-092",
    doi = "10.1088/1475-7516/2024/11/015",
    journal = "JCAP",
    volume = "11",
    pages = "015",
    year = "2024"
}

@article{Hamaguchi:2018oqw,
    author = "Hamaguchi, Koichi and Nagata, Natsumi and Yanagi, Keisuke and Zheng, Jiaming",
    title = "{Limit on the Axion Decay Constant from the Cooling Neutron Star in Cassiopeia A}",
    eprint = "1806.07151",
    archivePrefix = "arXiv",
    primaryClass = "hep-ph",
    reportNumber = "UT-18-13, IPMU 18-0111, IPMU-18-0111",
    doi = "10.1103/PhysRevD.98.103015",
    journal = "Phys. Rev. D",
    volume = "98",
    number = "10",
    pages = "103015",
    year = "2018"
}

@article{Sedrakian:2015krq,
    author = "Sedrakian, Armen",
    title = "{Axion cooling of neutron stars}",
    eprint = "1512.07828",
    archivePrefix = "arXiv",
    primaryClass = "astro-ph.HE",
    doi = "10.1103/PhysRevD.93.065044",
    journal = "Phys. Rev. D",
    volume = "93",
    number = "6",
    pages = "065044",
    year = "2016"
}

@article{Leinson:2014ioa,
    author = "Leinson, L. B.",
    title = "{Axion mass limit from observations of the neutron star in Cassiopeia A}",
    eprint = "1405.6873",
    archivePrefix = "arXiv",
    primaryClass = "hep-ph",
    doi = "10.1088/1475-7516/2014/08/031",
    journal = "JCAP",
    volume = "08",
    pages = "031",
    year = "2014"
}

@article{Bollig:2020xdr,
    author = "Bollig, Robert and DeRocco, William and Graham, Peter W. and Janka, Hans-Thomas",
    title = "{Muons in Supernovae: Implications for the Axion-Muon Coupling}",
    eprint = "2005.07141",
    archivePrefix = "arXiv",
    primaryClass = "hep-ph",
    doi = "10.1103/PhysRevLett.125.051104",
    journal = "Phys. Rev. Lett.",
    volume = "125",
    number = "5",
    pages = "051104",
    year = "2020",
    note = "Erratum \href{https://doi.org/10.1103/PhysRevLett.126.189901}{{\em Phys. Rev. Lett.} {\bf 126}, 189901 (2021)}"
}

@article{Alda:2024cxn,
    author = "Alda, Jorge and Levati, Gabriele and Paradisi, Paride and Rigolin, Stefano and Selimovi\'c, Nudzeim",
    title = "{Collider and astrophysical signatures of light scalars with enhanced {\ensuremath{\tau}} couplings}",
    eprint = "2407.18296",
    archivePrefix = "arXiv",
    primaryClass = "hep-ph",
    doi = "10.1007/JHEP06(2025)008",
    journal = "JHEP",
    volume = "06",
    pages = "008",
    year = "2025"
}

@article{Caputo:2021rux,
    author = "Caputo, Andrea and Raffelt, Georg and Vitagliano, Edoardo",
    title = "{Muonic boson limits: Supernova redux}",
    eprint = "2109.03244",
    archivePrefix = "arXiv",
    primaryClass = "hep-ph",
    reportNumber = "MPP-2021-154",
    doi = "10.1103/PhysRevD.105.035022",
    journal = "Phys. Rev. D",
    volume = "105",
    number = "3",
    pages = "035022",
    year = "2022"
}

@article{Ferreira:2025qui,
    author = "Ferreira, Ricardo Z. and Marsh, M. C. David and Ravensburg, Eike",
    title = "{ALP couplings to muons and electrons: a comprehensive analysis of supernova bounds}",
    eprint = "2510.14469",
    archivePrefix = "arXiv",
    primaryClass = "hep-ph",
    month = "10",
    year = "2025"
}

@article{Carenza:2024ehj,
    author = "Carenza, Pierluca and Giannotti, Maurizio and Isern, Jordi and Mirizzi, Alessandro and Straniero, Oscar",
    title = "{Axion astrophysics}",
    eprint = "2411.02492",
    archivePrefix = "arXiv",
    primaryClass = "hep-ph",
    reportNumber = "BARI-TH/66-24",
    doi = "10.1016/j.physrep.2025.02.002",
    journal = "Phys. Rept.",
    volume = "1117",
    pages = "1--102",
    year = "2025"
}

\end{document}